\documentclass[aps,prd,preprintnumbers,superscriptaddress,nofootinbib]{revtex4}

\usepackage{hyperref}
\hypersetup{colorlinks=true,linkcolor=purple,anchorcolor=blue,citecolor=blue, filecolor=blue,urlcolor=red,bookmarksnumbered=true,
pdfview=FitB
}
\usepackage{color}
\usepackage{xcolor}
\colorlet{purple1}{blue!70!red}
\colorlet{darkred}{red!50!black}

\usepackage{graphicx}
\usepackage[bf,SL,BF]{subfigure}
\usepackage{psfrag}
\usepackage{color}
\usepackage{amssymb}
\usepackage{amsmath}
\usepackage{epstopdf}
\usepackage{bbold}
\usepackage[normalem]{ulem}

\newcommand{\be}{\begin{eqnarray}}
\newcommand{\ee}{\end{eqnarray}}

\newcommand{\bfb}{{\bf b}_{\perp}}

\newcommand{\bfp}{{\bf p}_{\perp}}  
\newcommand{\pp}{{p}_{\perp}}  
\newcommand{\dpi}{{\Delta}_{\perp}}  
\newcommand{\bfd}{{\bf \Delta}_{\perp}}

\newcommand{\bfD}{{\bf D}_{\perp}}

\newcommand{\Aodp}{A^\nu_1(x^{\prime\prime})}
\newcommand{\Aop}{A^\nu_1(x^{\prime})}
\newcommand{\Atdp}{A^\nu_2(x^{\prime\prime})}
\newcommand{\Atp}{A^\nu_2(x^{\prime})}
\newcommand{\zf}{\sqrt{1-\xi^2}}
\newcommand{\zfs}{(1-\xi^2)}
\newcommand{\exf}{\exp\big[-\tilde{a}(x^{\prime\prime})\bfp^{\prime\prime 2}-\tilde{a}(x^{\prime})\bfp^{\prime 2} \big]}

\begin{document}

\title{
Leading twist GTMDs at nonzero skewness and Wigner distributions in boost-invariant longitudinal position space
}

\author{Tanmay~Maji}
\email{tanmay@fudan.edu.cn} 
\affiliation{Key Laboratory of Nuclear Physics and Ion-beam Application (MOE) and Institute of Modern Physics, Fudan University, Shanghai, China 200433}

\author{Chandan~Mondal}
\email{mondal@impcas.ac.cn} 
\affiliation{Institute of Modern Physics, Chinese Academy of Sciences, Lanzhou 730000, China}
\affiliation{School of Nuclear Science and Technology, University of Chinese Academy of Sciences, Beijing 100049, China}

\author{Daekyoung~Kang}
\email{dkang@fudan.edu.cn} 
\affiliation{Key Laboratory of Nuclear Physics and Ion-beam Application (MOE) and Institute of Modern Physics, Fudan University, Shanghai, China 200433}

\date{\today}

\begin{abstract}

We investigate the leading twist quark generalized transverse momentum distributions (GTMDs) at nonzero skewness in a light-front quark-diquark model for the nucleon motivated by soft-wall AdS/QCD. The boost-invariant longitudinal coordinate, $\sigma=\frac{1}{2} b^- P^+$, is identified as the Fourier conjugate of the skewness. The Fourier transform of the GTMDs with respect to  the skewness variable $\xi$ can be employed to provide the Wigner distributions in the boost-invariant longitudinal position space $\sigma$, the coordinate conjugate to light-front time, $\tau=t+z/c$. The Wigner distributions in the longitudinal position space exhibit diffraction patterns, which are analogous to the diffractive scattering of a wave in optics.



\end{abstract}

\maketitle

\section{Introduction\label{intro}}
A key tool for revealing hadronic structure is the deep
inelastic scattering (DIS) process, where individual quarks and gluons, together known as partons, are
resolved. One can extract the parton distribution functions (PDFs)~\cite{Collins:1981uw, Martin:1998sq, Gluck:1994uf, Gluck:1998xa} from such process.  The PDFs encode the distribution of longitudinal momentum and polarizations carried by the partons. Being functions of longitudinal momentum fraction ($x$) only, they provide one dimensional picture of the hadrons. They do not give knowledge about the transverse motion and spatial location of the constituents inside the hadrons. A more comprehensive structural information of hadrons is encoded in the transverse momentum dependent parton distribution functions (TMDs) and the generalized parton distributions (GPDs). The TMDs appear in the description of semi-inclusive reactions like semi-inclusive deep inelastic scattering (SIDIS) and Drell-Yan process~\cite{Mulders:1995dh,Barone:2001sp,Bacchetta:2006tn,Brodsky:2002cx,Bacchetta:2017gcc}, whereas the GPDs are accessible in the description of hard exclusive reactions like deeply virtual Compton scattering (DVCS) or deeply virtual meson production (DVMP)~\cite{Ji:1996nm,Diehl:2003ny,Belitsky:2005qn,Goeke:2001tz}. Both the distributions provide us with essential information about the momentum distribution and the orbital motion of partons inside the hadrons, and allow us to draw three-dimensional pictures of the hadrons. 

Meanwhile, the entire perspective of the hadronic structure can be achieved through the Wigner distributions, the quantum-mechanical counterpart of classical phase-space distributions, that unify the momentum and the position distributions and give subtle details of the partons inside the hadron. The Winger distributions were introduced in quantum chromodynamics (QCD) by Ji~\cite{Ji:2003ak} and have been investigated extensively in recent times to
understand the multi-dimensional partonic imaging of the hadrons~\cite{Lorce:2011ni,Lorce:2011kd,Lorce:2011dv,Mukherjee:2014nya,Mukherjee:2015aja,More:2017zqq,Liu:2015eqa,Chakrabarti:2016yuw,Chakrabarti:2017teq,Chakrabarti:2019wjx,Gutsche:2016gcd,Kaur:2019lox,Kumar:2017xcm,Kanazawa:2014nha,Ma:2018ysi,Kaur:2019jow,Kaur:2019kpi,Zhang:2021tnr}. The Wigner
distributions are six-dimensional phase-space distributions, which do not have a probabilistic interpretation, but after some phase-space
reductions, they reduce to the TMDs and
the GPDs. The angular momentum of a parton can be extracted from Wigner distributions by
taking the phase-space average~\cite{Lorce:2011kd}. Through Fourier transformations, the Wigner distributions are linked to the generalized transverse momentum distributions (GTMDs), which are functions of the light-cone three momenta of the parton as well as the momentum transfer to
the hadron. They are often denoted as the  `mother distributions'  since several GTMDs, in certain kinematical limits, reduce to the TMDs and the GPDs. The physical process,
which gives access to the quark GTMDs is the exclusive double Drell–Yan process~\cite{Bhattacharya:2017bvs}, while the gluon GTMDs are measurable in diffractive di-jet production in deep-inelastic lepton-nucleon and lepton-nucleus scattering~\cite{Hatta:2016dxp,Ji:2016jgn,Hatta:2016aoc,Bhattacharya:2022vvo} and ultra-peripheral proton-nucleus collisions~\cite{Hagiwara:2017fye}, as well as in virtual photon-nucleus quasi-elastic scattering~\cite{Zhou:2016rnt}. 

At leading-twist, there are sixteen GTMDs for the nucleon. They are characterized by different spin-orbit and spin-spin correlations between the nucleon and a parton inside the nucleon. Two of the GTMDs, $F_{1,4}$ and $G_{1,1}$~\cite{Meissner:2009ww,Kanazawa:2014nha}, play an important role in understanding the nucleon spin structure and describe the strength of spin-orbit interactions similar to spin-orbit interactions in atomic systems like hydrogen~\cite{Lorce:2011kd,Lorce:2014mxa}. The first complete classification of  various parton distributions and their connection with the GTMDs and/or the Wigner distributions has been reported in Refs.~\cite{Meissner:2008ay,Meissner:2009ww}. Regarding the GTMDs and the Wigner distributions of spin-$1/2$ composite systems, notable analyses exist, using different theoretical models, e.g., in the light-cone constituent quark model~\cite{Lorce:2011ni,Lorce:2011kd,Lorce:2011dv}, the light-front dressed quark model~\cite{Mukherjee:2014nya,Mukherjee:2015aja,More:2017zqq}, the chiral soliton model~\cite{Lorce:2011ni,Lorce:2011kd}, light-cone spectator model~\cite{Liu:2015eqa}, the light-front quark-diquark model~\cite{Chakrabarti:2016yuw,Chakrabarti:2017teq,Chakrabarti:2019wjx,Gutsche:2016gcd,Kaur:2019lox,Kumar:2017xcm}, quark target model~\cite{Kanazawa:2014nha}, etc.  These distributions for spin-$0$ hadrons have also been investigated using different theoretical approaches~\cite{Ma:2018ysi,Kaur:2019jow,Kaur:2019kpi,Zhang:2021tnr}. Meanwhile, the scale evolution of the GTMDs has been studied in Refs.~\cite{Echevarria:2016mrc, Mukherjee:2014nya}.

It is well known that the skewness variable ($\xi$) represents the longitudinal momentum transfer in a physical process and in particular $\xi=0$ corresponds to the momentum transfer only in the transverse direction. It should be noted that the most of the previous analyses for the nucleon GTMDs have been made by assuming the momentum transfer in the process only in the transverse direction. However, the experiments always probe $\xi\ne 0$. Thus, it becomes desirable to develop a deeper understanding of GTMDs at nonzero skewness. In this work, we investigate all the leading twist quark GTMDs at nonzero skewness within the Dokshitzer Gribov Lipatov Altarelli Parisi (DGLAP) region in a light-front quark-diquark model (LFQDM) for the nucleon~\cite{Maji:2016yqo}. In this model, both the scalar and the axial vector diquarks are considered and the light-front wave functions (LFWFs) are constructed from the two
particle effective wave functions obtained in soft-wall Anti-de Sitter (AdS)/QCD. So far, this model has been successfully employed to describe many interesting properties of the nucleon e.g., electromagnetic form factor, PDFs, GPDs, TMDs, Wigner distributions at zero-skewness, spin asymmetries, etc.,~\cite{Maji:2016yqo,Maji:2017bcz,Maji:2017ill,Chakrabarti:2017teq,Maji:2017zbx,Maji:2017wwd,Kumar:2017dbf}. We obtain the GTMDs at nonzero skewness for unpolarized as well as longitudinally and transversely polarized nucleons. Our work is therefore suited for the direct analysis of experimental data. One can map out the Wigner distributions as the Fourier transform (FT) of the GTMDs. We then investigate the Wigner distributions in the longitudinal position space by taking the FT of the GTMDs with respect to $\xi$. We illustrate that the FT of the GTMDs in $\xi$ reveals the structure of a nucleon in a longitudinal impact parameter space, $\sigma=\frac{1}{2}b^-P^+$~\cite{Brodsky:2006in}, where the three-dimensional (3D) coordinate $\vec{b}=(b_{\perp},b^-)$ is conjugate to the momentum transfer $\vec{\Delta}$, provide a light-front image of the target nucleon in a frame-independent 3D light-front coordinate space. In this context, the DVCS amplitudes and the GPDs  of a relativistic spin-$\frac{1}{2}$ composite system in the boost-invariant longitudinal position space have been investigated in Refs.~\cite{Brodsky:2006in,Brodsky:2006ku,Chakrabarti:2008mw,Manohar:2010zm,Kumar:2015fta,Mondal:2015uha,Chakrabarti:2015ama,Mondal:2017wbf}. The results were analogous to the diffractive scattering of a wave in optics.
 
The paper is organized as follows. In Sec.~\ref{model}, we give brief introductions to the nucleon LFWFs of the quark-diquark model motivated by soft-wall AdS/
QCD. The leading twist nucleon GTMDs at nonzero skewness have been evaluated
in this model  and discussed the numerical results in Sec.~\ref{Results}. We study the Wigner distributions in the longitudinal boost-invariant space in Sec.~\ref{sec_WD_sigma}. Summary is given in Sec.~\ref{con}.

\section{Light-front quark-diquark model for nucleon\label{model}}

The proton state is written as superposition of the quark-diquark states allowed under $SU(4)$ spin-flavor symmetry as~\cite{Jakob:1997wg,Bacchetta:2008af,Maji:2016yqo} 
\be 
|P; \pm\rangle = C_S|u~ S^0\rangle^\pm + C_V|u~ A^0\rangle^\pm + C_{VV}|d~ A^1\rangle^\pm\, \label{PS_state}
\ee
where $|u~ S^0\rangle$, $|u~ A^0\rangle$  and $|d~ A^1\rangle$  are the isoscalar-scalar diquark singlet state, isoscalar-vector diquark state and isovector-vector diquark state, respectively. 

The two-particle Fock-state expansion for $J^z =\pm1/2$ with spin-0 diquark  is given by
\be
|u~ S\rangle^\pm & =& \int \frac{dx~ d^2\bfp}{2(2\pi)^3\sqrt{x(1-x)}} \bigg[ \psi^{\pm(u)}_{+}(x,\bfp)|+\frac{1}{2}~s; xP^+,\bfp\rangle \nonumber \\
 &+& \psi^{\pm(u)}_{-}(x,\bfp)|-\frac{1}{2}~s; xP^+,\bfp\rangle\bigg]\,,\label{fock_PS}
\ee
where the LFWFs $\psi^{\lambda_N(u)}_{\lambda_q}(x,\bfp)$ with nucleon helicities $\lambda_N=\pm$ and for quark $\lambda_q=\pm$; plus and minus correspond to $+\frac{1}{2}$ and $-\frac{1}{2}$, respectively, are \cite{Lepage:1980fj}
\be 
\psi^{+(u)}_+(x,\bfp)&=& N_S~ \varphi^{(u)}_{1}(x,\bfp)\,,\nonumber \\
\psi^{+(u)}_-(x,\bfp)&=& N_S\bigg(- \frac{p^1+ip^2}{xM} \bigg)\varphi^{(u)}_{2}(x,\bfp)\,, \label{LFWF_S}\\
\psi^{-(u)}_+(x,\bfp)&=& N_S \bigg(\frac{p^1-ip^2}{xM}\bigg) \varphi^{(u)}_{2}(x,\bfp)\,,\nonumber \\
\psi^{-(u)}_-(x,\bfp)&=&  N_S~ \varphi^{(u)}_{1}(x,\bfp)\,,\nonumber
\ee
and $|\lambda_q~\lambda_S; xP^+,\bfp\rangle$ represents the two-particle state having the scalar diquark  of helicity $\lambda_S=0$ (singlet). Meanwhile, the state with spin-1 diquark is expressed as  \cite{Ellis:2008in}
\be
|\nu~ A \rangle^\pm & =& \int \frac{dx~ d^2\bfp}{2(2\pi)^3\sqrt{x(1-x)}} \bigg[ \psi^{\pm(\nu)}_{++}(x,\bfp)|+\frac{1}{2}~+1; xP^+,\bfp\rangle \nonumber\\
 &+& \psi^{\pm(\nu)}_{-+}(x,\bfp)|-\frac{1}{2}~+1; xP^+,\bfp\rangle +\psi^{\pm(\nu)}_{+0}(x,\bfp)|+\frac{1}{2}~0; xP^+,\bfp\rangle \nonumber \\
 &+& \psi^{\pm(\nu)}_{-0}(x,\bfp)|-\frac{1}{2}~0; xP^+,\bfp\rangle + \psi^{\pm(\nu)}_{+-}(x,\bfp)|+\frac{1}{2}~-1; xP^+,\bfp\rangle \nonumber\\
 &+& \psi^{\pm(\nu)}_{--}(x,\bfp)|-\frac{1}{2}~-1; xP^+,\bfp\rangle  \bigg]\,,\label{fock_PS}
\ee
with $|\lambda_q~\lambda_D; xP^+,\bfp\rangle$ being the two-particle state with the axial-vector diquark helicities $\lambda_D=\pm 1,0$~(triplet).
For $J=+1/2$, the LFWFs $\psi^{\lambda_N(u)}_{\lambda_q, \lambda_D}(x,\bfp)$ are,  
\be 
\psi^{+(\nu)}_{+~+}(x,\bfp)&=& N^{(\nu)}_1 \sqrt{\frac{2}{3}} \bigg(\frac{p^1-ip^2}{xM}\bigg) \varphi^{(\nu)}_{2}(x,\bfp)\,,\nonumber \\
\psi^{+(\nu)}_{-~+}(x,\bfp)&=& N^{(\nu)}_1 \sqrt{\frac{2}{3}} \varphi^{(\nu)}_{1}(x,\bfp)\,,\nonumber \\
\psi^{+(\nu)}_{+~0}(x,\bfp)&=& - N^{(\nu)}_0 \sqrt{\frac{1}{3}} \varphi^{(\nu)}_{1}(x,\bfp)\,,\label{LFWF_Vp}\\
\psi^{+(\nu)}_{-~0}(x,\bfp)&=& N^{(\nu)}_0 \sqrt{\frac{1}{3}} \bigg(\frac{p^1+ip^2}{xM} \bigg)\varphi^{(\nu)}_{2}(x,\bfp)\,,\nonumber \\
\psi^{+(\nu)}_{+~-}(x,\bfp)&=& 0\,,\nonumber \\
\psi^{+(\nu)}_{-~-}(x,\bfp)&=&  0\,, \nonumber 
\ee
and for $J=-1/2$
\be 
\psi^{-(\nu)}_{+~+}(x,\bfp)&=& 0\,,\nonumber \\
\psi^{-(\nu)}_{-~+}(x,\bfp)&=& 0\,,\nonumber \\
\psi^{-(\nu)}_{+~0}(x,\bfp)&=& N^{(\nu)}_0 \sqrt{\frac{1}{3}} \bigg( \frac{p^1-ip^2}{xM} \bigg) \varphi^{(\nu)}_{2}(x,\bfp)\,,\label{LFWF_Vm}\\
\psi^{-(\nu)}_{-~0}(x,\bfp)&=& N^{(\nu)}_0\sqrt{\frac{1}{3}} \varphi^{(\nu)}_{1}(x,\bfp)\,,\nonumber \\
\psi^{-(\nu)}_{+~-}(x,\bfp)&=& - N^{(\nu)}_1 \sqrt{\frac{2}{3}} \varphi^{(\nu)}_{1}(x,\bfp)\,,\nonumber \\
\psi^{-(\nu)}_{-~-}(x,\bfp)&=& N^{(\nu)}_1 \sqrt{\frac{2}{3}} \bigg(\frac{p^1+ip^2}{xM}\bigg) \varphi^{(\nu)}_{2}(x,\bfp)\,,\nonumber
\ee
having flavor index $\nu=u,d$. The wave functions are normalized according to the quark counting rules \cite{Maji:2016yqo}. 
The LFWFs $\varphi^{(\nu)}_i(x,\bfp)$ are the modified form of the soft-wall AdS/QCD prediction for the two particle effective wave functions 
\be
\varphi_i^{(\nu)}(x,\bfp)=\frac{4\pi}{\kappa}\sqrt{\frac{\log(1/x)}{1-x}}x^{a_i^\nu}(1-x)^{b_i^\nu}\exp\bigg[-\delta^\nu\frac{\bfp^2}{2\kappa^2}\frac{\log(1/x)}{(1-x)^2}\bigg]\,.
\label{LFWF_phi}
\ee
The wave functions $\varphi_i^\nu ~(i=1,2)$ reduce to the original AdS/QCD wavefunction \cite{Brodsky:2007hb,deTeramond:2012rt} for the parameters $a_i^\nu=b_i^\nu=0$  and $\delta^\nu=1.0$. We use the AdS/QCD scale parameter $\kappa =0.4$ GeV~\cite{Chakrabarti:2013dda,Chakrabarti:2013gra} and the quarks are  assumed  to be  massless. The parameters of this model are determined from the fitting of the flavor decomposed Dirac and Pauli form factors data and listed in Refs.~\cite{Maji:2016yqo,Maji:2017bcz}. This model wave function with the parameters provide a reasonably good agreement with the proton electric and magnetic charge radius data as well as parton distribution data.

%
\section{GTMDs with non-zero skewness in LFQDM}\label{Results}
In this section, we present the detail calculations of the leading twist GTMDs in the LFQDM.  The bilinear decomposition of the fully unintegrated quark-quark correlator for a spin-$1/2$ hadron is presented and parameterized in terms of GTMDs in Ref.~\cite{Meissner:2009ww}. 
In the fixed light-cone time $ z^+=0$, the quark-quark correlator for GTMDs is defined as~\cite{Meissner:2008ay,Meissner:2009ww}
\be
W^{\nu [\Gamma]}_{[\lambda^{\prime\prime}\lambda^{\prime}]}(x,\xi,\bfd, \bfp)=\frac{1}{2}\int \frac{dz^-}{(2\pi)} \frac{d^2z_T}{(2\pi)^2} e^{ip.z} 
\langle P^{\prime\prime}; \lambda^{\prime\prime} |\bar{\psi}^\nu (-z/2)\Gamma \mathcal{W}_{[-z/2,z/2]} \psi^\nu (z/2) |P^\prime;\lambda^{\prime}\rangle \bigg|_{z^+=0}\,,
\label{Wdef}
\ee
where $|P^\prime;\lambda^{\prime}\rangle $ and $|P^{\prime\prime}; \lambda^{\prime\prime}\rangle$ are the initial and final states of the proton with helicities $\lambda^\prime$ and $\lambda^{\prime\prime}$, respectively and $\psi\, (\bar{\psi})$ is the quark field. The $\Gamma$ denotes the leading twist Dirac $\gamma$-matrices, i.e., $\Gamma=\{\gamma^+,\, \gamma^+\gamma^5,\, i\sigma^{j+} \gamma^5\}$ corresponding to unpolarized, longitudinally polarized and transversely polarized quarks, respectively. The gauge link, $\mathcal{W}_{[-z/2,z/2]}$, ensures
the $SU(3)$ color gauge invariance of the bilocal quark operator.
Here, we follow the convention $x^\pm=(x^0 \pm x^3)$ and the kinematics are given by  
\be 
P &\equiv& \bigg(P^+,\frac{M^2+\bfd^2/4}{(1-\xi^2)P^+},\textbf{0}_\perp\bigg)\,,\\
p &\equiv& \bigg(xP^+, p^-,\bfp \bigg)\,,\\
\Delta &\equiv& \left(-2\xi P^+, \frac{t + \bfd^2}{-2 \xi P^+},\bfd \right)\,,
\ee
where the skewness is defined as $\xi=- \Delta^+/2P^+$. In the symmetric frame, the average momentum of proton $P= \frac{1}{2} (P^{\prime\prime}+P^\prime)$, while momentum transfer $\Delta=(P^{\prime\prime}-P^\prime)$. The initial and final four momenta of the proton are then given by
\be
P^{\prime} &\equiv& \bigg((1+\xi)P^+,\frac{M^2+\bfd^2/4}{(1+\xi)P^+},-\bfd/2\bigg)\,.\label{Pp}\\
P^{\prime\prime} &\equiv& \bigg((1-\xi)P^+,\frac{M^2+\bfd^2/4}{(1-\xi)P^+},\bfd/2\bigg)\,. \label{Ppp}
\ee
Note that the  square of the total momentum transfer, $t= \Delta^2$, and one can derive the following relation explicitly using $\Delta^-=( P^{\prime \prime -} - P^{\prime  -})$
\be 
- t= \frac{4 \xi^2 M^2 + \bfd^2}{(1-\xi^2)}\,. \label{mt_def}
\ee 
Here, we define  $\xi$ following the convention in Ref.~\cite{Meissner:2009ww}, which differs by a minus sign with respect to that in Ref.~\cite{Brodsky:2000xy}.  The bilinear decomposition of the quark-quark correlator, Eq.~(\ref{Wdef}), relates to the leading twist GTMDs as given in Appendix~\ref{App}.
Meanwhile, the correlator $W^{\nu [\Gamma]}_{[\lambda^{\prime\prime}\lambda^{\prime}]}$ defined in Eq.~(\ref{Wdef})  can be expressed in terms of overlaps of the LFWFs given in Eqs.~(\ref{LFWF_S}), (\ref{LFWF_Vp}), and (\ref{LFWF_Vm}).
We obtain for the scalar diquark
\be
W^{ [\gamma^+](S)}_{[\lambda^{\prime\prime} \lambda^{\prime}]}(x,\bfp,\bfd)&=&\frac{1}{16\pi^3} \sum_{\lambda_q} \psi^{\lambda^{\prime\prime}\dagger}_{\lambda_q}(x^{\prime\prime},\bfp^{\prime\prime})\psi^{\lambda^{\prime}\dagger}_{\lambda_q}(x^{\prime},\bfp^{\prime})\,, \label{WVs} \\
W^{ [\gamma^+\gamma^5](S)}_{[\lambda^{\prime\prime} \lambda^{\prime}]}(x,\bfp,\bfd)&=&\frac{1}{16\pi^3} \sum_{\lambda_q} (2 \lambda_q)~ \psi^{\lambda^{\prime\prime}\dagger}_{\lambda_q}(x^{\prime\prime},\bfp^{\prime\prime})\psi^{\lambda^{\prime}\dagger}_{\lambda_q}(x^{\prime},\bfp^{\prime})\label{WAs}\,, \\
W^{ [i\sigma^{j+}\gamma^5](S)}_{[\lambda^{\prime\prime} \lambda^{\prime}]}(x,\bfp,\bfd)&=&\frac{1}{16\pi^3} \sum_{\lambda^{\prime\prime}_q}\sum_{\lambda^\prime_q} (2 \lambda^\prime_q i)^i~ \psi^{\lambda^{\prime\prime}\dagger}_{\lambda^{\prime\prime}_q}(x^{\prime\prime},\bfp^{\prime\prime})\psi^{\lambda^{\prime}\dagger}_{\lambda^\prime_q}(x^{\prime},\bfp^{\prime})\,, \label{WTs}
\ee 
while for the axial-vector diquark
\be
W^{[\gamma^+](A)}_{[\lambda^{\prime\prime} \lambda^{\prime} ]}(x,\bfp,\bfd) &=&  \frac{1}{16\pi^3} \sum_{\lambda_q}\sum_{\lambda_D} \psi^{\lambda^{\prime\prime}\dagger}_{\lambda_q\lambda_D}(x^{\prime\prime},\bfp^{\prime\prime})\psi^{\lambda^{\prime}\dagger}_{\lambda_q\lambda_D}(x^{\prime},\bfp^{\prime})\,,\label{WVA} \\
W^{ [\gamma^+\gamma^5](A)}_{[\lambda^{\prime\prime} \lambda^{\prime} ]}(x,\bfp,\bfd) &=&  \frac{1}{16\pi^3} \sum_{\lambda_q}\sum_{\lambda_D} (2\lambda_q)  ~ \psi^{\lambda^{\prime\prime}\dagger}_{\lambda_q\lambda_D}(x^{\prime\prime},\bfp^{\prime\prime})\psi^{\lambda^{\prime}\dagger}_{\lambda_q\lambda_D}(x^{\prime},\bfp^{\prime})\,, \label{WAA}\\
W^{ [i\sigma^{j+}\gamma^5](A)}_{[\lambda^{\prime\prime} \lambda^{\prime} ]}(x,\bfp,\bfd) &=&  \frac{1}{16\pi^3} \sum_{\lambda^{\prime\prime}_q}\sum_{\lambda^{\prime}_q}\sum_{\lambda_D} \epsilon^{ij}_\perp (2\lambda^\prime_q i)^i  ~ \psi^{\lambda^{\prime\prime}\dagger}_{\lambda^{\prime\prime}_q\lambda_D}(x^{\prime\prime},\bfp^{\prime\prime})\psi^{\lambda^{\prime}\dagger}_{\lambda^{\prime}_q\lambda_D}(x^{\prime},\bfp^{\prime})\,, \label{WTA}
\ee
with the Dirac structures $\Gamma=\gamma^+,~\gamma^+\gamma^5$, and $i\sigma^{j+}\gamma^5$. 
The initial and final transverse momenta of the struck quark are given by
\be 
\bfp^{\prime}=\bfp-(1-x^\prime)\frac{\bfd}{2},& \quad & {\rm with} \quad x^{\prime}=\frac{x+\xi}{1+\xi}\,,\\
\bfp^{\prime\prime}=\bfp+(1-x^{\prime\prime})\frac{\bfd}{2}\,, & \quad &  {\rm with} \quad  x^{\prime\prime}=\frac{x-\xi}{1-\xi}\,,
\ee
respectively. With the scalar and the axial-vector diquark components, the correlator in the LFQDM model is written as
\be 
W^{\nu[\Gamma]}_{[\lambda^{\prime\prime}\lambda^\prime]}(x, \bfp, \bfd) &=& C^2_S ~W^{\nu[\Gamma](S)}_{[\lambda^{\prime\prime}\lambda^\prime]}(x, \bfp, \bfd) +  C^2_A ~W^{\nu[\Gamma](A)}_{[\lambda^{\prime\prime}\lambda^\prime]}(x, \bfp, \bfd)\,,
\ee
where, $C_A=C_V, C_{VV}$ for $u$ and $d$ quarks respectively.

Following the bilinear decomposition of the correlator given in Eqs.~(\ref{WV_def}), (\ref{WA_def}), and (\ref{WT_def}), we express the GTMDs in terms of the correlators $W^{\nu [\Gamma]}_{[\lambda^{\prime\prime}\lambda^{\prime}]}$ with proper helicity combinations and Dirac structure. Using the LFWFs  given in Eqs.~(\ref{WVs})-(\ref{WTA}),  we end up with the results of
leading twist GTMDs in the LFQDM model and the explicit expressions of the GTMDs are \\
(i) for unpolarised quark with Dirac matrix structure $\Gamma=\gamma^+$:
\be 
F^\nu_{1,1}(x,\xi,\bfd^2,\bfp^2,\bfd.\bfp) &=& N^\nu_{F11}\frac{1}{16\pi^3} \zf  \bigg[\Aodp \Aop+ \bigg\{  \bfp^2 - \frac{\bfd^2}{4}\frac{(1-x)^2}{(1-\xi^2)}  \nonumber \\
&& + \frac{\xi (1-x)}{(1-\xi^2)}(\bfp.\bfd)\bigg\} \frac{\Atdp \Atp}{x^{\prime\prime} x^{\prime} M^2}\bigg] \exf\,, \label{F11} \\
F^\nu_{1,2}(x,\xi,\bfd^2,\bfp^2,\bfd.\bfp) &=& - N^\nu_{F12}\frac{1}{16\pi^3}\frac{1}{\zf} \bigg[ \frac{\Aodp \Atp}{x^\prime} - \frac{\Atdp \Aop}{x^{\prime\prime}} \bigg] \nonumber \\
&&  \times \exf   \nonumber \\ 
&& - \frac{\bfd^2}{2M^2} \frac{\xi}{(1-\xi^2)} \frac{ N^\nu_{F12} }{ N^\nu_{F14} } F_{1,4}(x,\xi,\bfd^2,\bfp^2,\bfd.\bfp)\,,\label{F12} 
\ee
\be 
F^\nu_{1,3}(x,\xi,\bfd^2,\bfp^2,\bfd.\bfp) &=& N^\nu_{F13} \frac{1}{16\pi^3} \frac{(1-x)}{\zf}\frac{1}{2} \bigg[ \frac{\Aodp \Atp}{x^\prime(1+\xi)} + \frac{\Atdp \Aop}{x^{\prime\prime}(1-\xi)} \bigg] \nonumber \\
&&  \times \exf   \nonumber \\
&& +  \frac{1}{2(1-\xi^2)} \frac{ N^\nu_{F13} }{ N^\nu_{F11} } F_{1,1}(x,\xi,\bfd^2,\bfp^2,\bfd.\bfp) \nonumber \\
&&+ \frac{1}{2M^2} \frac{\xi}{(1-\xi^2)}(\bfp.\bfd)\frac{ N^\nu_{F13} }{ N^\nu_{F14} } F_{1,4}(x,\xi,\bfd^2,\bfp^2,\bfd.\bfp)\,,\label{F13} \\
F^\nu_{1,4}(x,\xi,\bfd^2,\bfp^2,\bfd.\bfp) &=& - N^\nu_{F14}\frac{1}{16\pi^3}  \frac{(1-x)}{\zf} \frac{1}{x^{\prime\prime}x^\prime}\Atdp\Atp \exf\,, \nonumber \\
\label{F14} 
\ee
(ii) for longitudinally polarized quark with Dirac matrix structure $\Gamma=\gamma^+ \gamma^5$: 
\be
G^\nu_{1,1}(x,\xi,\bfd^2,\bfp^2,\bfd.\bfp) &=&  - N^\nu_{G11}\frac{1}{16\pi^3}  \frac{(1-x)}{\zf} \frac{1}{x^{\prime\prime}x^\prime}\Atdp\Atp \exf\,,\label{G11}\\
G^\nu_{1,2}(x,\xi,\bfd^2,\bfp^2,\bfd.\bfp) &=&  N^\nu_{G12}\frac{1}{16\pi^3}  \frac{1}{\zf} \bigg[ \frac{\Aodp \Atp}{x^\prime} + \frac{\Atdp \Aop}{x^{\prime\prime}} \bigg] \nonumber \\
&& \times \exf \,,\label{G12}\\
G^\nu_{1,3}(x,\xi,\bfd^2,\bfp^2,\bfd.\bfp) &=& N^\nu_{G13} \frac{1}{16\pi^3} \frac{\xi(1-x)}{\zfs^{3/2}} \frac{1}{2}   \bigg[ \frac{\Aodp \Atp}{x^\prime} + \frac{\Atdp \Aop}{x^{\prime\prime}} \bigg] \nonumber \\
&& \times \exf ] \nonumber \\ 
&& + \frac{\xi}{(1-\xi^2)} \frac{N^\nu_{G13}}{N^\nu_{G14}} G_{1,4}(x,\xi,\bfd^2,\bfp^2,\bfd.\bfp) \,,\label{G13}\\
G^\nu_{1,4}(x,\xi,\bfd^2,\bfp^2,\bfd.\bfp) &=& N^\nu_{G14}\frac{1}{16\pi^3} \zf  \frac{1}{2}\bigg[ \Aodp \Aop - \bigg\{ \bfp^2 - \frac{\bfd^2}{4}  \frac{(1-x)^2}{(1-\xi^2)} \nonumber \\
&& +  \frac{\xi (1-x)}{(1-\xi^2)} (\bfp.\bfd) \bigg\} \frac{\Atdp \Atp}{x^{\prime\prime}x^\prime  M^2 }  \bigg] \exf\,,\nonumber \\
\label{G14}
\ee
 (iii) for transversely polarized quark with Dirac matrix structure $\Gamma=i\sigma^{j+} \gamma^5$: 
\be
H^\nu_{1,1}(x,\xi,\bfd^2,\bfp^2,\bfd.\bfp) &=&  - N^\nu_{H11}\frac{1}{16\pi^3}  \zf \bigg[ \frac{1}{x^\prime}\Aodp\Atp - \frac{1}{x^{\prime \prime}}\Atdp\Aop \bigg] \nonumber \\
&&  \times \exf\,,\label{H11}\\
H^\nu_{1,2}(x,\xi,\bfd^2,\bfp^2,\bfd.\bfp) &=& N^\nu_{H12}\frac{1}{16\pi^3}  \zf \frac{1}{2}\bigg[ \frac{1-x^\prime}{x^\prime}\Aodp\Atp + \frac{1-x^{\prime \prime}}{x^{\prime \prime}}\Atdp\Aop \bigg] \nonumber \\
&&  \times \exf \,,\label{H12}
\ee
\be 
H^\nu_{1,3}(x,\xi,\bfd^2,\bfp^2,\bfd.\bfp) &=&N^\nu_{H13}\frac{1}{16\pi^3} \frac{1}{\zf} \bigg[ \Aodp \Aop \nonumber\\
&& + \bigg(\bfp^2 - \frac{(1-x)^2}{(1-\xi^2)} \frac{\bfd^2}{4} \bigg) \frac{\Atdp \Atp}{x^\prime x^{\prime\prime} M^2}  \bigg] \exf  \nonumber \\
&& -  \frac{1}{2M^2(1-\xi^2)}\frac{N^\nu_{H13} }{N^\nu_{H12} }  \bfd^2 H_{1,2}(x,\xi,\bfd^2,\bfp^2,\bfd.\bfp) \,,\label{G13}\\
H^\nu_{1,4}(x,\xi,\bfd^2,\bfp^2,\bfd.\bfp) &=& -N^\nu_{H14}\frac{1}{16\pi^3}  \frac{1}{ \zf} \frac{2}{x^{\prime\prime}x^\prime} \Atdp \Atp  \exf\,,\nonumber \\ \label{H14}
\ee
\be 
H^\nu_{1,5}(x,\xi,\bfd^2,\bfp^2,\bfd.\bfp) &=& N^\nu_{H15} \frac{1}{16\pi^3} \bigg[ \frac{\xi}{ \zfs^{3/2}} \frac{(1-x)}{x^{\prime\prime}x^\prime} \Atdp \Atp \bigg] \nonumber \\
&& \times  \exf \nonumber \\
&& + \frac{\xi}{(1-\xi^2)} \frac{N^\nu_{H15} }{N^\nu_{H17} } H^\nu_{1,7}(x,\xi,\bfd^2,\bfp^2,\bfd.\bfp) \,, \label{H15}\\
H^\nu_{1,6}(x,\xi,\bfd^2,\bfp^2,\bfd.\bfp) &=& N^\nu_{H16} \frac{1}{16\pi^3} \bigg[ \frac{1}{ \zfs^{3/2}} \frac{(1-x)^2}{2 x^{\prime\prime}x^\prime} \Atdp \Atp \bigg] \nonumber \\
&& \times   \exf  \nonumber \\
&& + \frac{1}{2(1-\xi^2)}\frac{N^\nu_{H16} }{N^\nu_{H12} }  H^\nu_{1,2}(x,\xi,\bfd^2,\bfp^2,\bfd.\bfp)  \nonumber \\
&&  + \frac{\xi}{(1-\xi^2)}  \frac{N^\nu_{H16} }{N^\nu_{H18} } H^\nu_{1,8}(x,\xi,\bfd^2,\bfp^2,\bfd.\bfp)\,, \label{H16}\\
H^\nu_{1,7}(x,\xi,\bfd^2,\bfp^2,\bfd.\bfp) &=&  - N^\nu_{H17}\frac{1}{16\pi^3}  \zf  \frac{1}{2}\bigg[ \frac{1}{x^\prime}\Aodp\Atp + \frac{1}{x^{\prime \prime}}\Atdp\Aop \bigg] \nonumber \\
&&  \times \exf\,,\label{H17}\\
H^\nu_{1,8}(x,\xi,\bfd^2,\bfp^2,\bfd.\bfp) &=& N^\nu_{H18}\frac{1}{16\pi^3}  \zf \frac{1}{4}\bigg[ \frac{(1-x^\prime)}{x^\prime}\Aodp\Atp - \frac{(1-x^{\prime \prime})}{x^{\prime \prime}}\Atdp\Aop \bigg] \nonumber \\
&&  \times \exf \,,\label{H18}
\ee
with 
\be 
A^{\nu}_i(x)&=&\frac{4\pi}{\kappa}\sqrt{\frac{\log(1/x)}{(1-x)}}x^{a^{\nu}_i}(1-x)^{b^{\nu}_i}\,,\\
\tilde{a}(x)&=& \frac{\log(1/x)}{2\kappa^2(1-x)^2}\,\\
\tilde{q}_\perp^2(x,\xi,\bfd^2,\bfp^2,\bfd.\bfp) &=& \tilde{a}(x^{\prime\prime})\bfp^{\prime\prime 2}+\tilde{a}(x^{\prime})\bfp^{\prime 2}\,.
\ee
The normalization constants $N^\nu_{\Lambda\lambda}$ are 
\be 
N^\nu_{F11},N^\nu_{G11} N^\nu_{H11}, N^\nu_{H12} 
&=&  \bigg(C^2_SN^2_S+C^2_A \big(\frac{1}{3}N^2_0+\frac{2}{3}N^2_1\big)\bigg)^\nu\,, \nonumber\\
N^\nu_{F14}, N^\nu_{G14} ,  N^\nu_{H17}, N^\nu_{H18}
&=&\bigg(C^2_SN^2_S+C^2_A\big(\frac{1}{3}N^2_0-\frac{2}{3}N^2_1\big)\bigg)^\nu\,, \nonumber\\
N^\nu_{F12}, N^\nu_{F13}, N^\nu_{G12}, N^\nu_{G13} , N^\nu_{H13}, N^\nu_{H14}\,, N^\nu_{H15},N^\nu_{H16} 
&=&  \bigg(C^2_SN^2_S-C^2_A \frac{1}{3}N^2_0\bigg)^\nu\,,
\ee
where  $C_A=C_V, C_{VV}$  for the $u$ and $d$ quarks respectively. Note that, $N_S=0$ for $d$ quark.

There are altogether 16 GTMDs at the leading twist. At $\xi=0$ limit, $x^{\prime\prime}=x^\prime=x$ and all the expressions for the GTMDs, Eqs.~(\ref{F11})-(\ref{H18}), are consistent with the results presented in Ref.~\cite{Chakrabarti:2017teq}. At $\Delta=0$ and $\xi=0$, the GTMDs reduce to the leading twist TMDs reported in Ref.~\cite{Maji:2017bcz}. For nonzero skewness, the GTMDs $F_{1,1}$ and $G_{1,4}$, Eqs.~(\ref{F11}) and (\ref{G14}), respectively have an additional term containing $\bfp.\bfd$, which breaks the axial symmetry of the distributions in the transverse momentum plane at a fixed $\bfd$. In the GTMDs $F_{1,2}$ and $F_{1,3}$, Eqs.~(\ref{F12}) and (\ref{F13}), an additional term is found for $\xi\neq0$ that involves the GTMD $F_{1,4}$. Similarly, $G_{1,3}$ has an additional term containing $G_{1,4}$, which vanishes at $\xi=0$. 
At the GPD limit, $t=\Delta^2$  and integrating over the quark transverse momentum $\bfp$, $F_{1,1},\, F_{1,2}$, and $F_{1,3}$ contribute to the unpolarized GPDs $H$ and $E$  \cite{Meissner:2009ww}, while $G_{1,2},\, G_{1,3}$, and $G_{1,4}$ contribute to the polarized GPDs $\widetilde{H}$ and 
$\widetilde{E                                                                                                                                                                                                                                                                                                                                                                          }$ as  shown in  Appendix~\ref{App}.
%

To illustrate the numerical results of the flavor dependent GTMDs, we emphasize on the $\xi$ dependence  since the other dependencies of the GTMDs with vanishing skewness have been investigated in several   studies~\cite{Lorce:2011ni,Lorce:2011kd,Lorce:2011dv,Mukherjee:2014nya,Mukherjee:2015aja,More:2017zqq,Liu:2015eqa,Chakrabarti:2016yuw,Chakrabarti:2017teq,Chakrabarti:2019wjx,Gutsche:2016gcd}. We consider the DGLAP region, $\xi< x <1$, for our discussion.  
Here, we present the numerical results of the GTMDs for the unpolarized and longitudinally polarized quarks evaluated in Eqs.~(\ref{F11})-(\ref{G14}). These eight GTMDs are related to several physical quantities like orbital angular momentum (OAM), axial and tensor charges,  etc., and also linked to the GPDs and the TMDs in certain kinematical limits.  Meanwhile, the GTMDs for transversely polarized quark are presented in the Appendix~\ref{AppB}.

\begin{figure}[t]
\includegraphics[scale=.32]{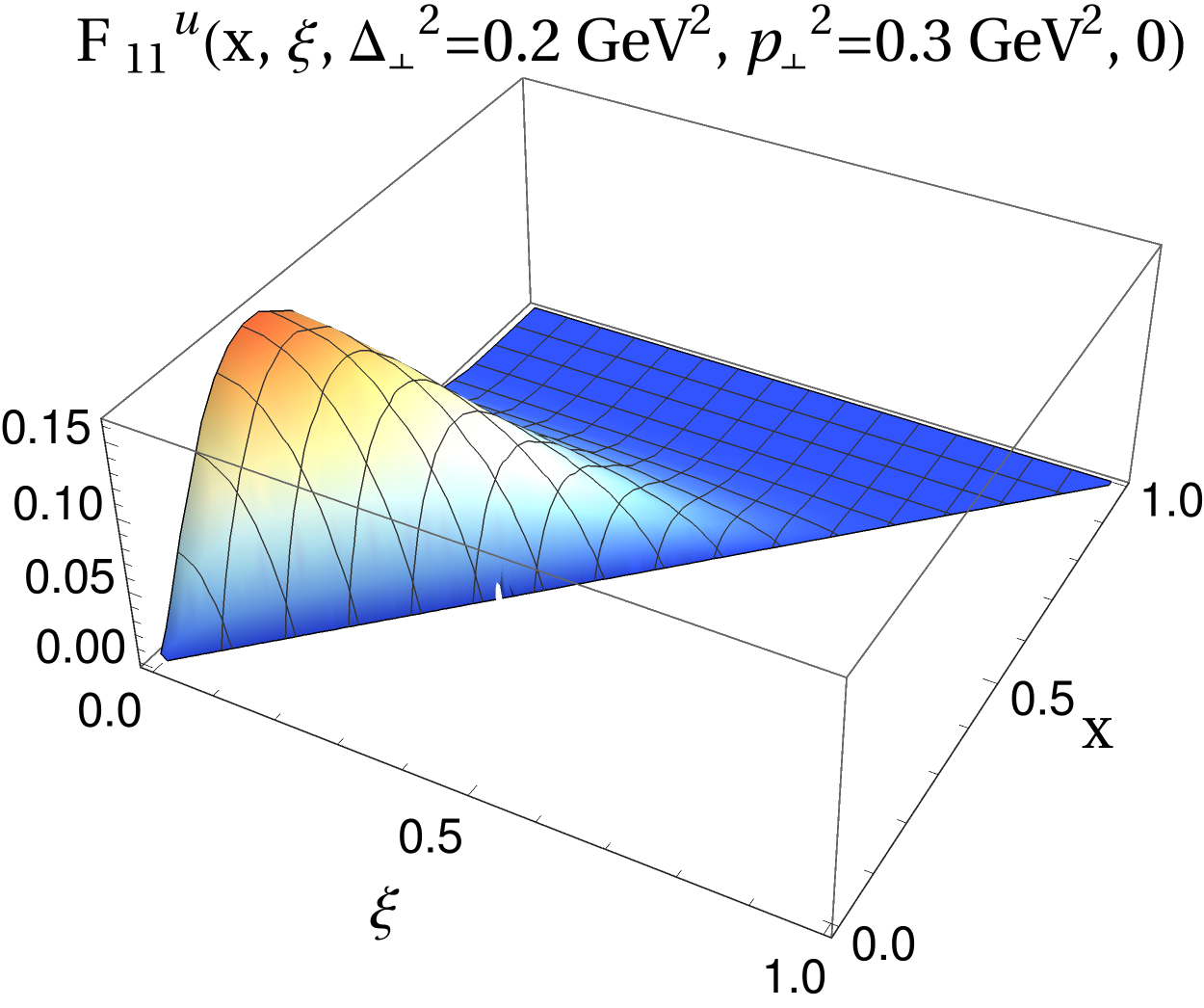}
\includegraphics[scale=.32]{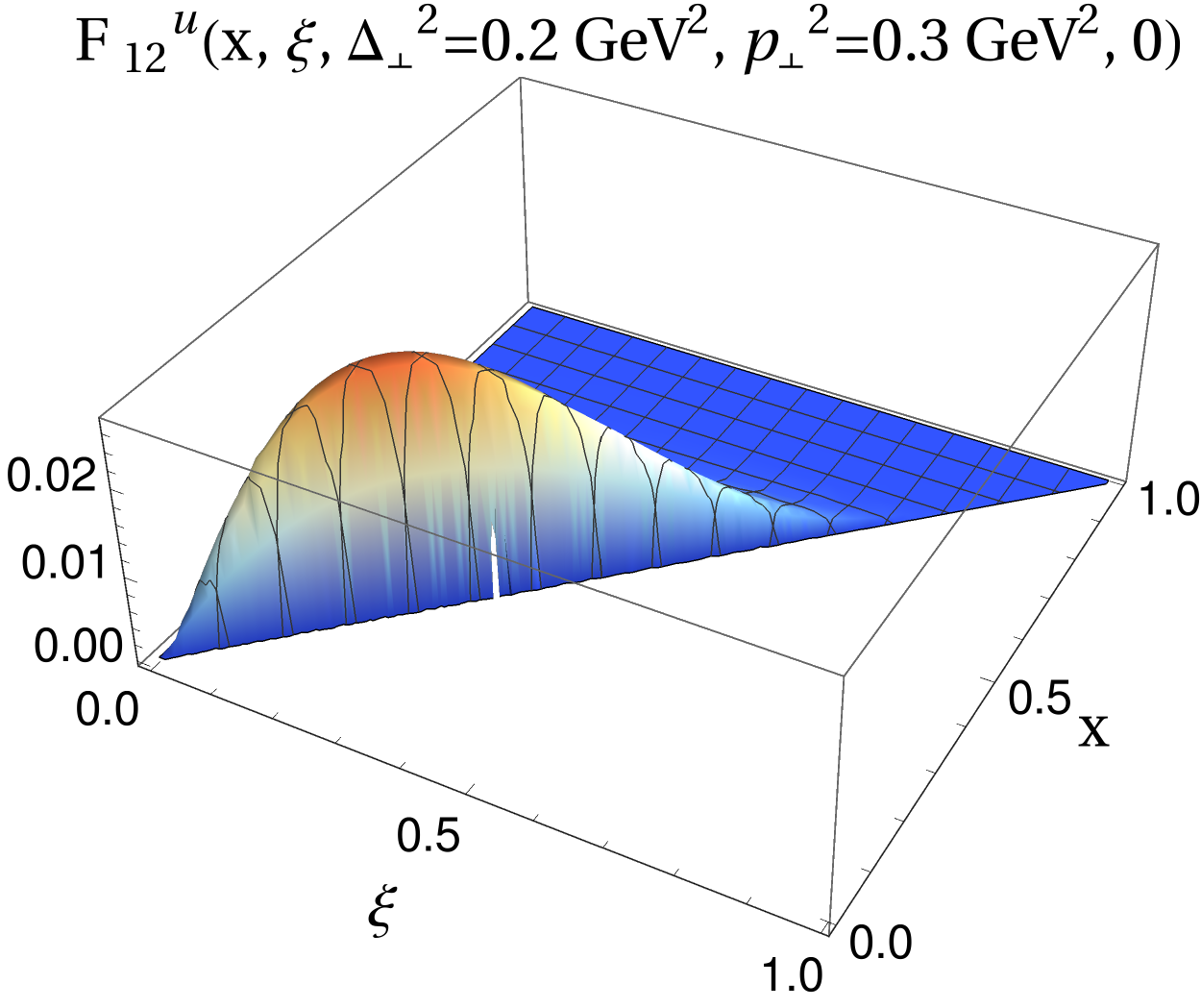} 
\includegraphics[scale=.32]{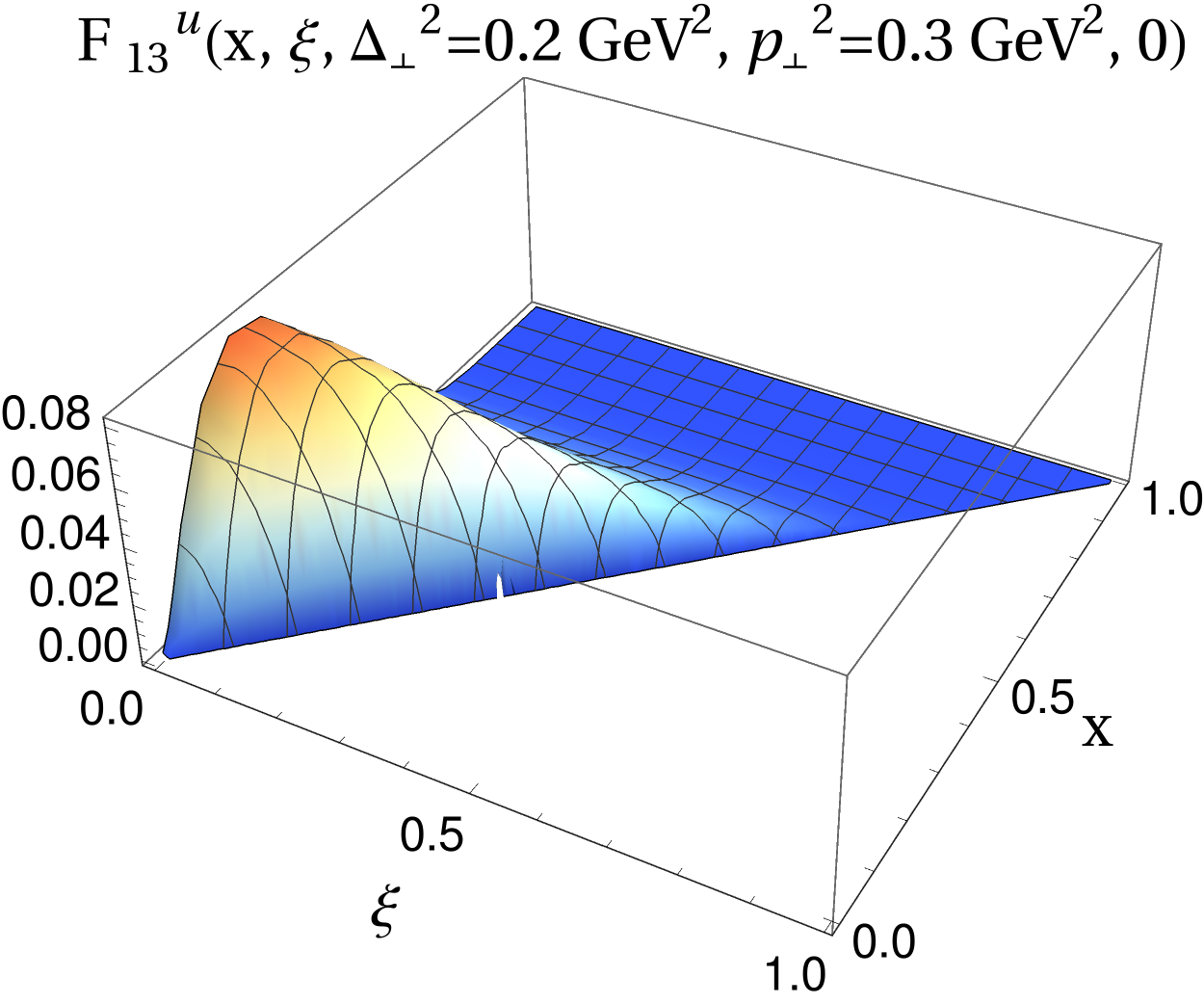} 
\includegraphics[scale=.32]{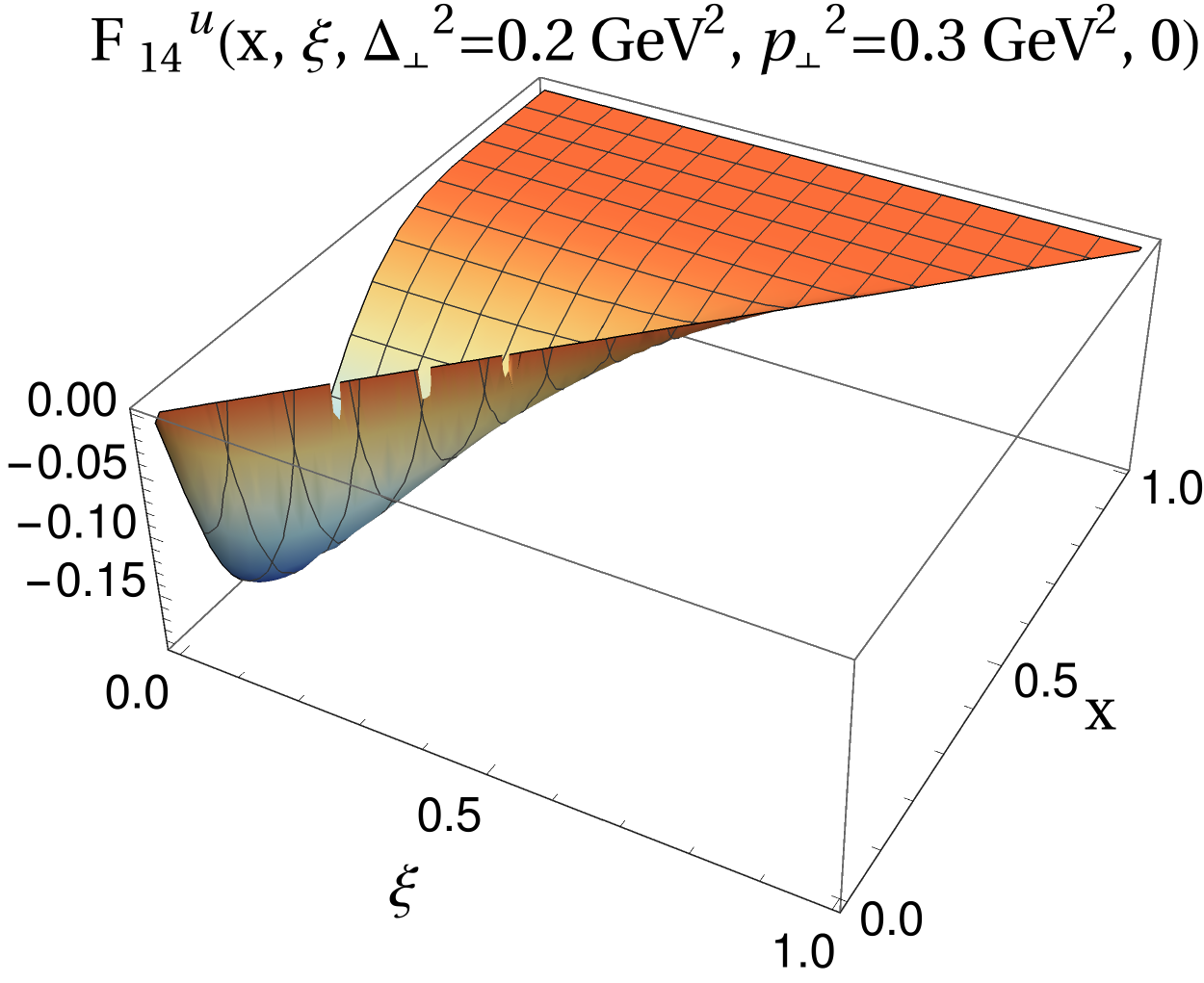} \\
\includegraphics[scale=.32]{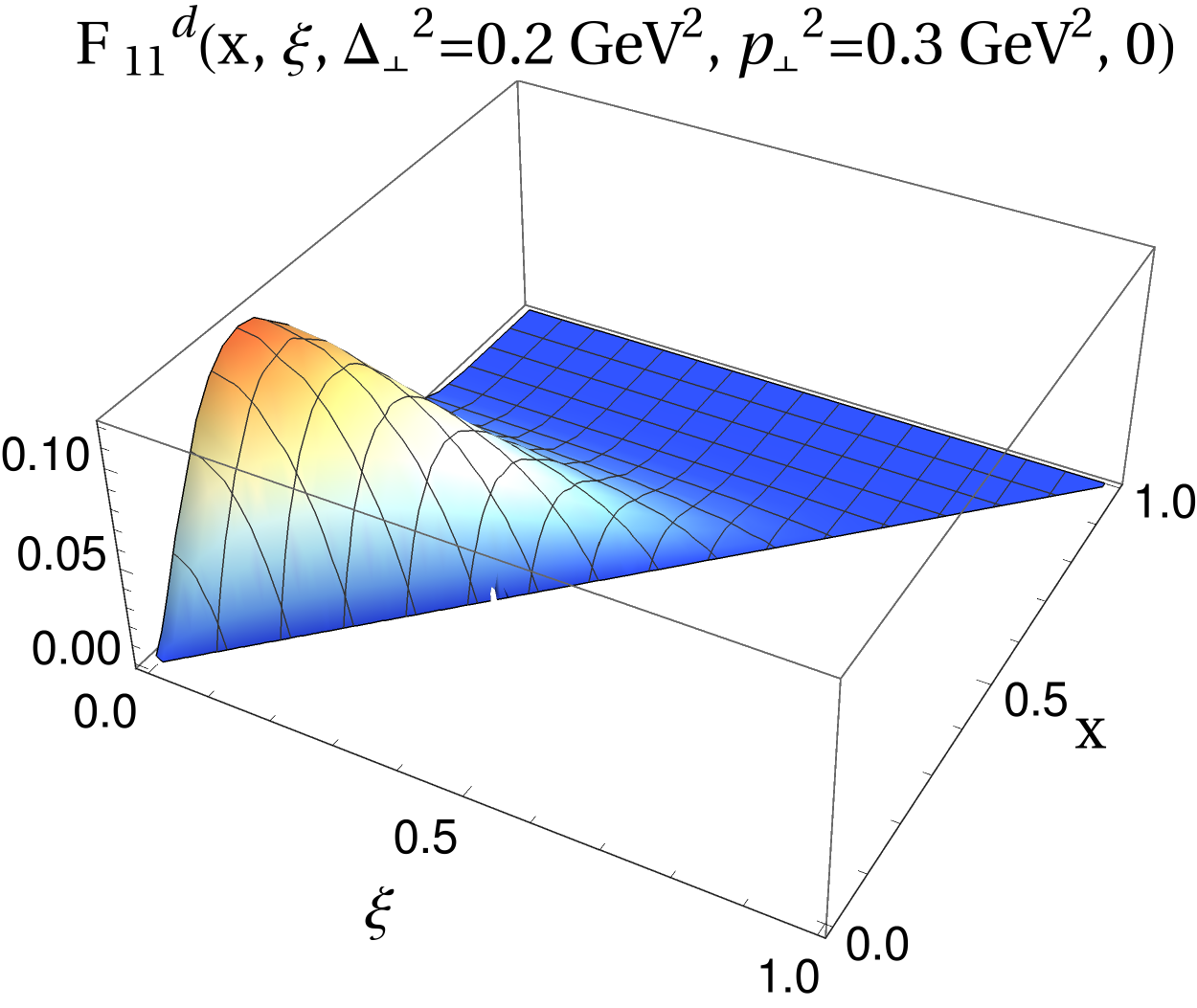} 
\includegraphics[scale=.32]{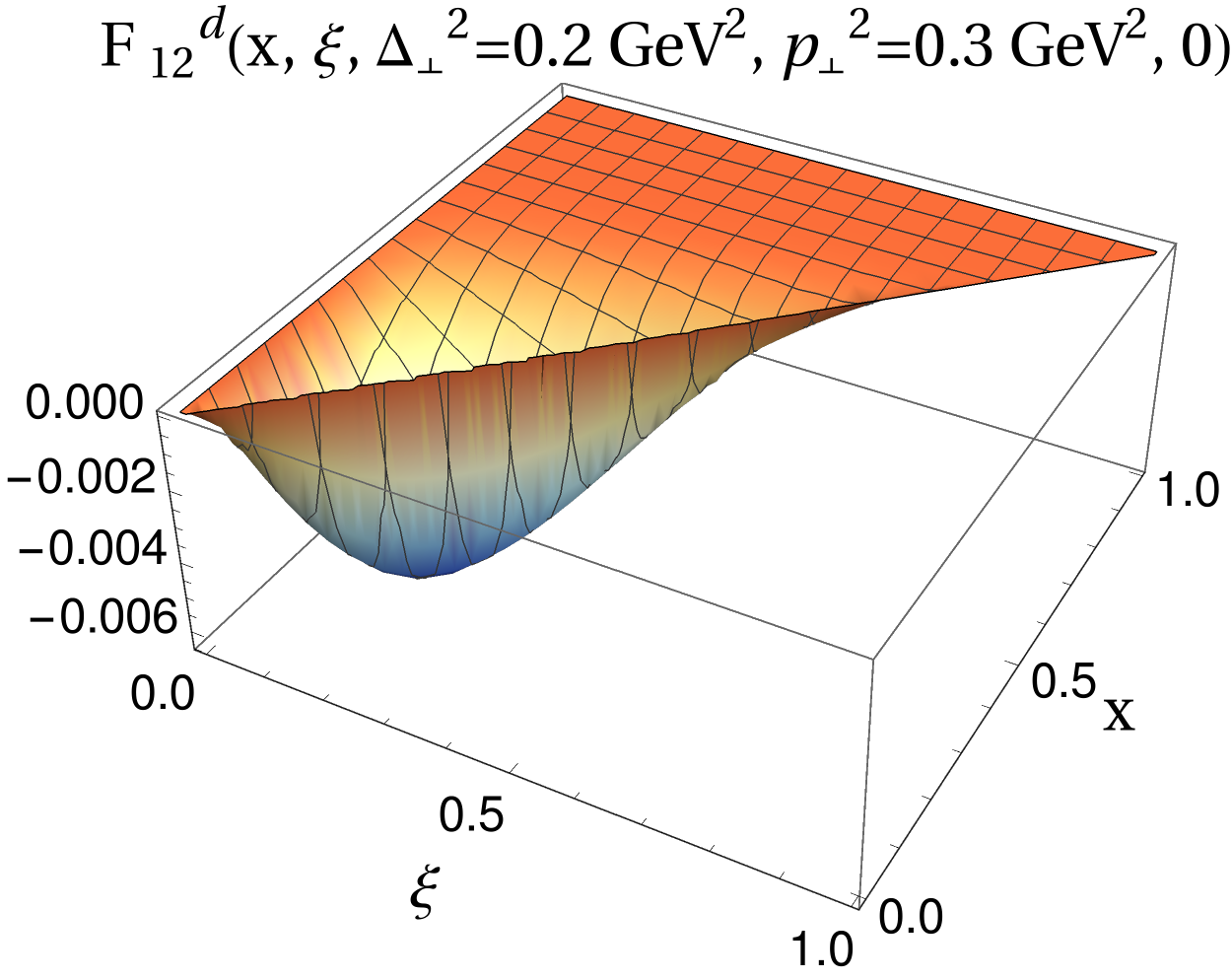} 
\includegraphics[scale=.32]{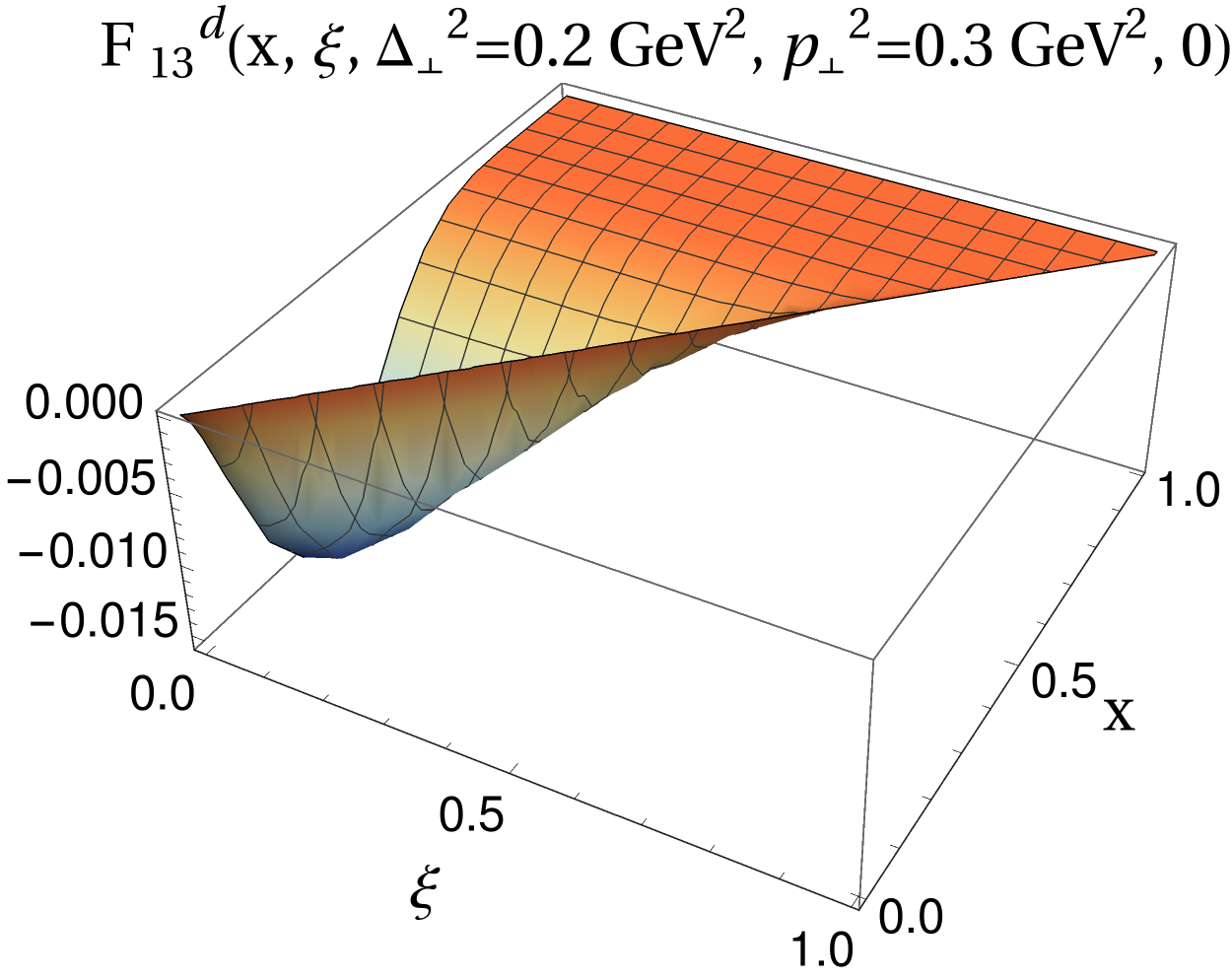} 
\includegraphics[scale=.32]{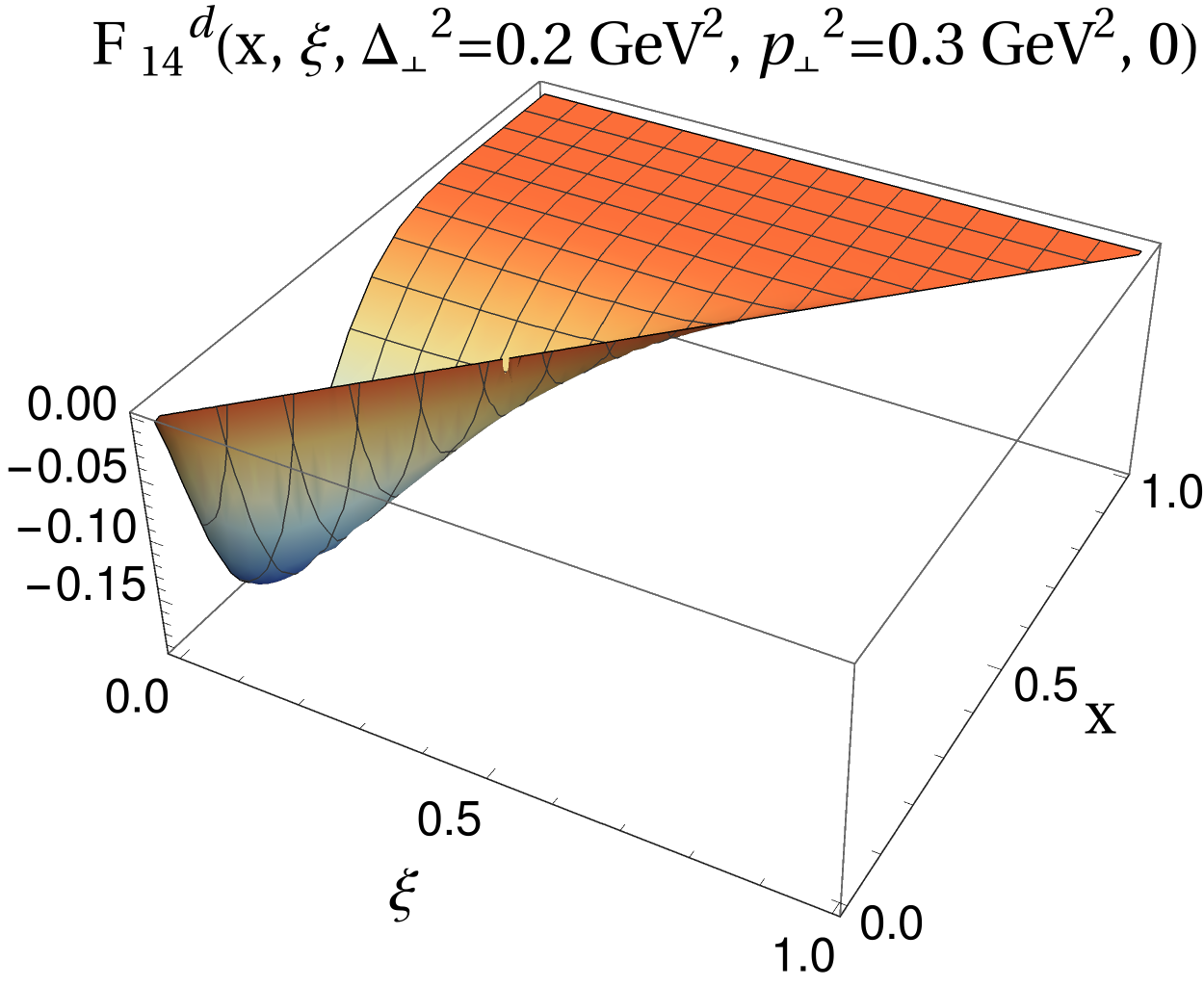} 
\caption{\label{unzx} The GTMDs as functions of $x$ and $\xi$ for an unpolarized quark. The upper panel is for the $u$ quark, while the lower panel represents the results for the $d$ quark. We fix $\bfd^2=0.2$~GeV$^2$, $\bfp^2=0.3$~GeV$^2$ and $\bfd \perp\bfp$. Left to right panels represent the GTMDs $F_{1,1},\, F_{1,2},\, F_{1,3}$, and  $F_{1,4}$, respectively. }
\end{figure}

\begin{figure}[b]
\includegraphics[scale=.32]{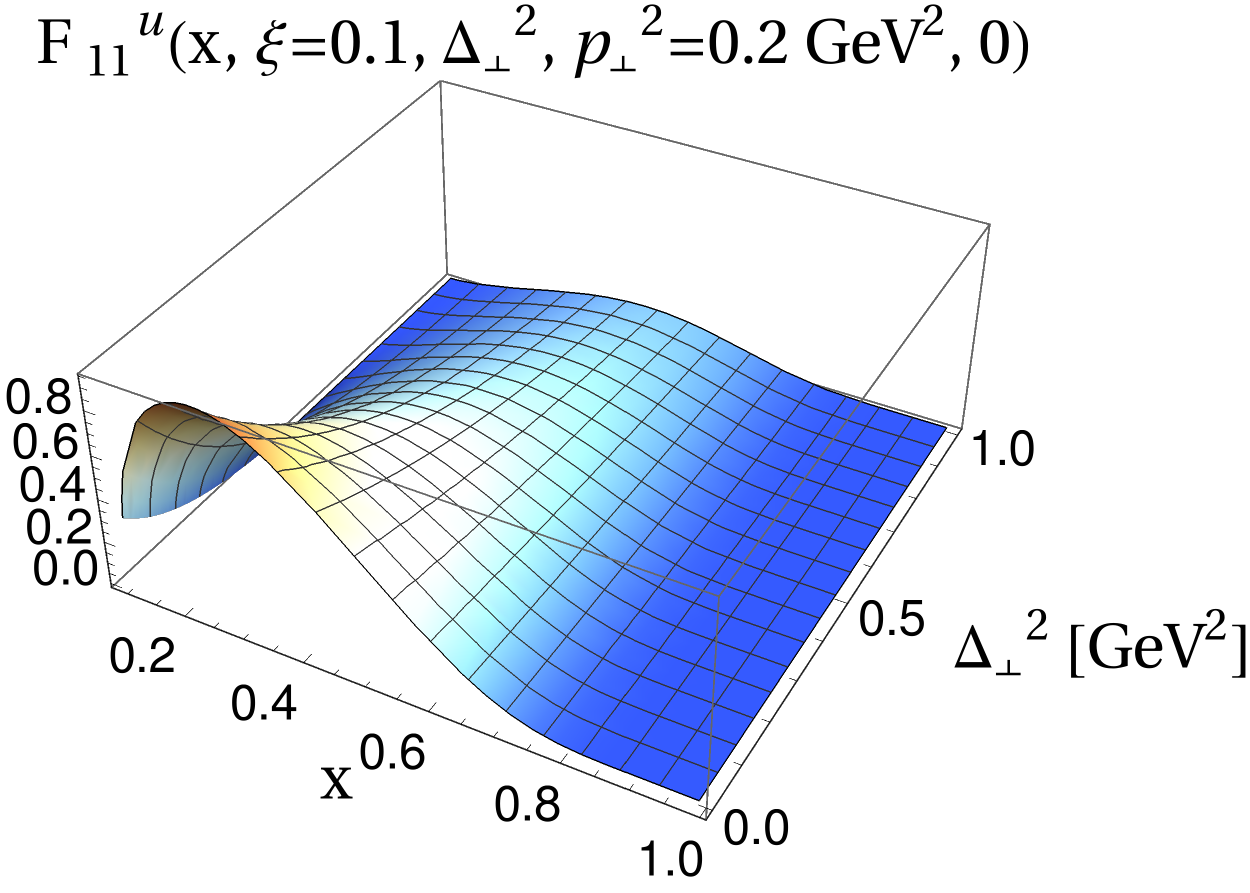}
\includegraphics[scale=.32]{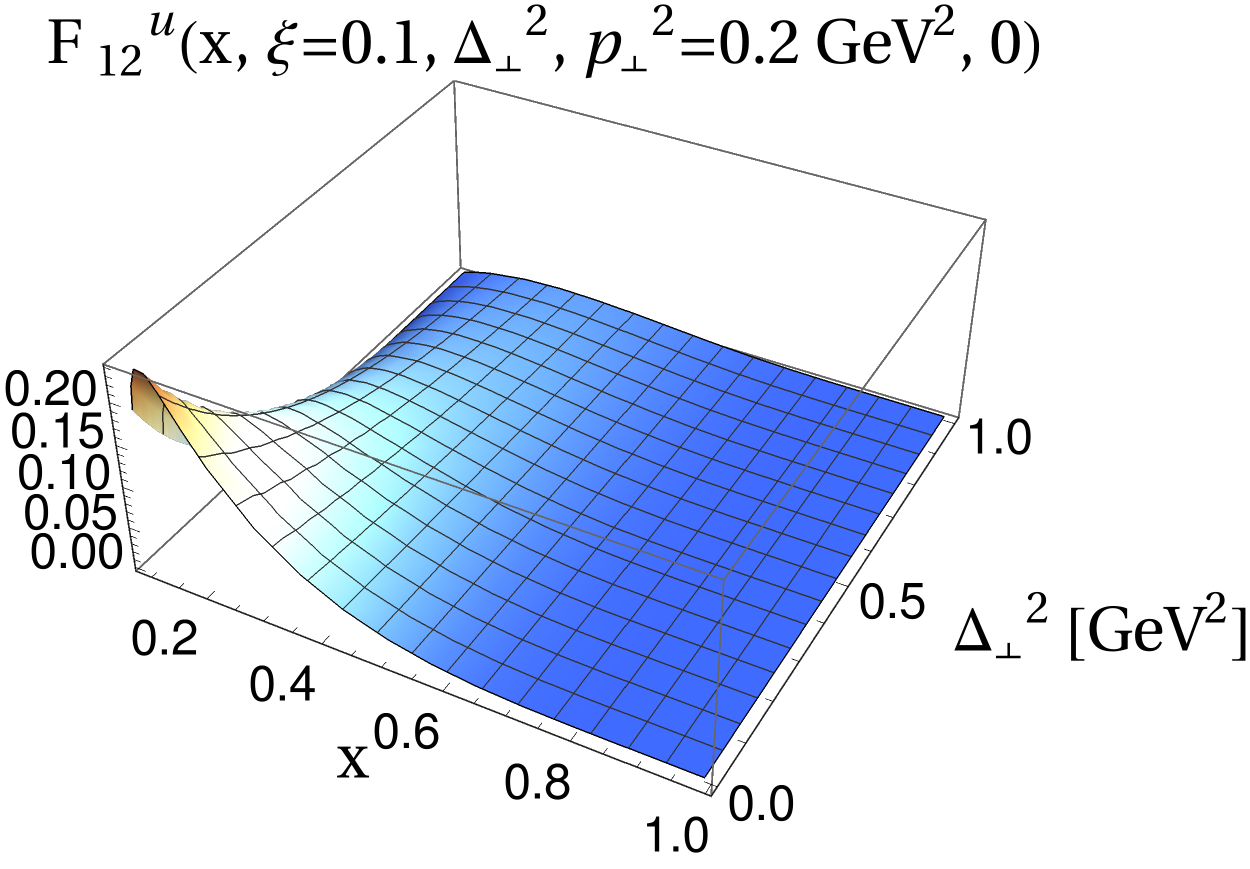} 
\includegraphics[scale=.32]{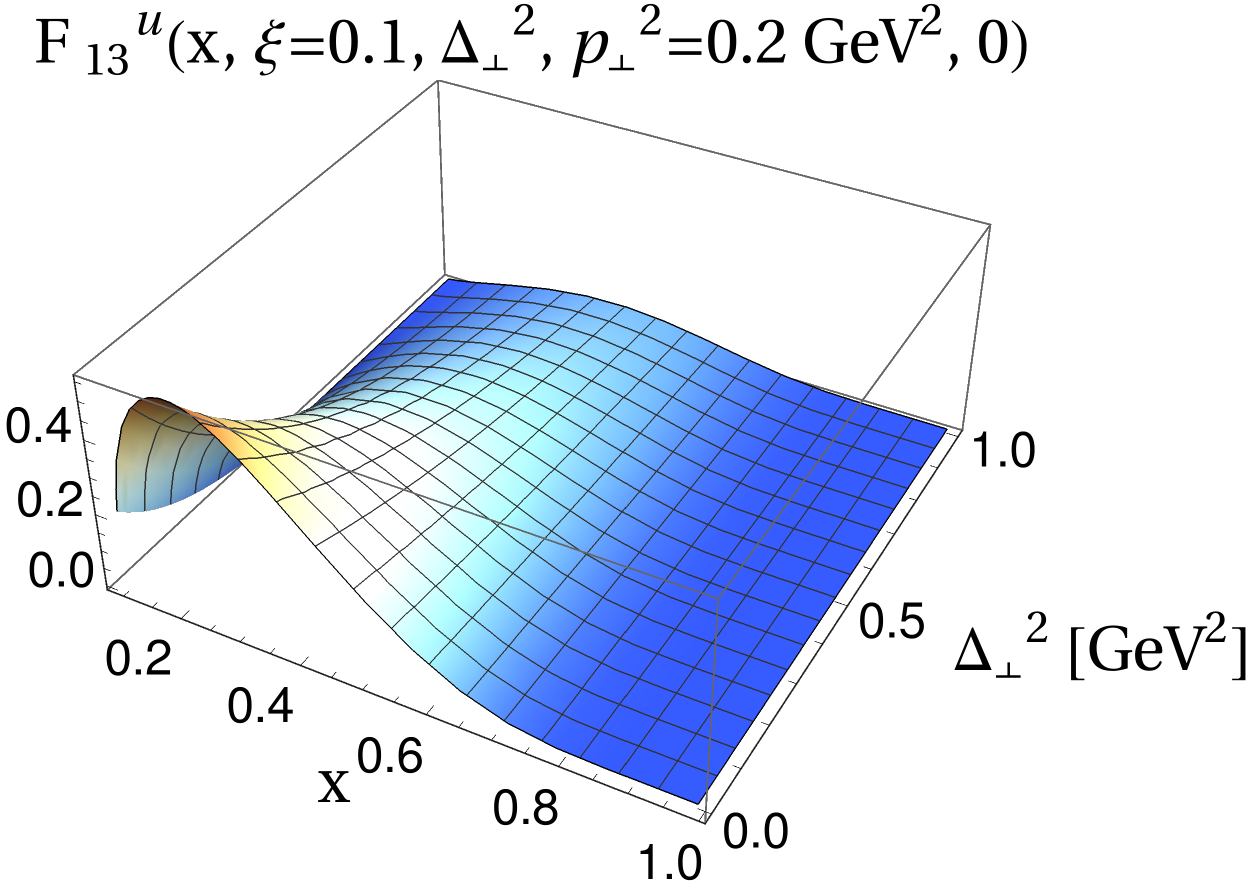} 
\includegraphics[scale=.32]{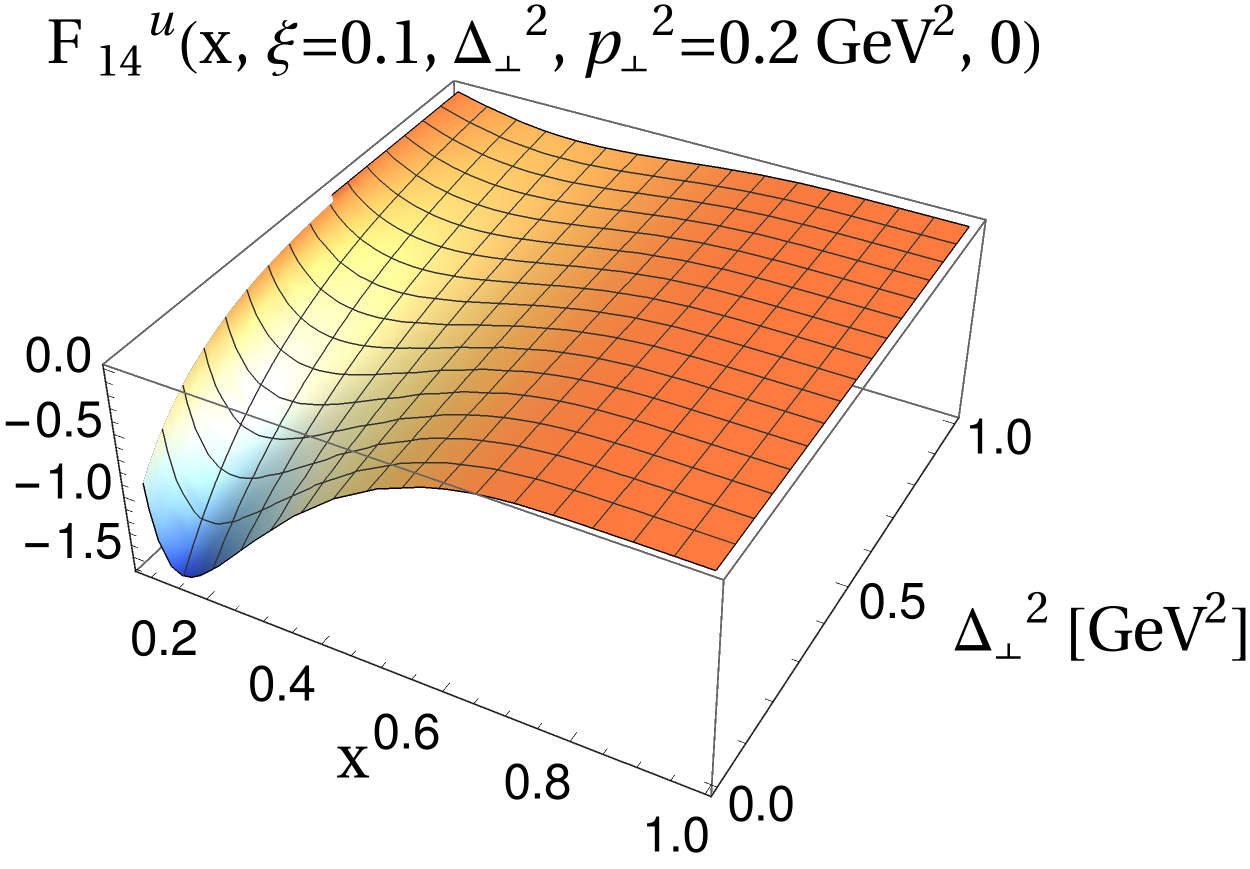} \\
\includegraphics[scale=.32]{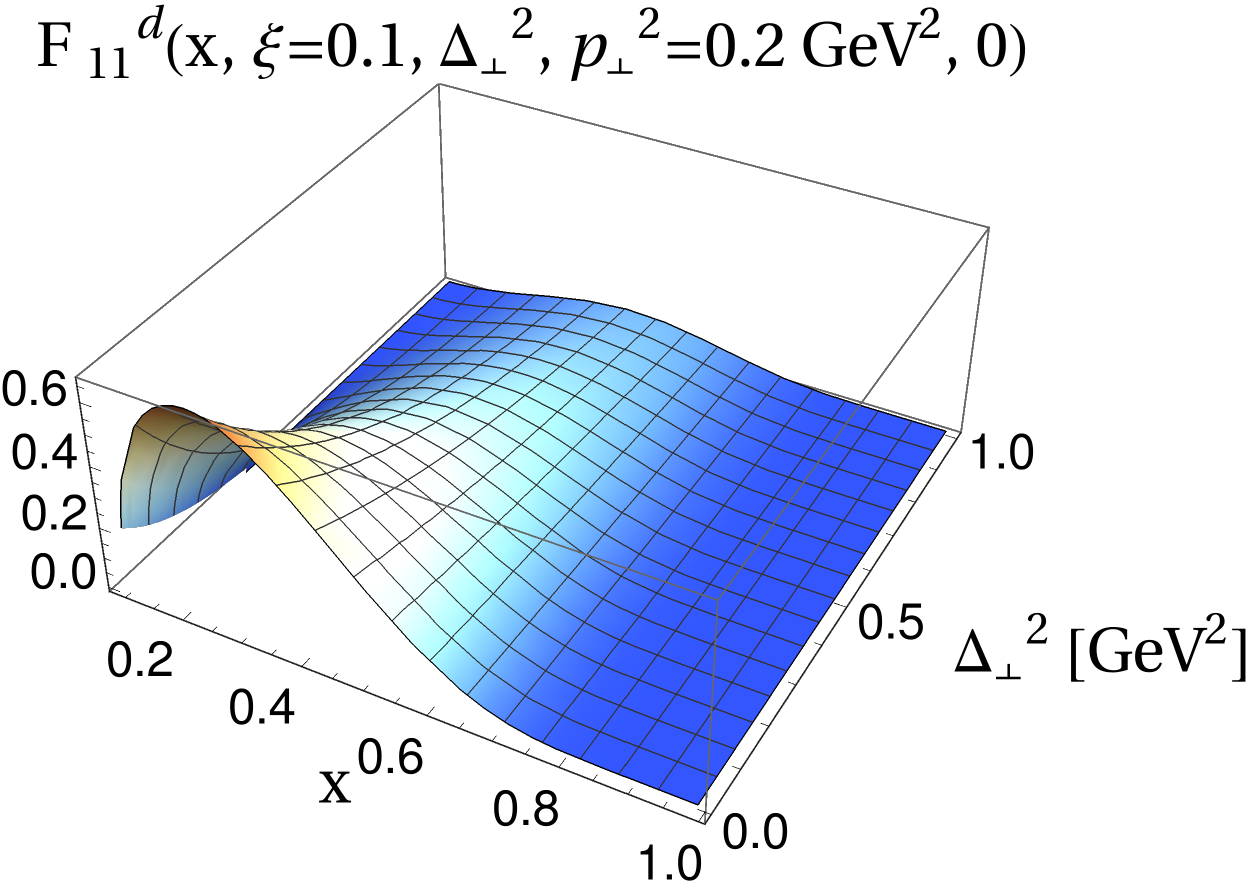} 
\includegraphics[scale=.32]{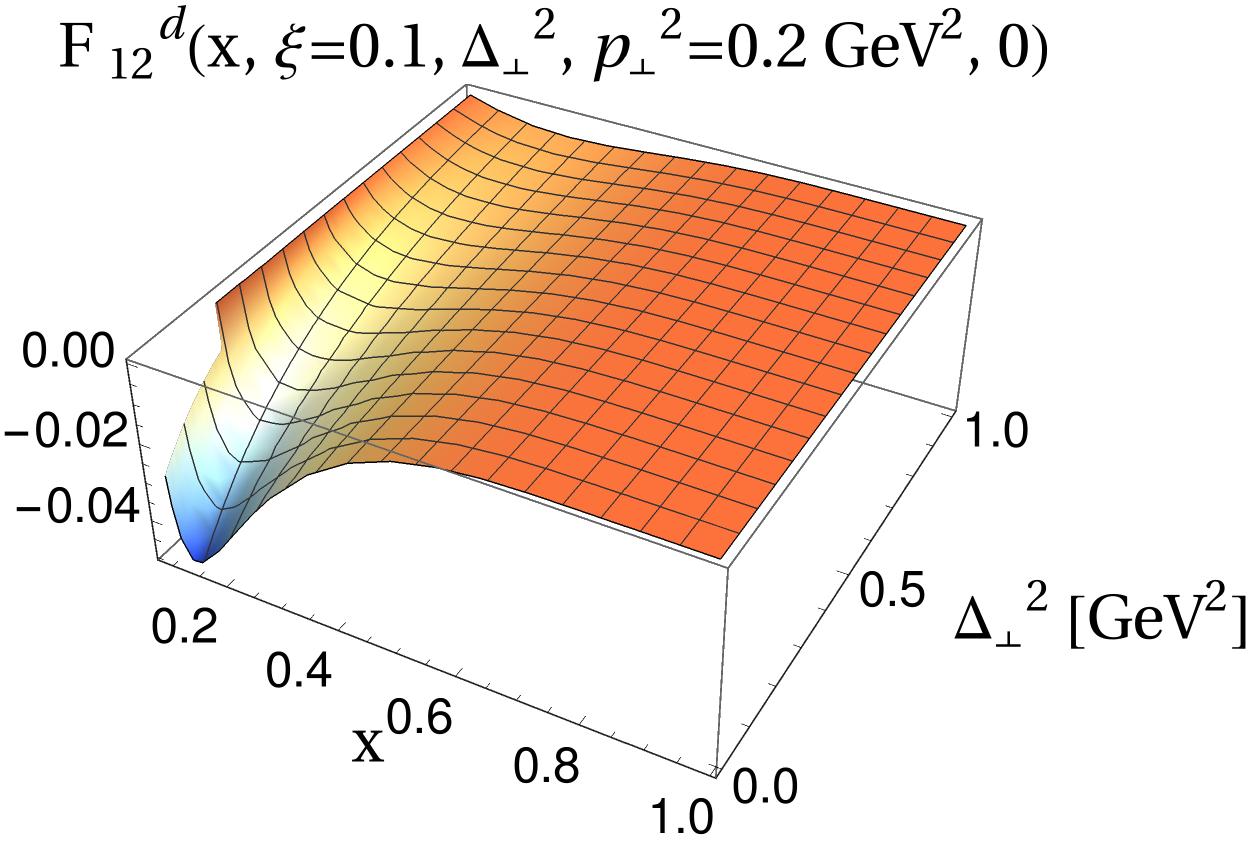} 
\includegraphics[scale=.32]{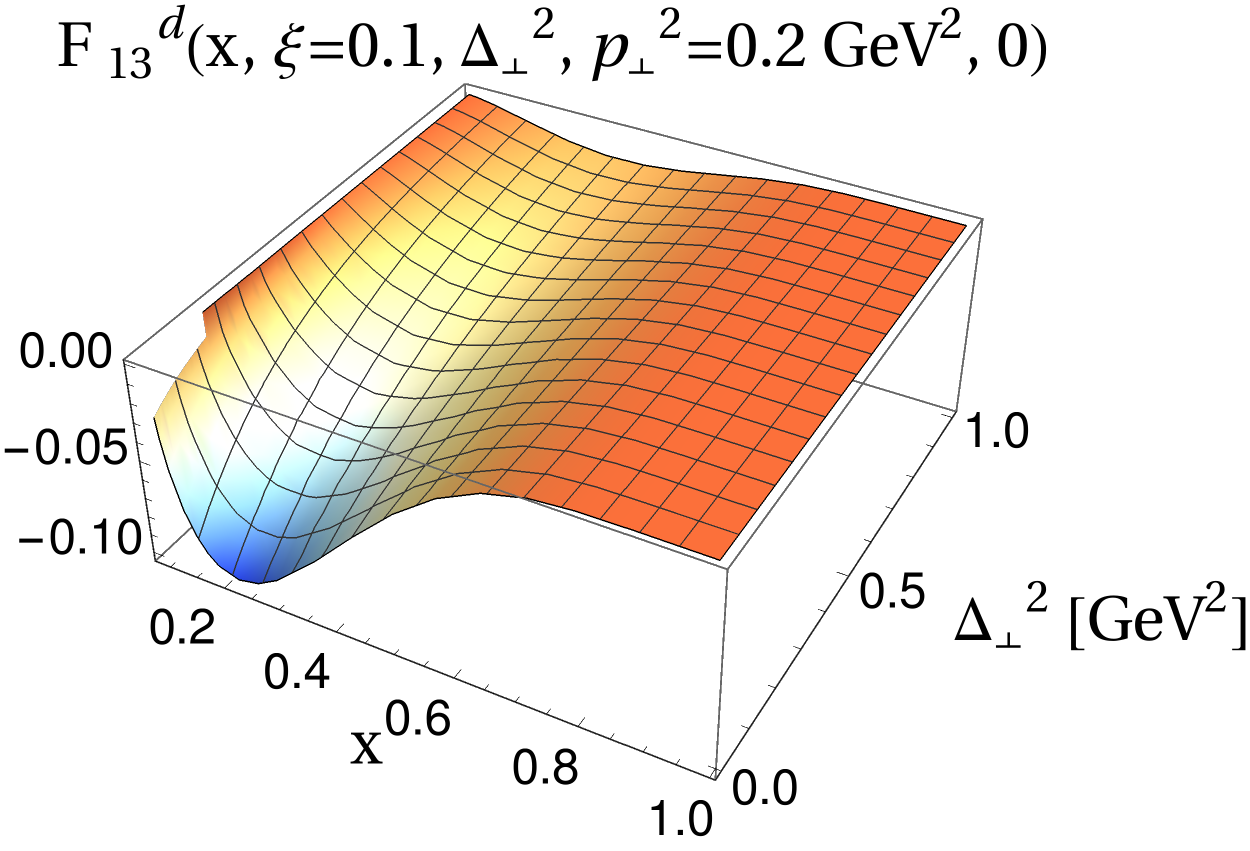} 
\includegraphics[scale=.32]{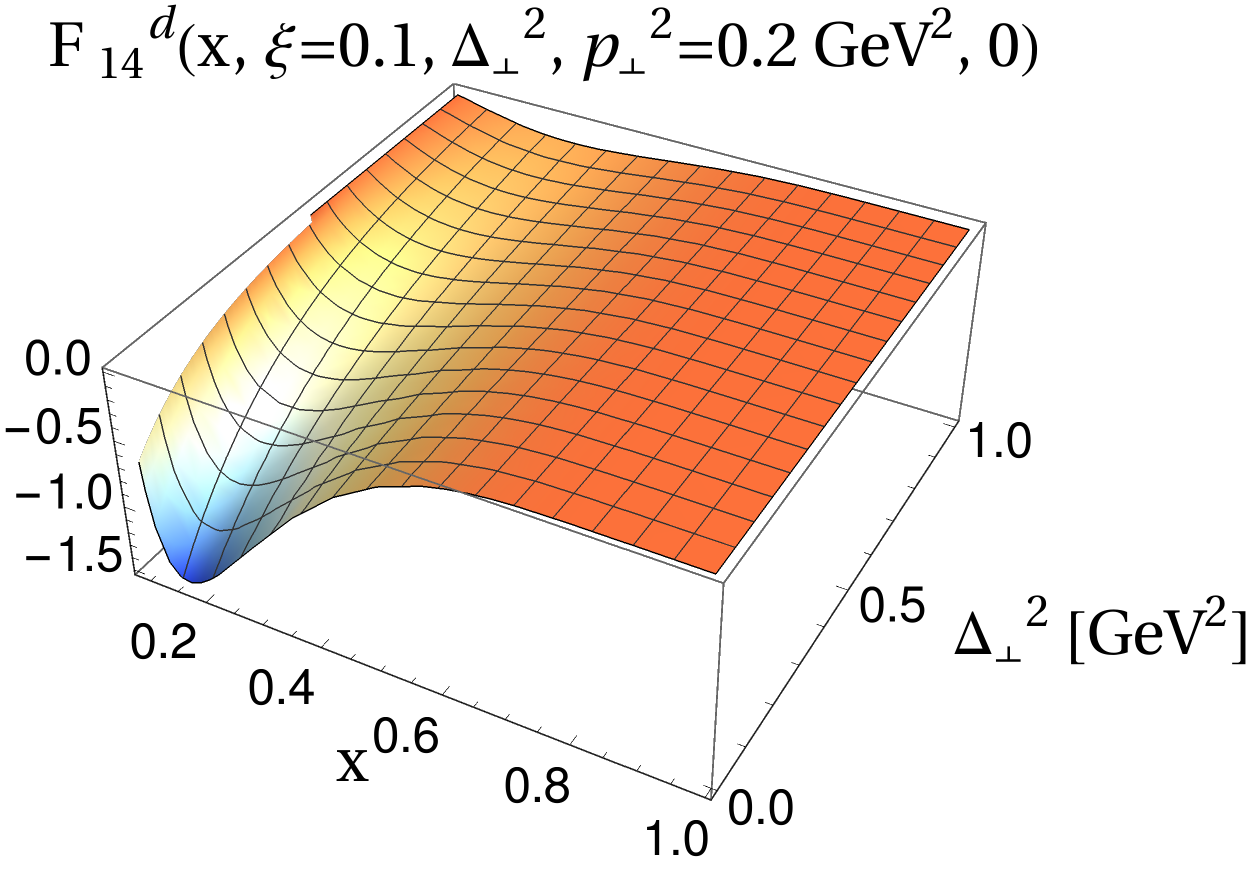} 
\caption{\label{unxD} The GTMDs as functions of $x$ and $\bfd^2$ for an unpolarized quark. The upper panel is for the $u$ quark, while the lower panel represents the results for the $d$ quark. We fix $\xi=0.1$, $\bfp^2=0.3$~GeV$^2$ and $\bfd \perp\bfp$. Left to right panels represent the GTMDs $F_{1,1},\, F_{1,2},\, F_{1,3}$, and  $F_{1,4}$, respectively.}
\end{figure}

\subsection{Unpolarized quark}
Figure~\ref{unzx} shows our model results of the GTMDs for an unpolarized quark in a proton as functions of $\xi$ and $x$ at fixed $\bfd^2=0.2$ GeV$^2$ and $\bfp^2=0.3$~GeV$^2$ with $\bfd$ being perpendicular to $\bfp$. The four columns represents the four GTMDs $F_{1,1},\, F_{1,2},\, F_{1,3}$, and  $F_{1,4}$. The upper and lower rows are for the $u$ and $d$ quarks, respectively. One notices that all the distributions exhibit the accessibility of the DGLAP region $x>\xi$. $F_{1,2}$ and $F_{1,3}$ for the $u$ quark show positive distributions, while they are negative for the $d$ quark. Meanwhile, we find that for both the quarks, $F_{1,1}$ is positive but $F_{1,4}$ shows negative distribution. 
In case of $F_{12}$ given in Eq.(\ref{F12}), the second term containing $F_{1,4}$ dominates and leads to the positive distribution for $u$ and negative for the $d$ quarks. We observe that the general features of  all the distributions are more or less similar. The GTMDs have their peaks at lower-$x \,(<0.5)$ and the peaks shift towards higher values of $x$ with decreasing the magnitude as the momentum transfer increases in the longitudinal direction. In the light-cone gauge, the canonical quark orbital angular momentum (OAM), $\ell_z$, has contributions from GTMDs $F_{1,4}$ at $\xi=0$ and $\bfd=0$  limit  \cite{Lorce:2011kd,Chakrabarti:2016yuw}:
\begin{align}
\ell_z^\nu=-\int {\rm d}x\,{\rm d}^2\bfp\, \frac{\bfp^2}{M^2}\,F_{1,4}^\nu(x,0,\bfp^2,0,0)\,.
\end{align}
The $\ell_z$  provides the correlation between proton spin and quark OAM. In our model, the negative polarity of $F_{1,4}$ for both the quarks indicates that the quark OAM tends to be aligned to the proton spin for both $u$ and $d$ quarks ($\ell_z^\nu>0$), which is consistent with the results reported in Ref.~\cite{Chakrabarti:2016yuw}. Meanwhile, it has been shown  in Ref.~\cite{Lorce:2011kd}, the quark OAM tends to be aligned to the proton spin for the $u$ quark ($\ell_z^u>0$), but anti-aligned for the $d$ quark ($\ell_z^d<0$).

In Fig.~\ref{unxD}, we present $x$ and $\bfd^2$ dependence of the unpolarized GTMDs at fixed $\xi=0.1$ and $\bfp^2=0.2$~GeV$^2$. Here again, we notice that the general feature of all the plots is almost same. The magnitudes of distributions decrease and the peaks along-$x$ move towards larger values of $x$ with increasing momentum transfer $\bfd^2$. As the total kinetic energy remains limited, the distributions in the transverse momentum broadens at higher-$x$ reflecting the trend
to carry a larger portion of the kinetic energy. These general features of the GTMDs are nearly model-independent properties of the GPDs and, indeed, they are observed in several theoretical studies of the GPDs~\cite{Ji:1997gm,Scopetta:2002xq,Petrov:1998kf,Penttinen:1999th,Boffi:2002yy,Vega:2010ns,Chakrabarti:2013gra,Mondal:2015uha,Chakrabarti:2015ama,Mondal:2017wbf,deTeramond:2018ecg,Xu:2021wwj} As we expect from the Eqs.(\ref{F11})--(\ref{F14}), $F_{1,2}$ and $ F_{1,3}$ distributions show opposite polarity for the $u$ and $d$ quarks, whereas the polarity of $F_{1,1}$ and $ F_{1,4}$ remain unchanged against the flavors. At the TMD limit, $\bfd=0$ and with vanishing skewness, the time reversal even (T-even) part of $F_{1,1}$ maps onto the unpolarized TMD $f^\nu_1(x,\bfp^2)$ and the T-odd part of $F_{1,2}$ is linked to the Sivers TMD $f^{\perp \nu}_{1T}(x,\bfp^2)$. In our model, the different polarities of $F_{1,2}$ for the $u$ and $d$ quarks lead to the Sivers effect \cite{Sivers:1989cc},
 where a quark in a transversely polarized target has the transverse momentum asymmetry in the perpendicular direction to the proton spin. This asymmetry for the $u$ quark is found to be in opposite momentum direction to that of the $d$ quark.

\begin{figure}[h]
\includegraphics[scale=.32]{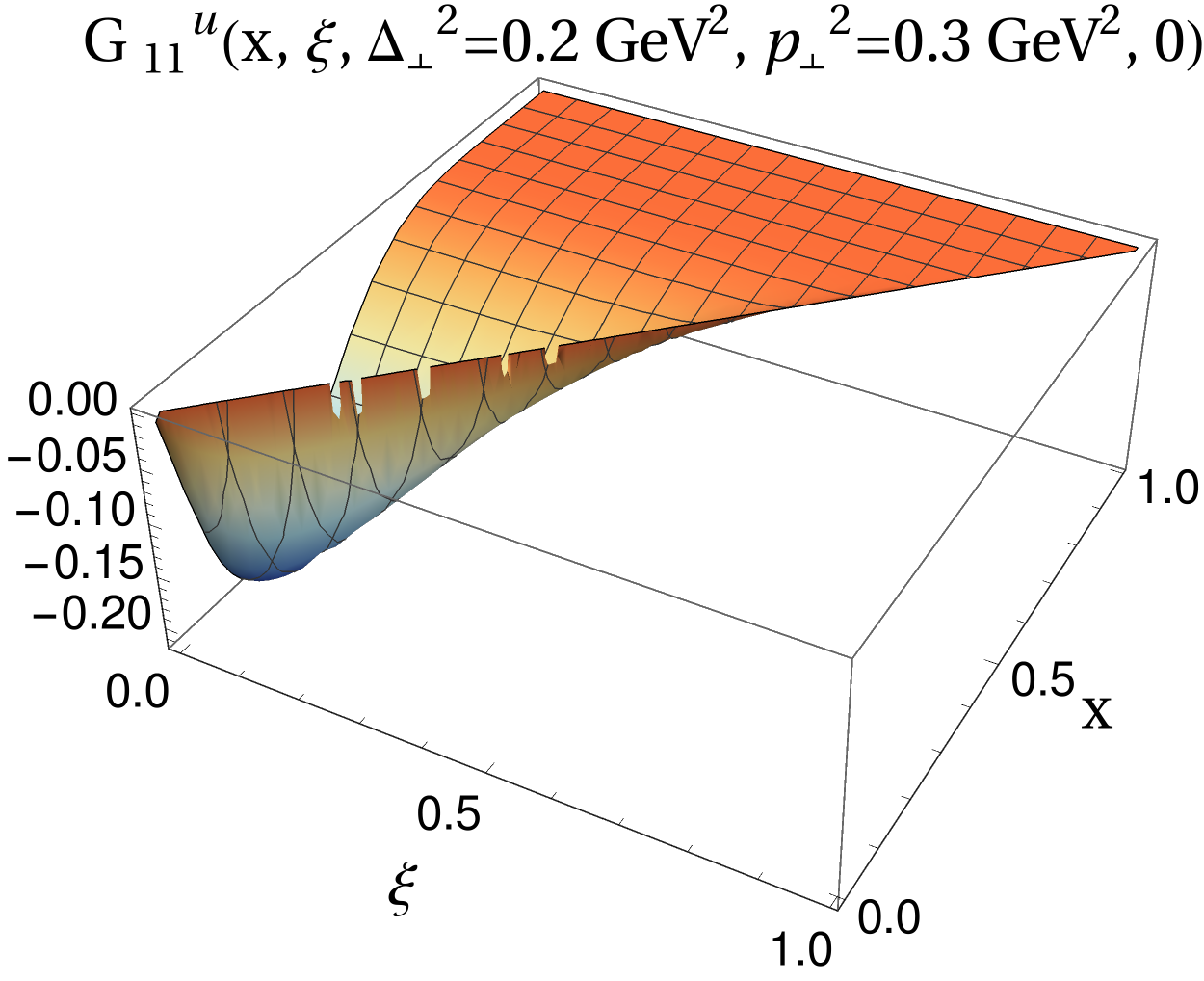}
\includegraphics[scale=.32]{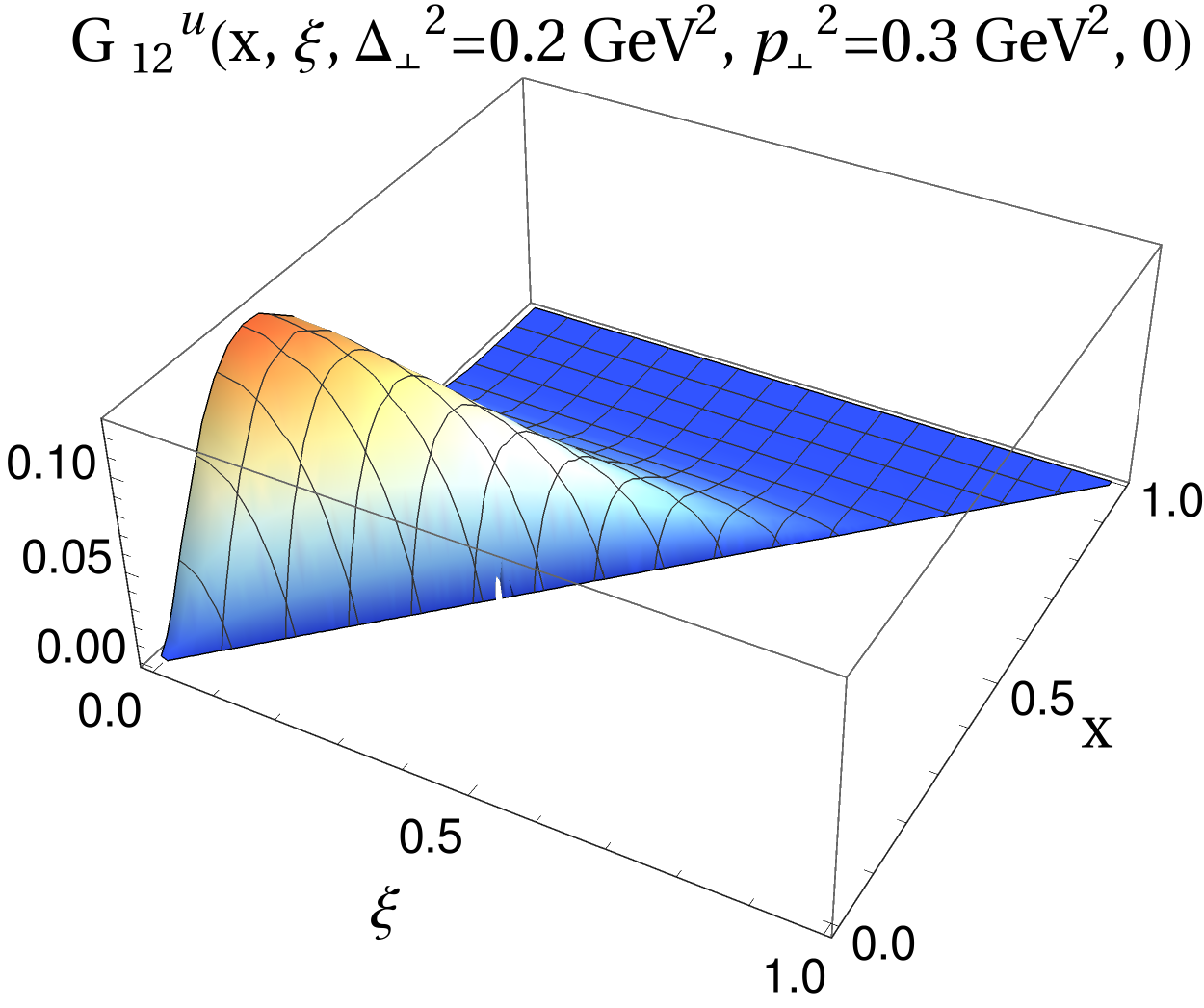} 
\includegraphics[scale=.32]{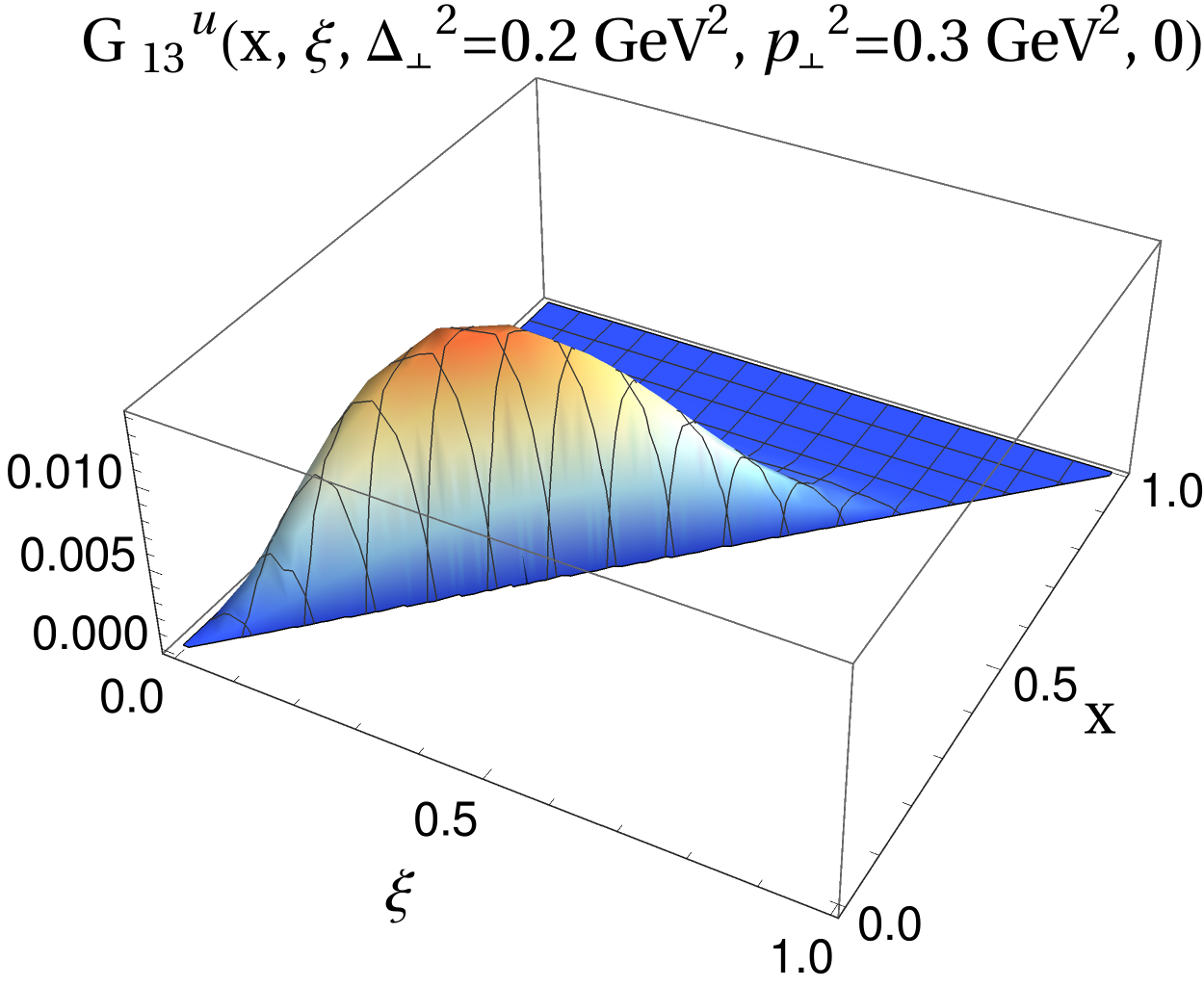} 
\includegraphics[scale=.32]{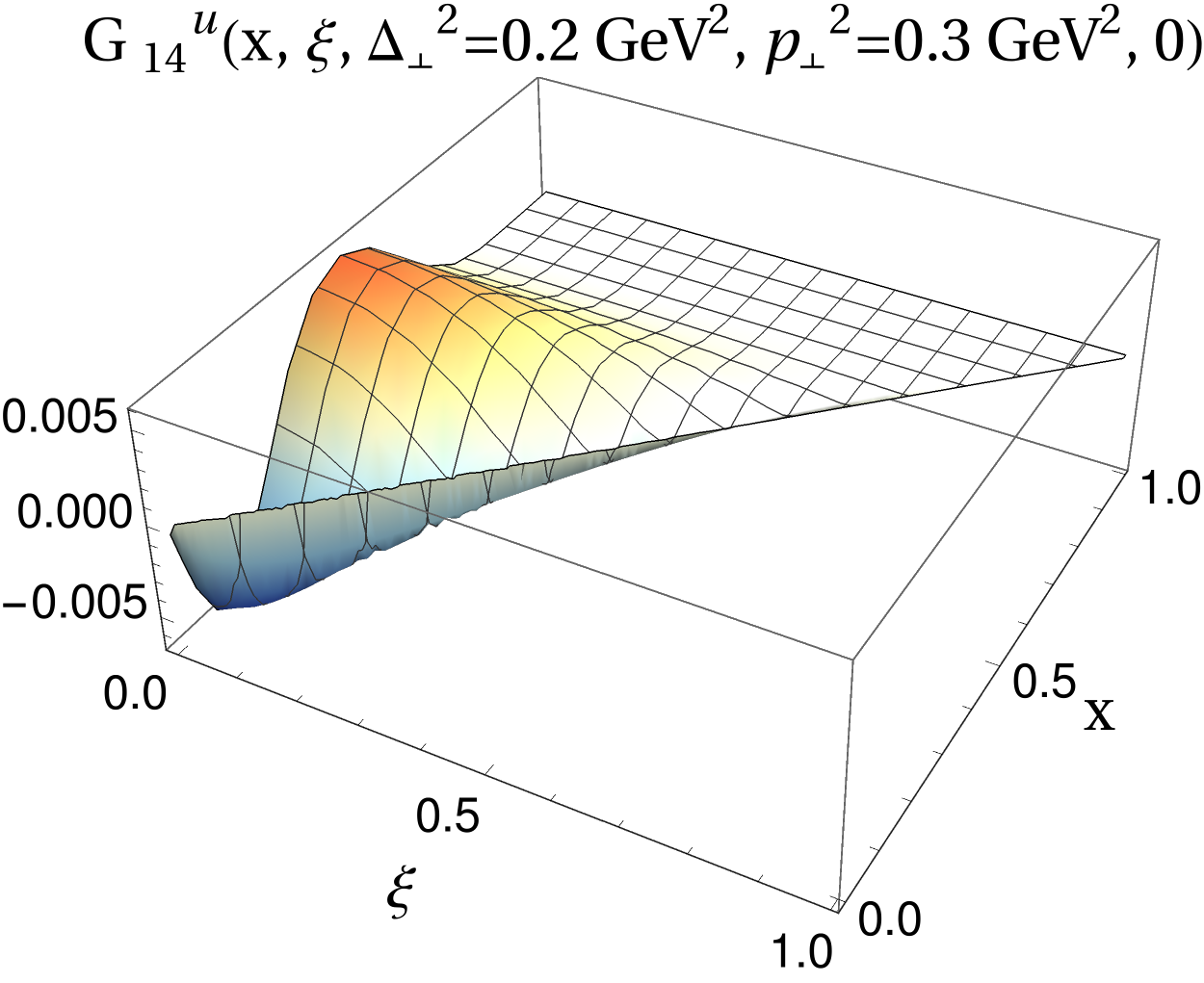} \\
\includegraphics[scale=.32]{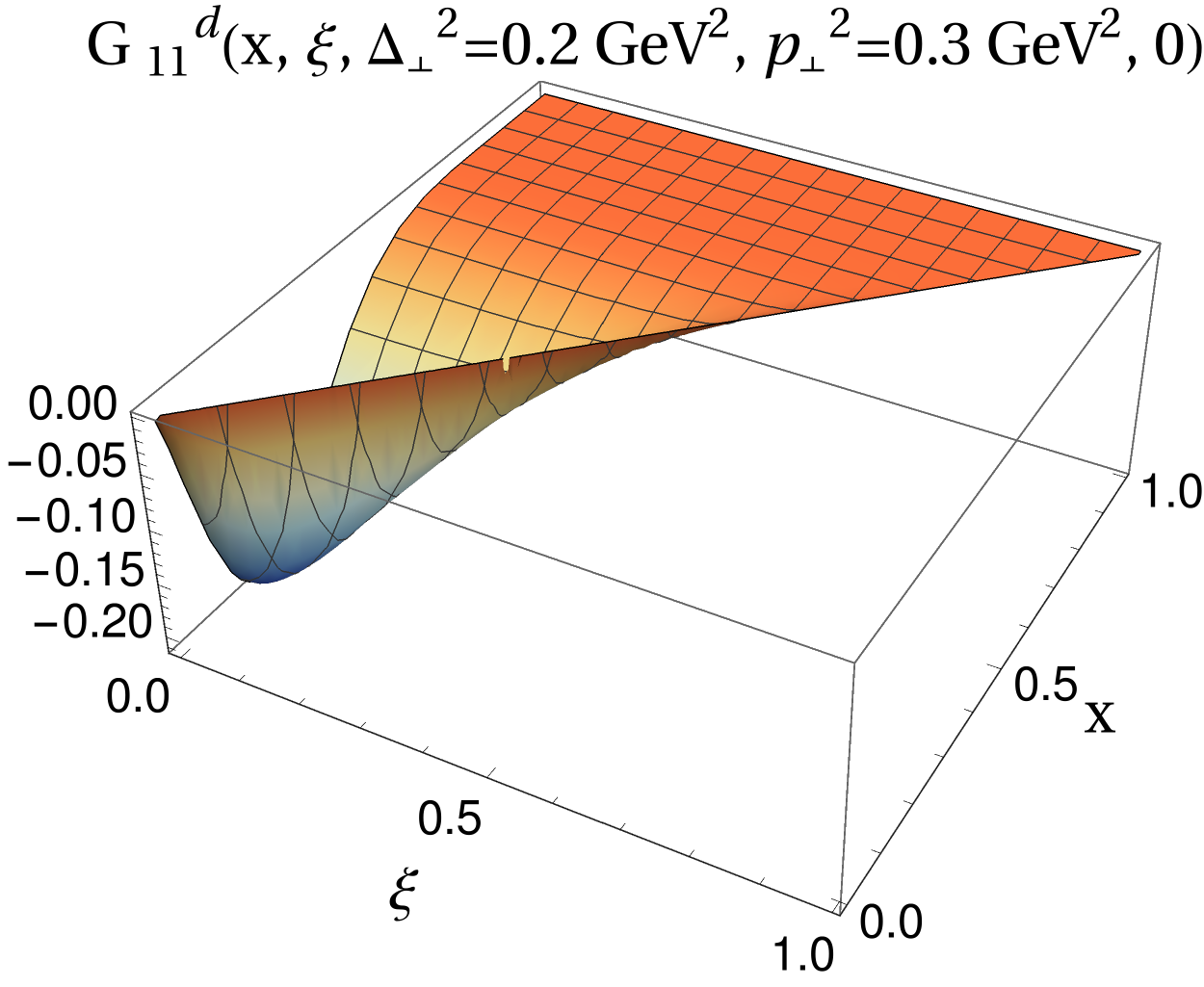} 
\includegraphics[scale=.32]{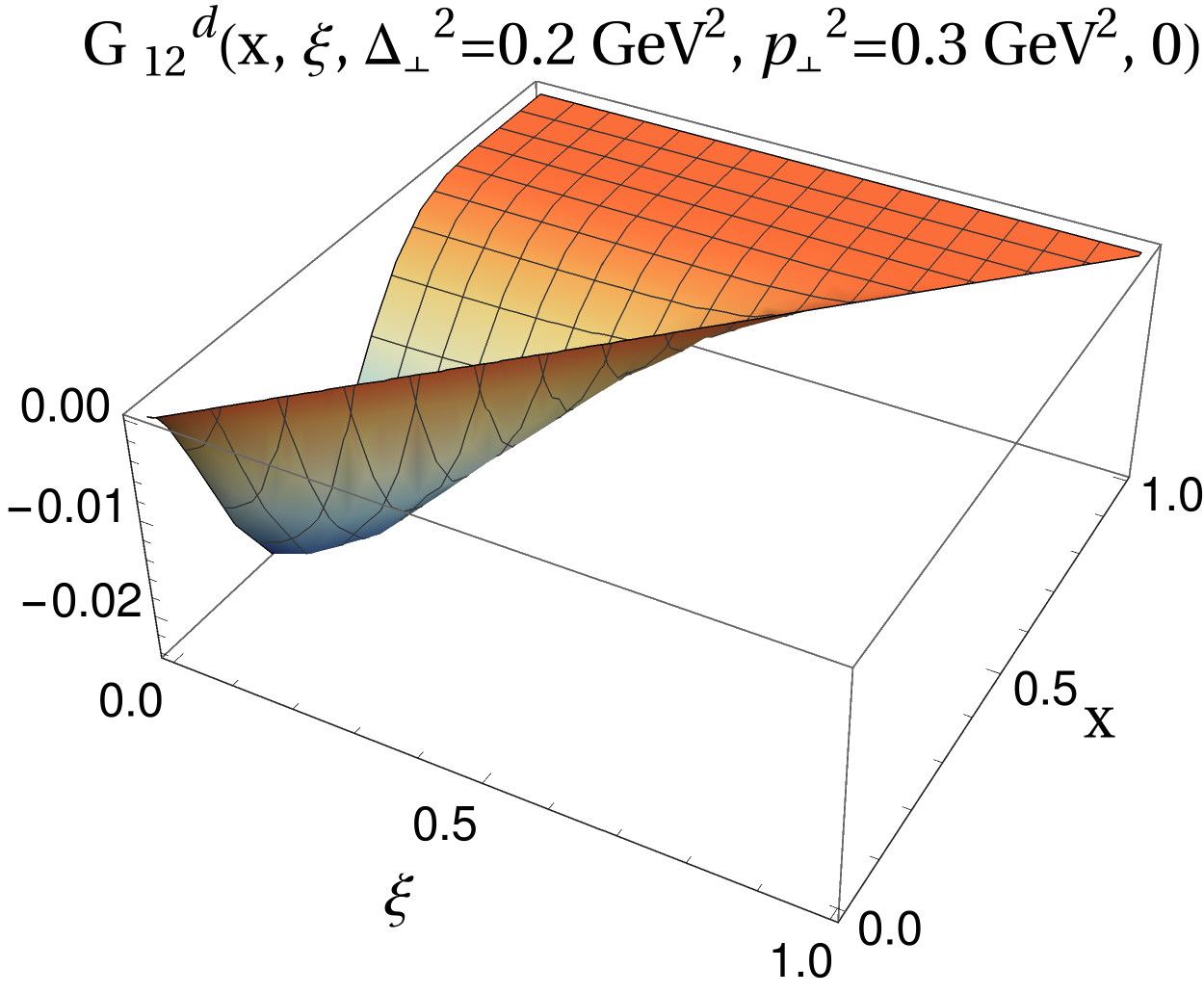} 
\includegraphics[scale=.32]{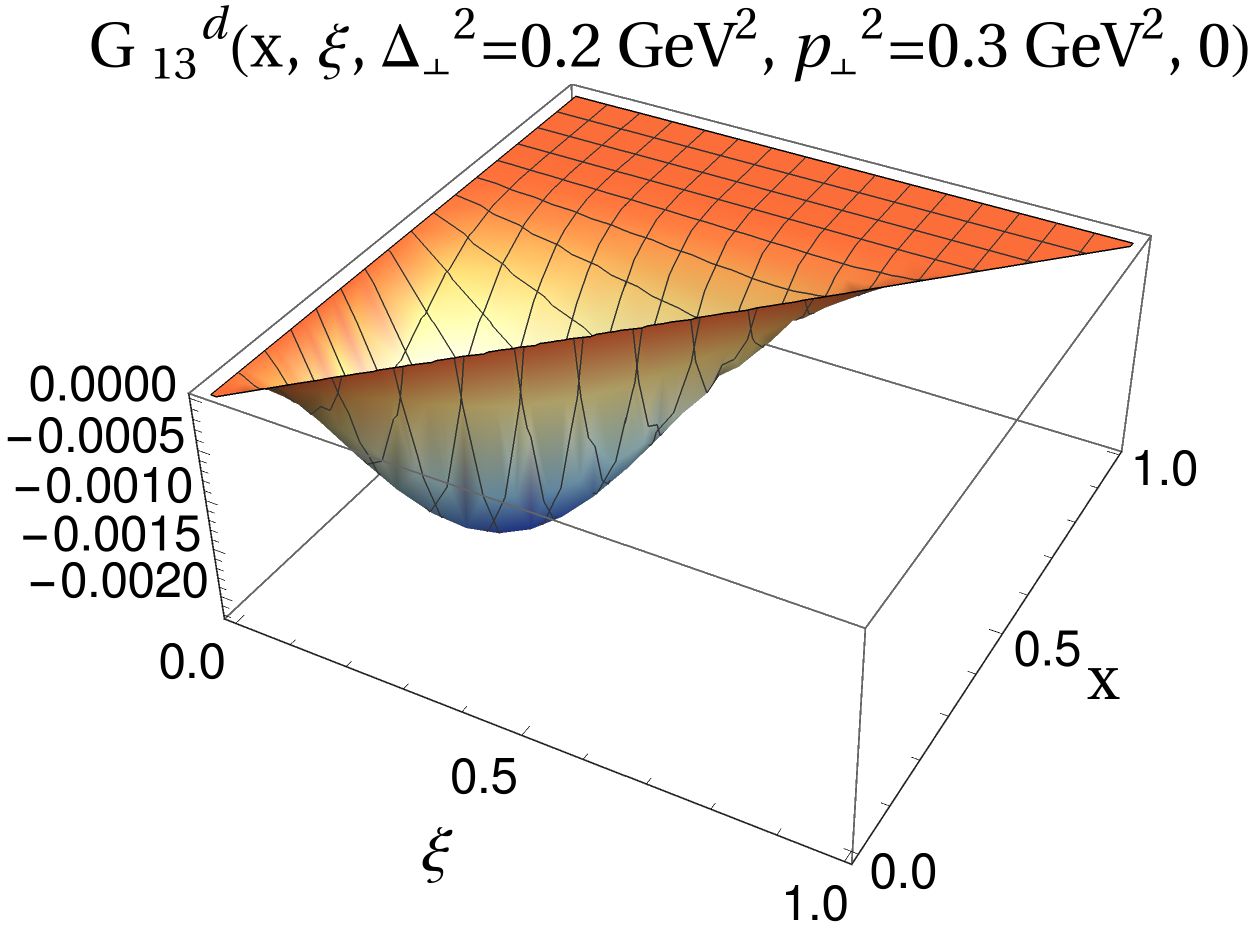} 
\includegraphics[scale=.32]{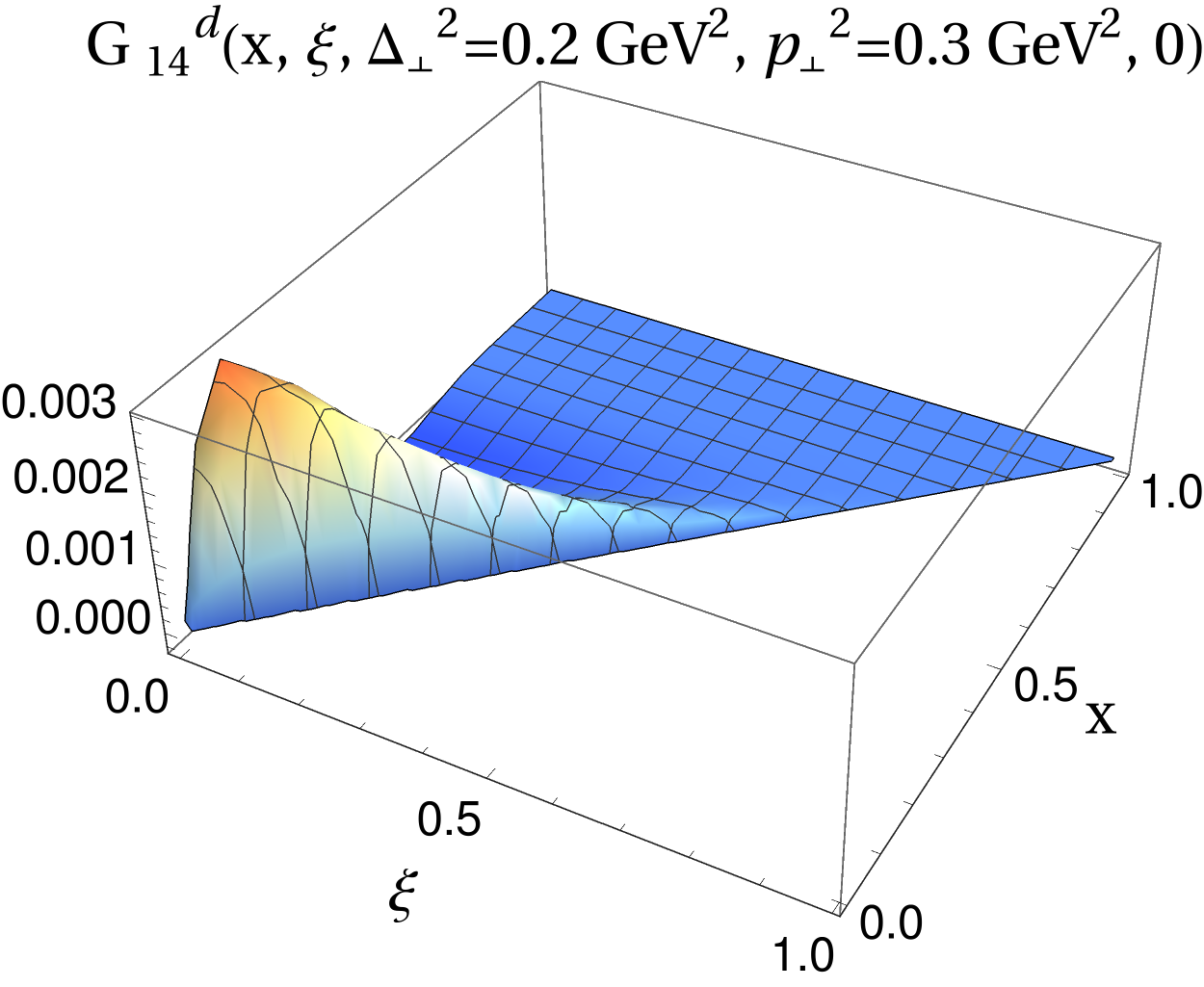} 
\caption{\label{longzx} The GTMDs as functions of $x$ and $\xi$ for a longitudinally polarized quark. The upper panel is for the $u$ quark, while the lower panel represents the results for the $d$ quark. We fix $\bfd^2=0.2$~GeV$^2$, $\bfp^2=0.3$~GeV$^2$ and $\bfd \perp\bfp$.  Left to right panels represent the GTMDs $G_{1,1},\, G_{1,2},\, G_{1,3}$, and  $G_{1,4}$, respectively.}
\end{figure}
\begin{figure}[b]
\includegraphics[scale=.32]{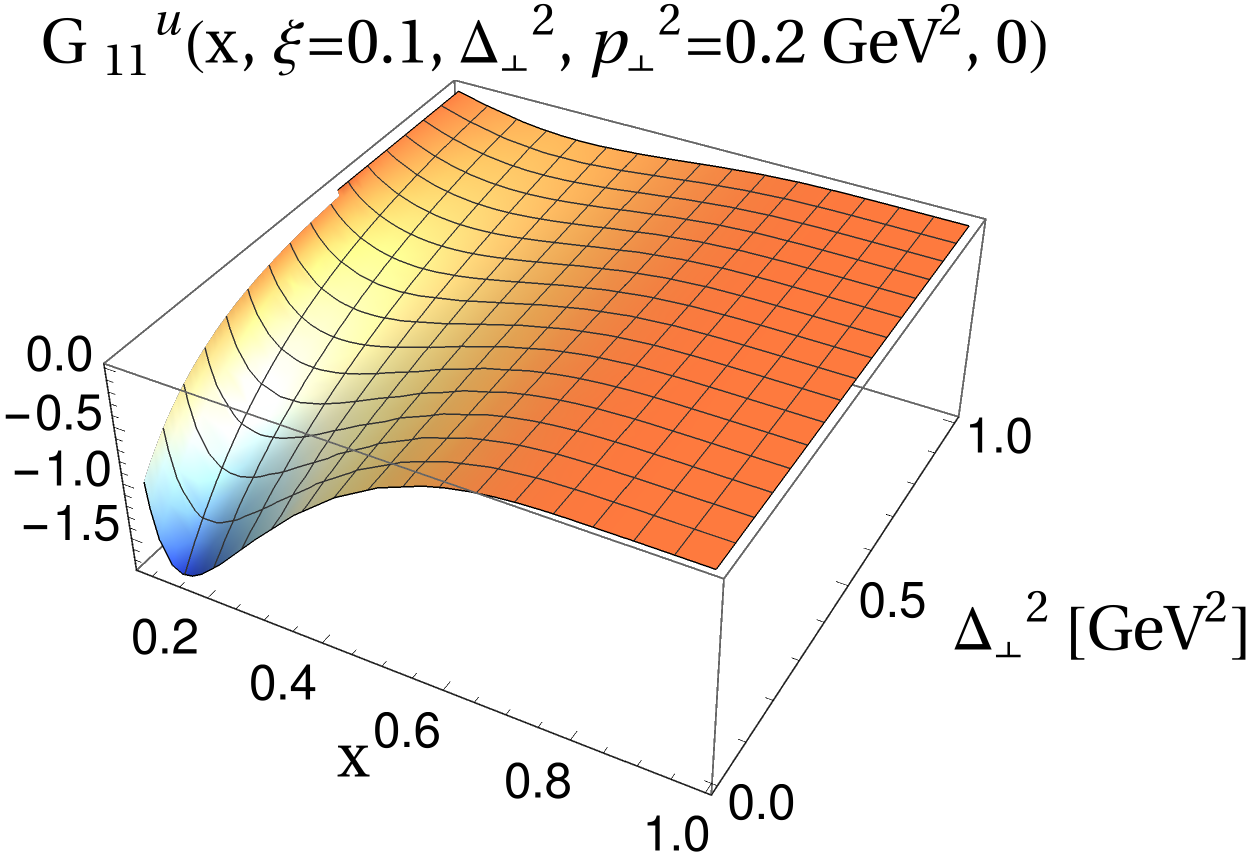}
\includegraphics[scale=.32]{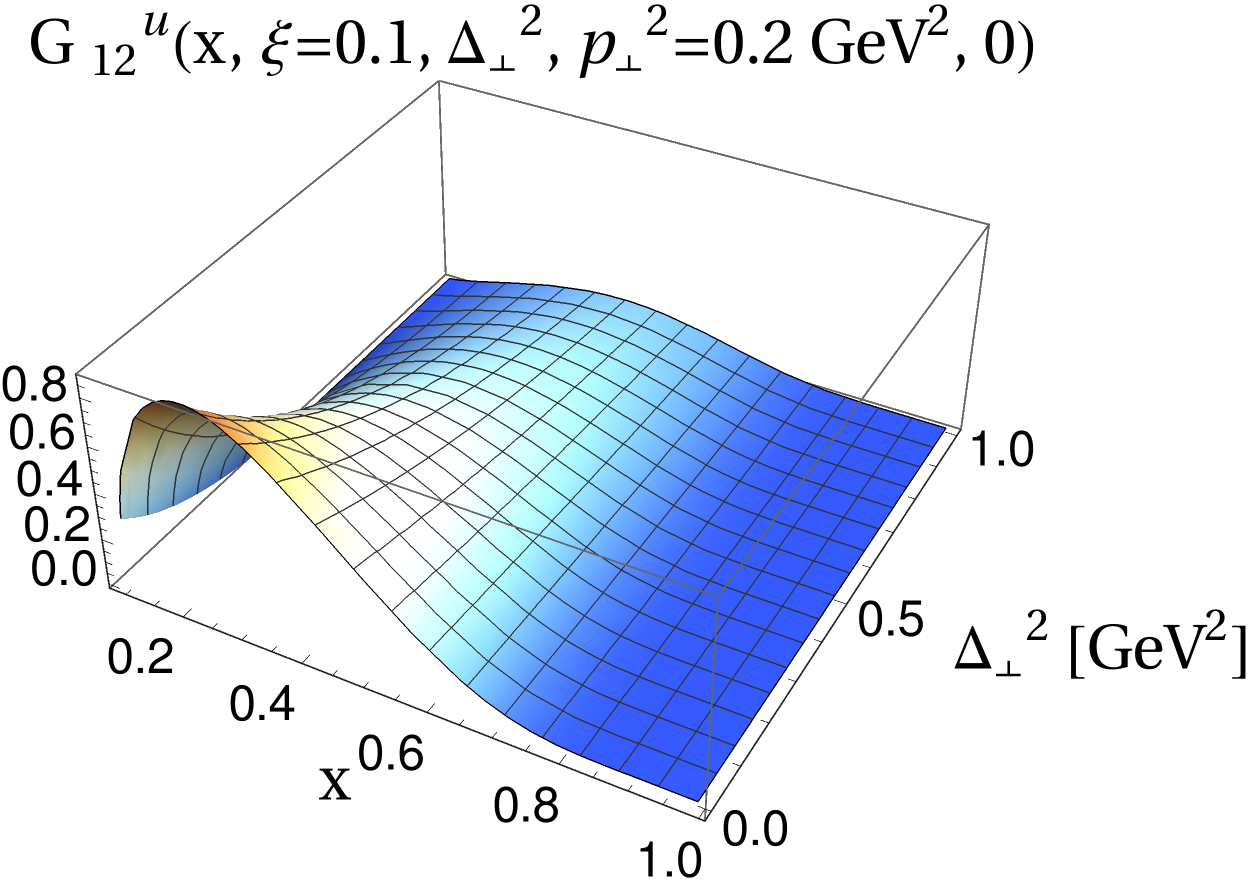} 
\includegraphics[scale=.32]{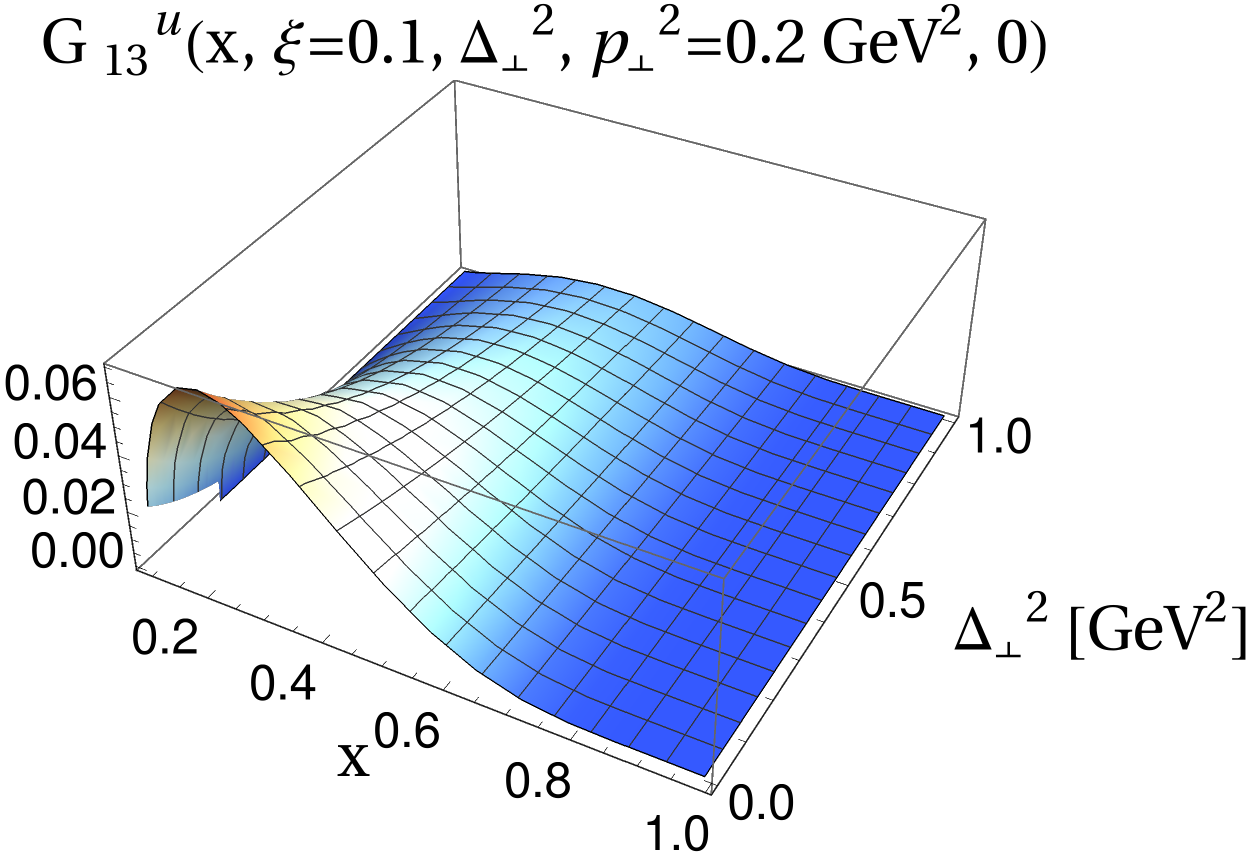} 
\includegraphics[scale=.32]{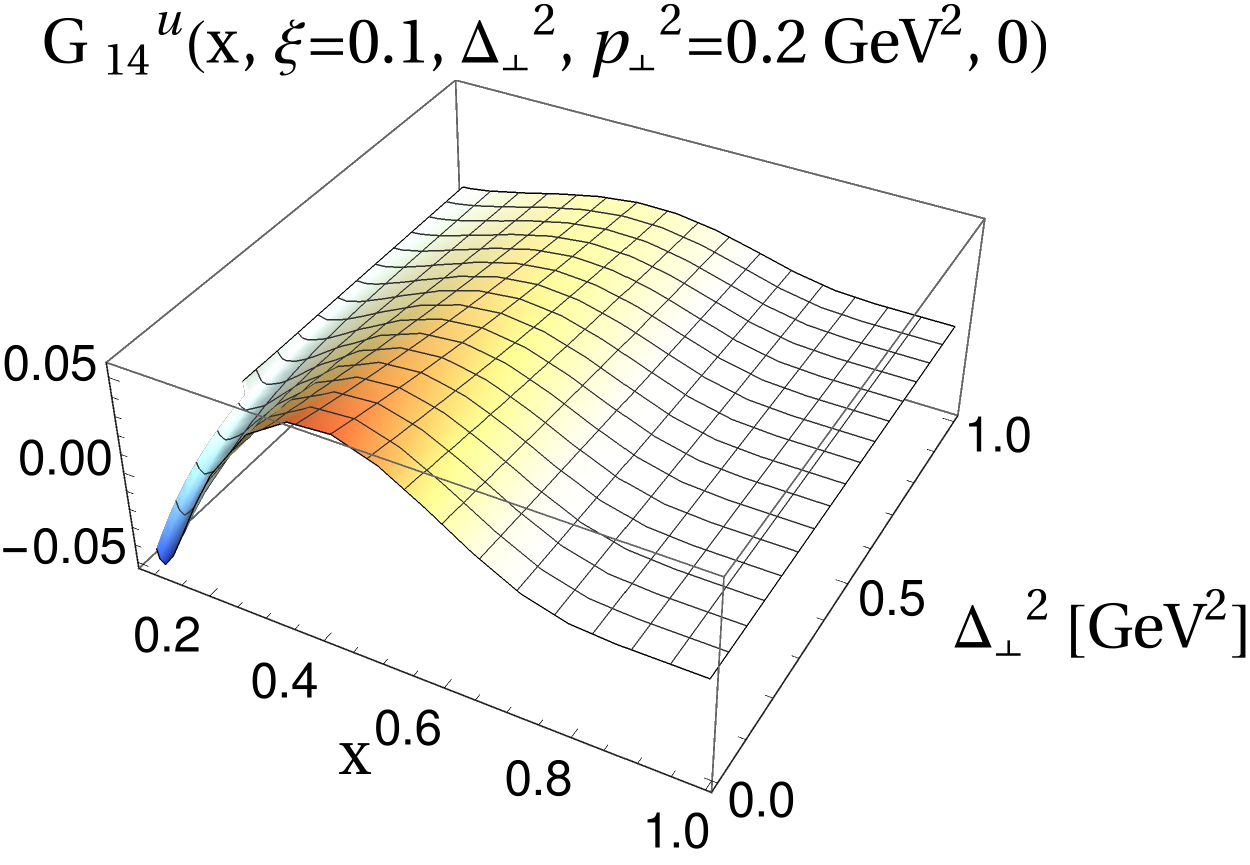} \\
\includegraphics[scale=.32]{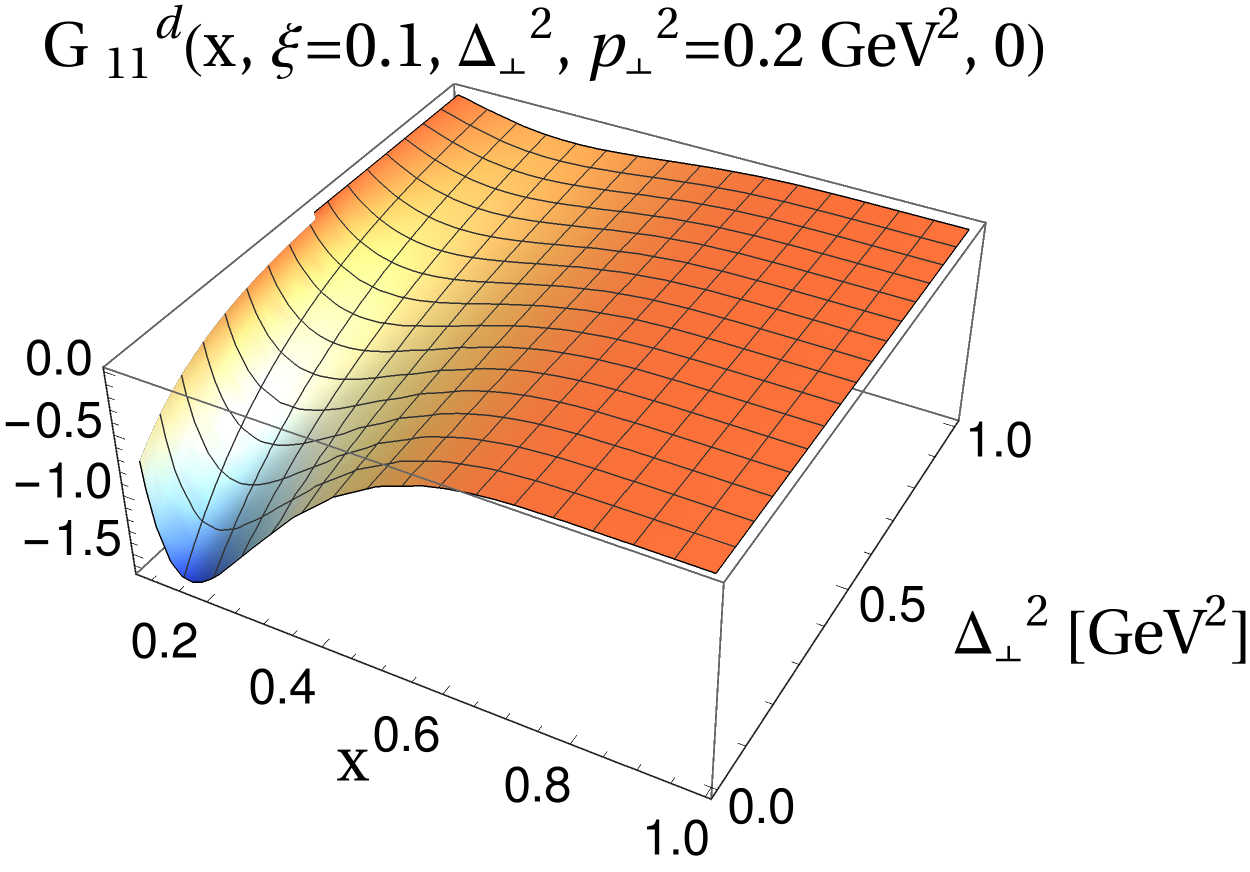} 
\includegraphics[scale=.32]{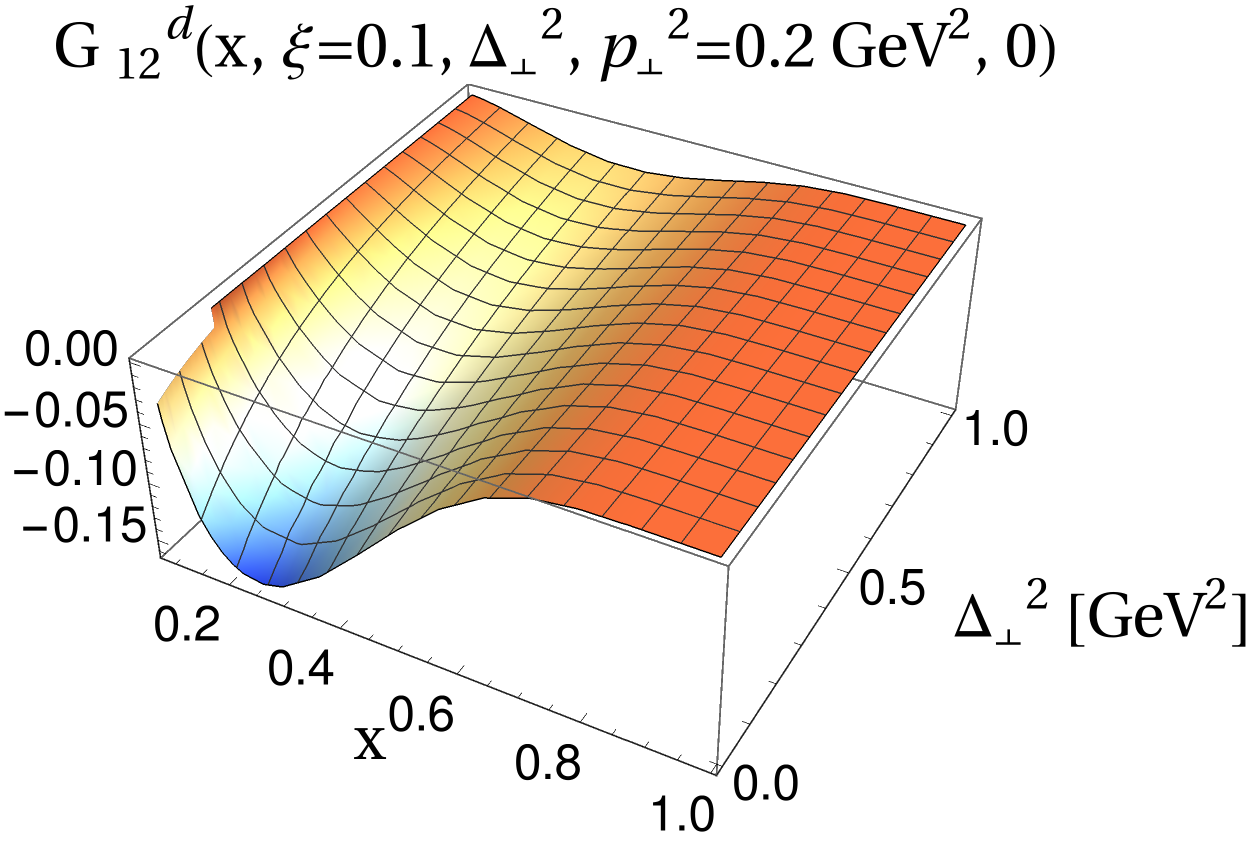} 
\includegraphics[scale=.32]{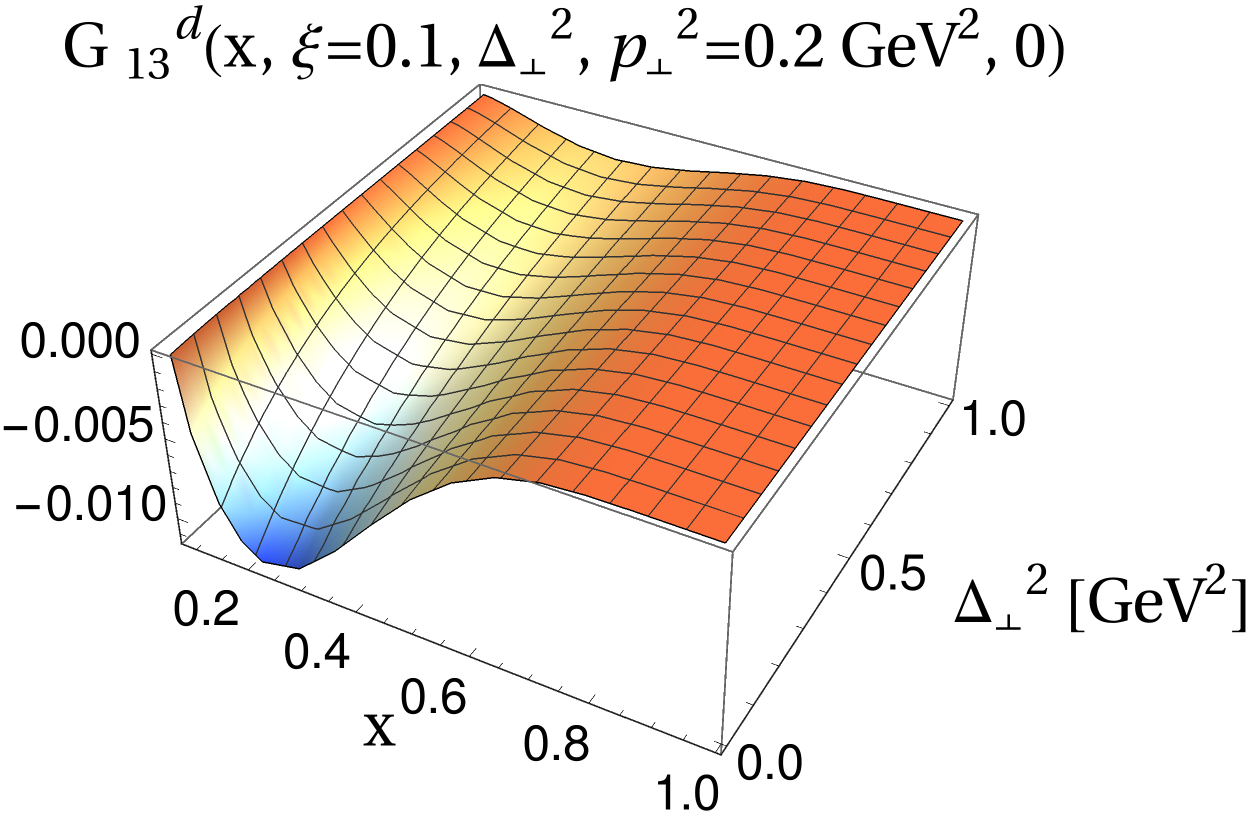} 
\includegraphics[scale=.32]{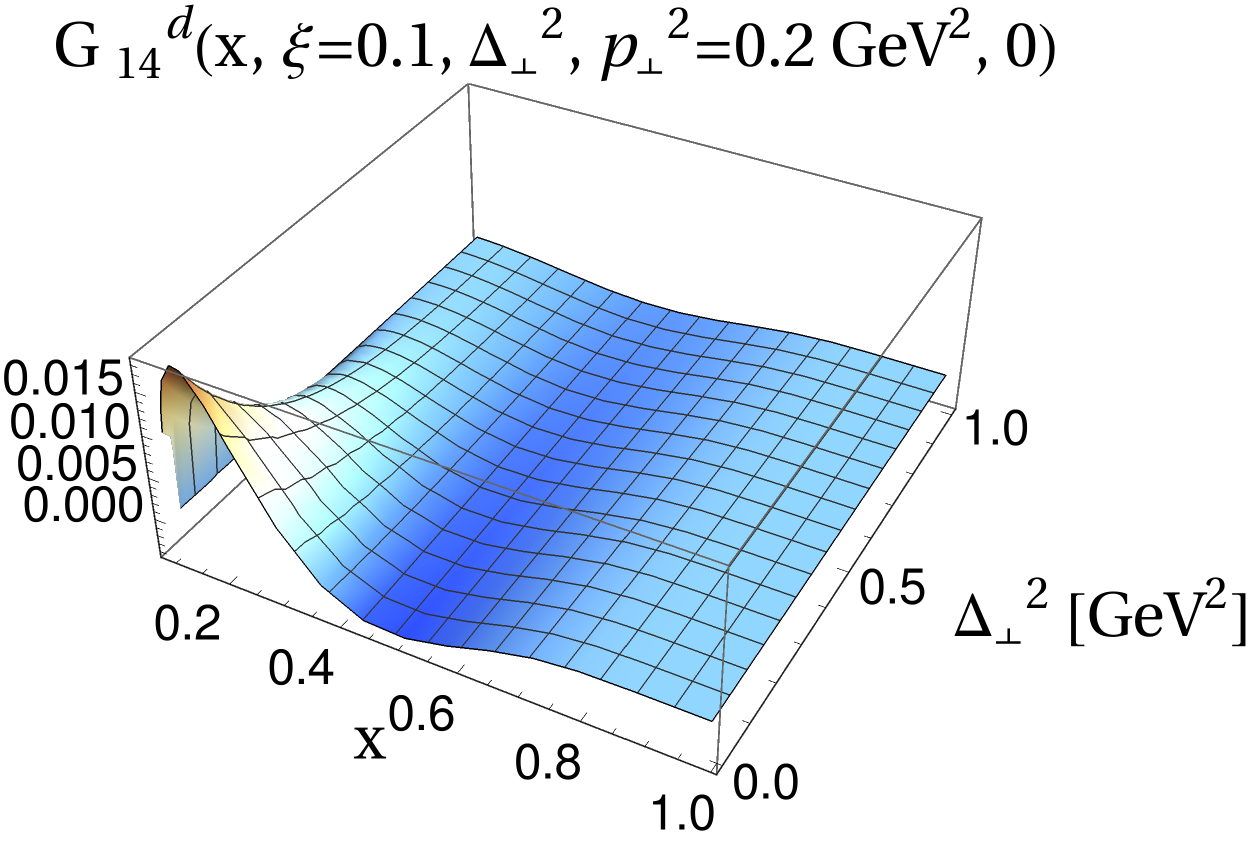} 
\caption{\label{longxD} The GTMDs as functions of $x$ and $\bfd^2$ for a longitudinally polarized. The upper panel is for the $u$ quark, while the lower panel represents the results for the $d$ quark. We fix $\xi=0.1$, $\bfp^2=0.3$~GeV$^2$ and $\bfd \perp\bfp$.  Left to right panels represent the GTMDs $G_{1,1},\, G_{1,2},\, G_{1,3}$, and  $G_{1,4}$, respectively.}
\end{figure}

\subsection{Longitudinally polarized quark}
The GTMDs for the longitudinally polarized quark, i.e, $G_{1,1},\, G_{1,2},\, G_{1,3}$, and $G_{1,4}$ as functions of $x$ and $\xi$ for fixed $\bfd=0.2$ GeV$^2$ and $\bfp=0.3$ GeV$^2$ are shown in Fig.~\ref{longzx}. As we mentioned earlier, the distributions are evaluated in the DGLAP region, $x> \xi$.  The distributions $G_{1,2}$ and  $G_{1,3}$ are positive for the $u$ quark and they are negative for the $d$ quark, whereas $G_{1,1}$ shows negative distribution for both the quarks. The $G_{1,4}$ for the $d$ quark is positive at low-$x$ and slightly negative around $x=0.5$, while for the $u$ quark it exhibits distinctly different behavior having a negative peak at lower-$x$ and a positive peak at larger-$x$. At the $\xi=0$ limit, the spin-orbit correlation of a quark can be expressed in terms of  $G_{1,1}$  \cite{Lorce:2011kd,Chakrabarti:2017teq}:
\begin{align}
C_z^\nu=-\int {\rm d}x\,{\rm d}^2\bfp\, \frac{\bfp^2}{M^2}\,G_{1,1}^\nu(x,0,\bfp^2,0,0)\,,
\end{align}
where $C_z^\nu>0$ indicates that the quark spin and OAM tend to be aligned and $C_z^\nu<0$ implies that they are antialigned. The negative $G_{1,1}^\nu$ distribution in our model indicates that $C_z^\nu>0$, reflecting quark spin and OAM tend to be aligned. The $G_{1,4}$ shows a dipolar behavior with  opposite polarity for the $u$ and $d$ quarks and this GTMD at vanishing skewness and $\bfd^2=0$ limit contributes to the axial charge $g_A $ defined as $g_A = \int {\rm d}x\,{\rm d}^2\bfp\, G_{1,4}(x,0,\bfp^2,0,0)  $, which is related to the spin as $s^\nu_z=\frac{1}{2} g^\nu_A$.  At the GPD limit ($t=-\bfd^2$ and integrating over $\bfp$), the GPDs $\tilde{H}$ and $\tilde{E}$ can be expressed in terms of $G_{1,2},\, G_{1,3},\, G_{1,4}$ as shown in Eqs.~({\ref{Ht}) and (\ref{Et}}). We illustrate the $x$ and $\bfd^2$ dependence of the longitudinally polarized quark GTMDs in Fig.\ref{longxD}. We find that the qualitative behavior of the polarized and unpolarized GTMDs are more or less similar. 

\section{Wigner Distributions in boost-invariant longitudinal space} \label{sec_WD_sigma}

The Wigner distributions in the transverse impact parameter space have been studied extensively in several models including the LFQDM model for zero skewness. The transverse impact parameter $\bfb$ is the Fourier conjugate to the variable $\bfD=\bfd/\zfs$~\cite{Diehl:2002he,Burkardt:2002hr,Ralston:2001xs,Kaur:2018ewq}, which simply reduces to $\bfd$ for zero skewness ($\xi=0$).
Meanwhile, the skewness variable $\xi$ is conjugate to the boost-invariant longitudinal impact parameter defined as $\sigma=\frac{1}{2}b^-P^+$. The Fourier transformation of the correlator $W^{\nu [\Gamma]}_{[\lambda^{\prime\prime}\lambda^{\prime}]}(x,\xi,\bfd, \bfp)$ with respect to skewness variable $\xi$ provides a distribution  in the boost-invariant longitudinal space $\sigma$.
Notably, the Fourier transform of the DVCS amplitude with respect to $\xi$ at fixed invariant momentum transfer provides an interesting diffraction pattern in the longitudinal impact-parameter space \cite{Brodsky:2006in,Brodsky:2006ku}. The results were analogous to the diffractive scattering of a wave in optics. On the other hand, the GPDs extracted in different phenomenological models~\cite{Chakrabarti:2008mw,Manohar:2010zm,Kumar:2015fta} and AdS/QCD inspired model~\cite{Mondal:2015uha,Chakrabarti:2015ama,Mondal:2017wbf,Kaur:2018ewq} exhibit an analogous behavior in longitudinal boost-invariant space. It is therefore interesting to study a more general distribution, Wigner distribution, in the longitudinal impact parameter space, which is defined as
\be
\tilde{\rho}^{\nu [\Gamma]}(x,\sigma,\bfd,\bfp;S)=\int^{\xi_{s}}_0 \frac{{\rm d} \xi}{2\pi}\, e^{i\sigma.\xi} \,W^{\nu [\Gamma]}(x, \xi, \bfd,\bfp;S)\,,
\label{wig_rho_sig}
\ee
where the upper limit of the integration, $\xi_{s}$, is equivalent to the slit width that provides a necessary condition for occurring of the diffraction pattern.
Since we are considering the region $\xi<x<1$, the upper limit of the integration $\xi_s$ is given by $\xi_{s}=x$ if $\xi_{\rm max}>x$; otherwise it is given by $\xi_{s}=\xi_{\rm max}$ if $\xi_{\rm max}<x$, where the maximum value of $\xi$ for a fixed value of $- t$ is given by~\cite{Brodsky:2006in,Brodsky:2006ku,Chakrabarti:2008mw,Manohar:2010zm}
\be 
\xi_{\rm max}=\frac{-t}{2M^2} \bigg( \sqrt{1+\frac{4M^2}{(-t)}}-1 \bigg)\,. \label{zmax_def}
\ee

Similar to the Wigner distributions in the $\bfb$ space for various polarization configurations of the proton and the quark~\cite{Lorce:2011kd,Liu:2015eqa,Chakrabarti:2017teq}, in the longitudinal position space they are defined as
\be 
\tilde{\rho}^{\nu }_{UY}(x,\sigma,\bfd,\bfp)&=&\frac{1}{2}[\tilde{\rho}^{\nu [\Gamma_Y]}(x,\sigma,\bfd,\bfp; +\hat{S}_z) + \tilde{\rho}^{\nu [\Gamma_Y]}(x,\sigma,\bfd,\bfp; -\hat{S}_z)]\,,\label{rho_UY_def}\\
\tilde{\rho}^{\nu }_{LY}(x,\sigma,\bfd,\bfp)&=&\frac{1}{2}[\tilde{\rho}^{\nu [\Gamma_Y]}(x,\sigma,\bfd,\bfp; +\hat{S}_z) - \tilde{\rho}^{\nu [\Gamma_Y]}(x,\sigma,\bfd,\bfp; -\hat{S}_z)]\,,\label{rho_LY_def}\\
\tilde{\rho}^{j \nu }_{TY}(x,\sigma,\bfd,\bfp)&=&\frac{1}{2}[\tilde{\rho}^{\nu [\Gamma_Y]}(x,\sigma,\bfd,\bfp; +\hat{S}_j) - \tilde{\rho}^{\nu [\Gamma_Y]}(x,\sigma,\bfd,\bfp; -\hat{S}_j)]\,,\label{rho_TY_def}
\ee
where the subscripts in the first place $U,\, L$, and $T$ represent the proton polarizations, i.e, unpolarized, longitudinally polarized, and transversely polarized, respectively and $Y=\{U,\,L,\,T\}$ defines the quark polarizations and the corresponding Dirac structures $\{\Gamma_Y=\gamma^+,\,\gamma^+\gamma^5,\, i\sigma^{j+}\gamma^5\}$. The longitudinal spin of the proton is represented by $\hat{S}_z$ and $\hat{S}_j$ is the transverse spin of proton along $x$ and $y$ axis with $j=1,\,2$, respectively. Thus, each of the Eqs.~(\ref{rho_UY_def})--(\ref{rho_TY_def}) stands for the three distributions for $Y=\{U,\,L,\,T\}$ and altogether, we have nine Wigner distributions for different polarization combinations of the quark and the proton. We have another polarization combination when the quark and the proton are polarized in right angle. For this polarization combination, the Wigner distribution, also known as pretzelous distribution, is defined as
\be 
\tilde{\rho}^{ \perp j \nu}_{TT}((x,\sigma,\bfd,\bfp)&=&\epsilon^{ij}_\perp(-1)^j\frac{1}{2} [\tilde{\rho}^{\nu [i\sigma^{j+}\gamma^5]}(x,\sigma,\bfd,\bfp; +\hat{S}_i) -\tilde{\rho}^{\nu [i\sigma^{j+}\gamma^5]}(x,\sigma,\bfd,\bfp; -\hat{S}_i)]\label{rho_TTprp_def}.
\ee 

Using the definition of the Wigner distributions in boost-invariant longitudinal impact parameter space, Eq.~\eqref{wig_rho_sig}, in Eqs.~(\ref{rho_UY_def})--(\ref{rho_TTprp_def}), the distributions can be parametrized in terms of leading twist GTMDs as: \\
(i) for unpolarized proton
\be
\tilde{\rho}^\nu_{UU}(x,\sigma,\bfd,\bfp)&=& \int^{\xi_{s}}_0 \frac{d \xi}{2\pi} e^{i\sigma.\xi}  \frac{1}{\zf} F^\nu_{1,1}\,, \label{rhoUU_F}\\
\tilde{\rho}^\nu_{UL}(x,\sigma,\bfd,\bfp)&=& \int^{\xi_{s}}_0 \frac{d \xi}{2\pi} e^{i\sigma.\xi}  \frac{-i}{M^2\zf}\epsilon^{ij}_\perp p^i_\perp \dpi^j G^\nu_{1,1}\,, \label{rhoUL_G}\\
\tilde{\rho}^{\nu j}_{UT}(x,\sigma,\bfd,\bfp)&=&  \int^{\xi_{s}}_0 \frac{d \xi}{2\pi} e^{i\sigma.\xi}  \frac{-i}{M\zf}\epsilon^{ij}_\perp\bigg[ p^i_\perp H^\nu_{1,1}+  \dpi^i H^\nu_{1,2}\bigg]\,, \label{rhoUT_H}
\ee
(ii) for longitudinally polarized proton
\be 
\tilde{\rho}^\nu_{LU}(x,\sigma,\bfd,\bfp)&=&  \int^{\xi_{s}}_0 \frac{d \xi}{2\pi} e^{i\sigma.\xi} \frac{i}{M^2\zf}\epsilon^{ij}_\perp p^i_\perp \dpi^j F^\nu_{1,4}\,,\label{rhoLU_F} \\
\tilde{\rho}^\nu_{LL}(x,\sigma,\bfd,\bfp)&=& \int^{\xi_{s}}_0 \frac{d \xi}{2\pi} e^{i\sigma.\xi} \frac{2}{\zf} G^\nu_{1,4}\,,\label{rhoLL_G}\\
\tilde{\rho}^{\nu j}_{LT}(x,\sigma,\bfd,\bfp)&=& \int^{\xi_{s}}_0 \frac{d \xi}{2\pi} e^{i\sigma.\xi} \frac{2 }{M\zf} \bigg[ p^j_\perp H^\nu_{1,7}+ \dpi^j H^\nu_{1,8}\bigg]\,, \label{rhoLT_H}
\ee
(iii) for transversely polarized proton
\be
\tilde{\rho}^{i\nu}_{TU}(x,\sigma,\bfd,\bfp)&=&  \int^{\xi_{s}}_0 \frac{d \xi}{2\pi} e^{i\sigma.\xi} \frac{-i}{2M\zf}\epsilon^{ij}_\perp \bigg[\dpi^j \bigg(F^\nu_{1,1}-2 \zfs F^\nu_{1,3}\bigg) \nonumber\\
&&- 2\zfs p^j_\perp F^\nu_{1,2} + \frac{\xi }{M^2}\epsilon^{kl}_\perp \pp^k \dpi^l \dpi^j F^\nu_{1,4} \bigg] \,,\label{rhoTU_F}\\
\tilde{\rho}^{i\nu}_{TL}(x,\sigma,\bfd,\bfp)&=&\int^{\xi_{s}}_0 \frac{d \xi}{2\pi} e^{i\sigma.\xi}  \bigg[\frac{-1}{2M^3\zfs^{3/2}} \epsilon^{ij}_\perp \epsilon^{kl}_\perp \pp^k \dpi^l \dpi^j G_{1,1} + \frac{\zf}{M}p^i_\perp G^\nu_{1,2} \nonumber \\
&&+ \frac{1}{M\zf} \dpi^i \bigg( \zfs G^\nu_{1,3}-\xi G^\nu_{1,4}\bigg) \bigg]\,, \label{rhoTL_G}\\
\tilde{\rho}^{j\nu}_{TT}(x,\sigma,\bfd,\bfp)&=& \int^{\xi_{s}}_0 \frac{d \xi}{2\pi} e^{i\sigma.\xi}  \epsilon^{ij}_\perp(-1)^j \bigg[\frac{1}{2M^2\zf} \bigg( \pp^i \dpi^i  H_{1,1} + (\dpi^i)^2 H_{1,2}\bigg)\nonumber\\
&& + \zf H^\nu_{1,3}+ \frac{\zf}{M^2}(\pp^j )^2 H^\nu_{1,4} + \frac{1}{M^2\zf} \pp^j \dpi^j \bigg(\zfs H^\nu_{1,5}  - \xi H^\nu_{1,7} \bigg)\nonumber \\
&&+ \frac{1}{M^2\zf} (\dpi^j)^2\bigg( \zfs H^\nu_{1,6} - \xi H^\nu_{1,8} \bigg) \bigg]\,. \label{rhoTT_H}
\ee
The pretzelous distribution is parametrized as
\be 
\tilde{\rho}^{\perp j \nu}_{TT}(x,\sigma,\bfd,\bfp)&=& \int^{\xi_{s}}_0 \frac{d \xi}{2\pi} e^{i\sigma.\xi}  \epsilon^{ij}_\perp \bigg[-\frac{1}{2M^2\zf}  \pp^i \dpi^j \bigg(H^\nu_{1,1} - 2 \zfs H^\nu_{1,5}\bigg)\nonumber\\
&&- \frac{1}{2M^2\zf} \dpi^i \dpi^j \bigg(H^\nu_{1,2}-2 \zfs H^\nu_{1,6}- \zeta  H^\nu_{1,8}\bigg) \nonumber\\
&&  + \frac{\zf}{M^2}  \pp^i \pp^j H^\nu_{1,4} + \frac{\xi}{2M^2\zf}  \pp^j \dpi^i H^\nu_{1,7}  \bigg]\,. \label{rhoTTperp_H}
\ee
All the GTMDs in Eqs.~(\ref{rhoUU_F})--(\ref{rhoTTperp_H}),  $ F^\nu_{1,m}, \, G^\nu_{1,m}$, and $H^\nu_{1,n}$ (for $m=1,2,3,4 $ and $n=1,2,3...8$) depend on the set of variables ($x,\xi,\bfp^2,\bfp.\bfd,\bfd^2$) and the Fourier transformation with respect to $\xi$ gives the Wigner distributions in the conjugate space $\sigma$.
Note that, each of the distributions carries flavor index $\nu$ and the flavor $u$ and $d$ are distinguished by the flavor dependent model parameters $a^\nu_i, b^\nu_i$ encoded in the LFWFs $\varphi^{(\nu)}_i$ of Eq.~\eqref{LFWF_phi}. 

\begin{figure}[h]
\includegraphics[scale=.4]{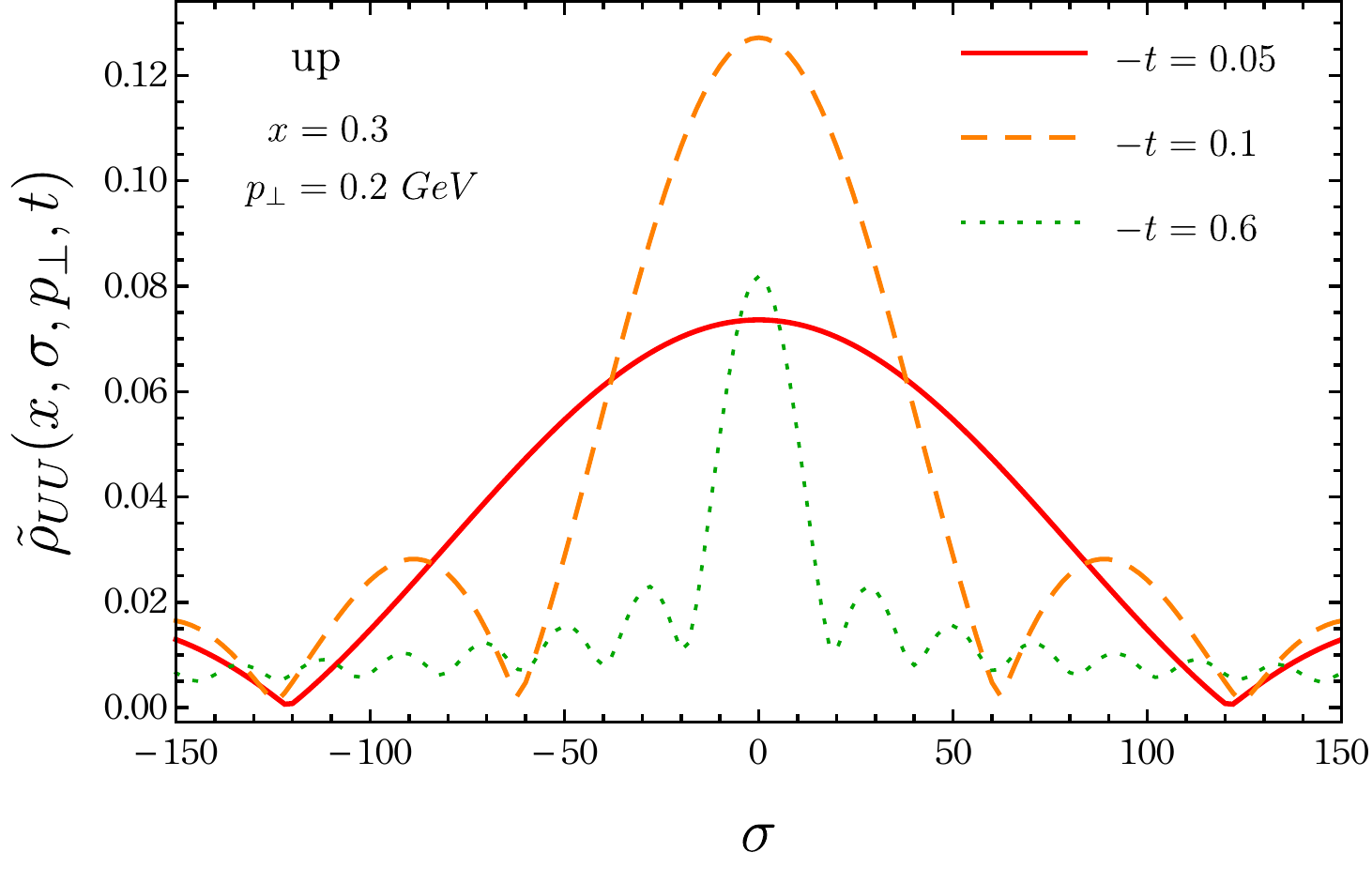} 
\includegraphics[scale=.4]{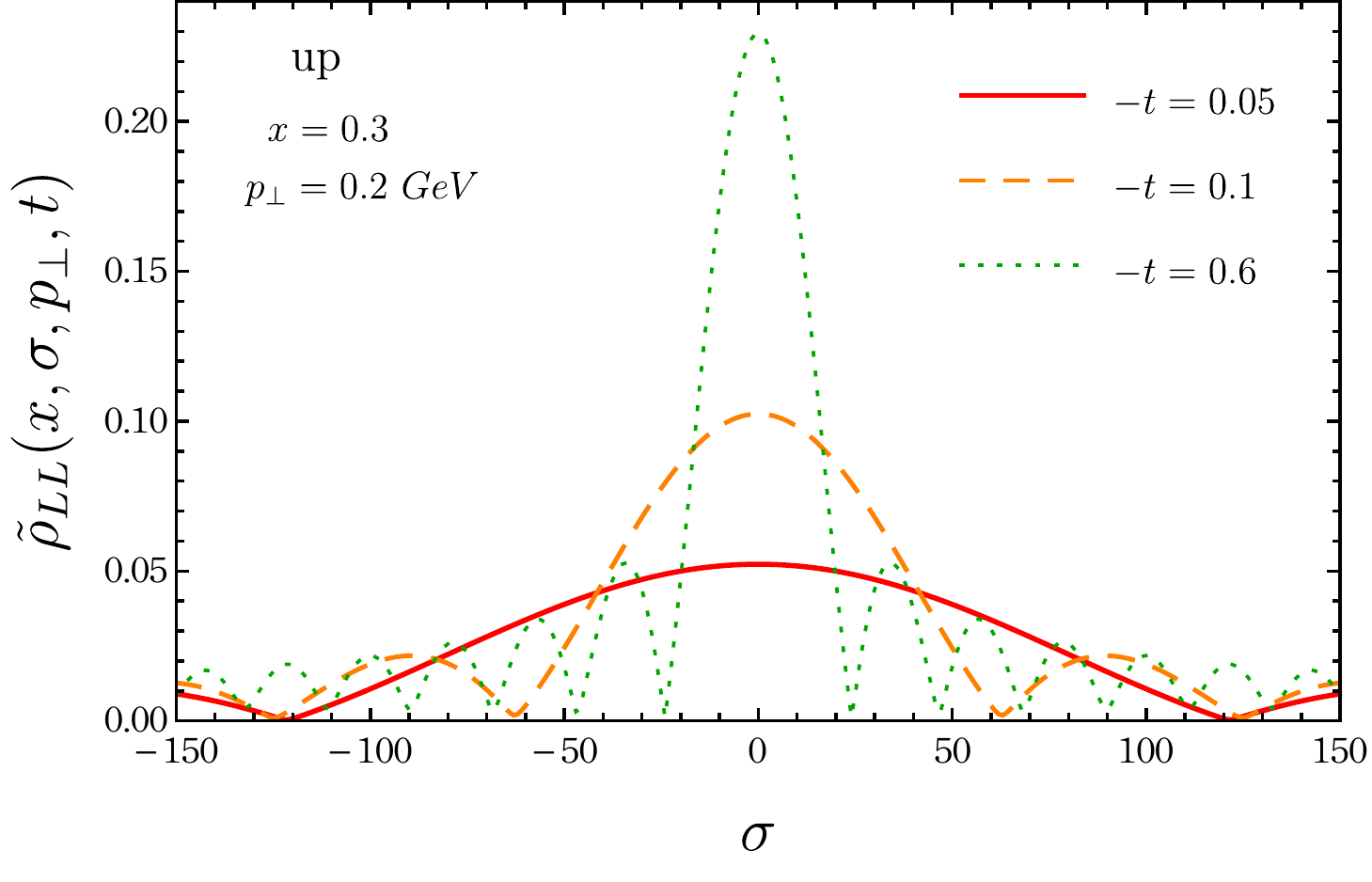} \\
\includegraphics[scale=.4]{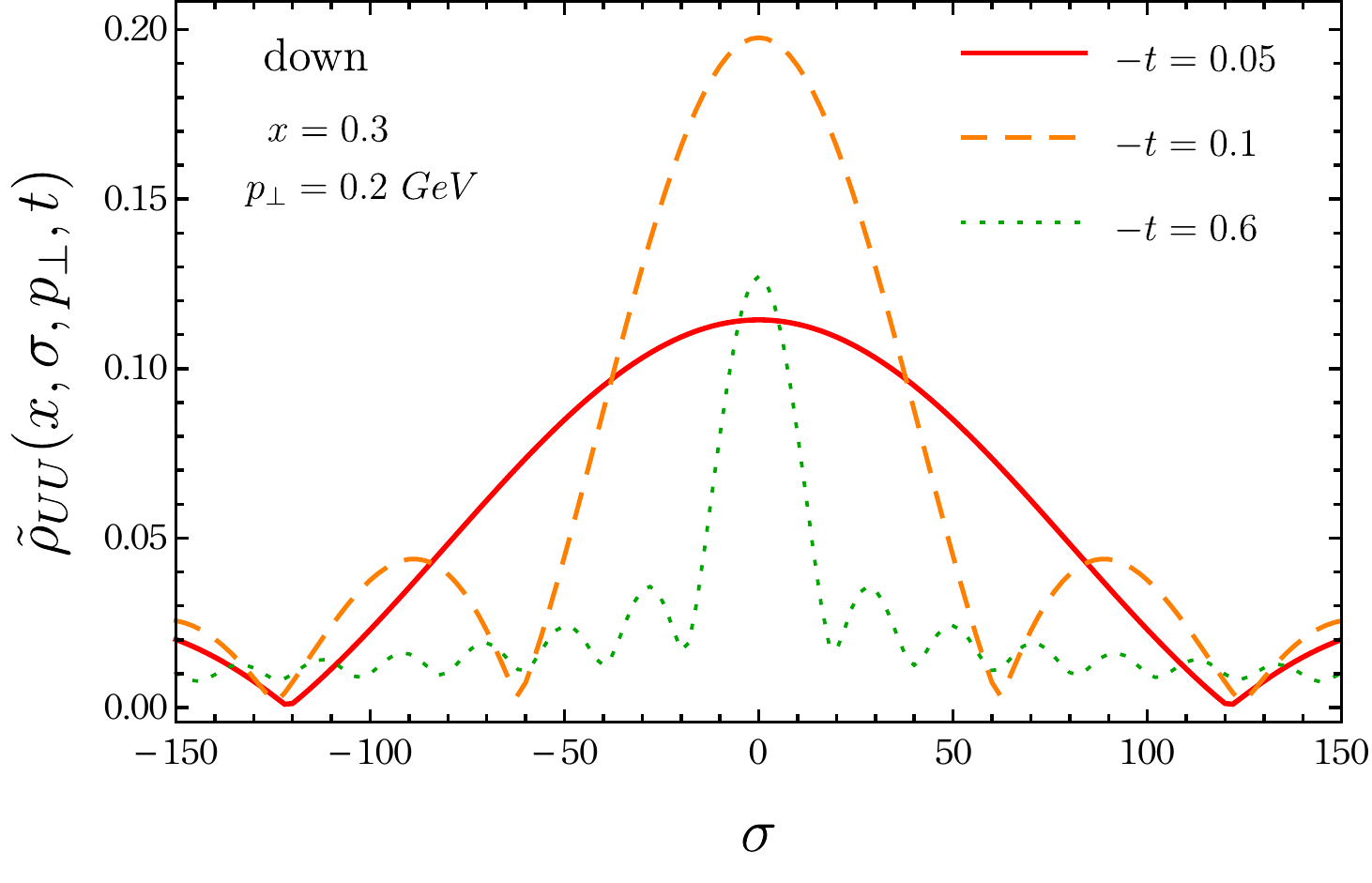} 
\includegraphics[scale=.4]{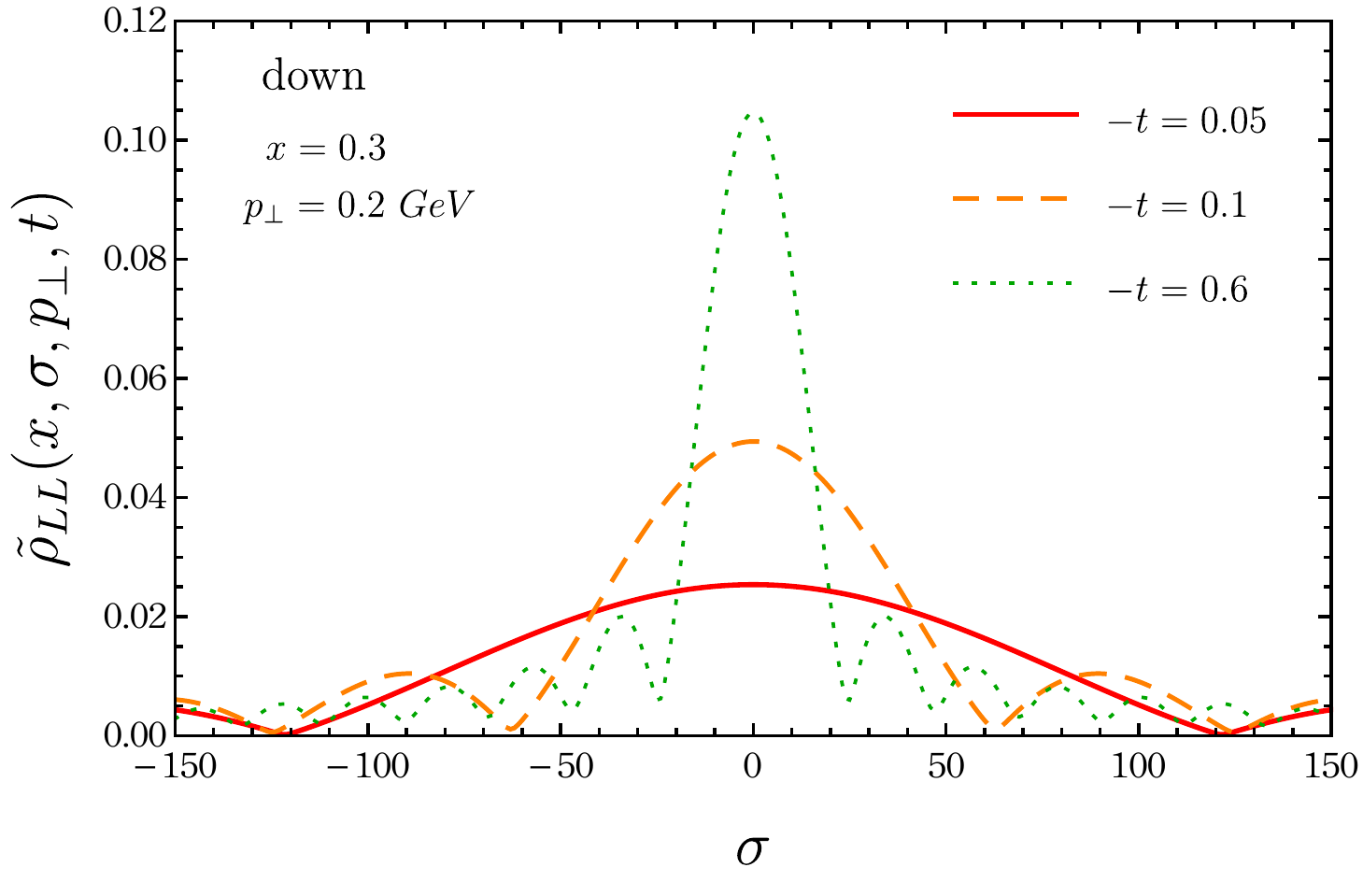} 
\caption{\label{rhoS3} The Wigner distribution  $\tilde{\rho}_{UU}$ (left panel) and $ \tilde{\rho}_{LL}$ (right panel)  in the boost invariant longitudinal position space at different values of $-t$ in GeV$^2$ for the $u$ (upper panel) and $d$ (lower panel) quarks.}
\end{figure}

\begin{figure}[h]
\includegraphics[scale=.4]{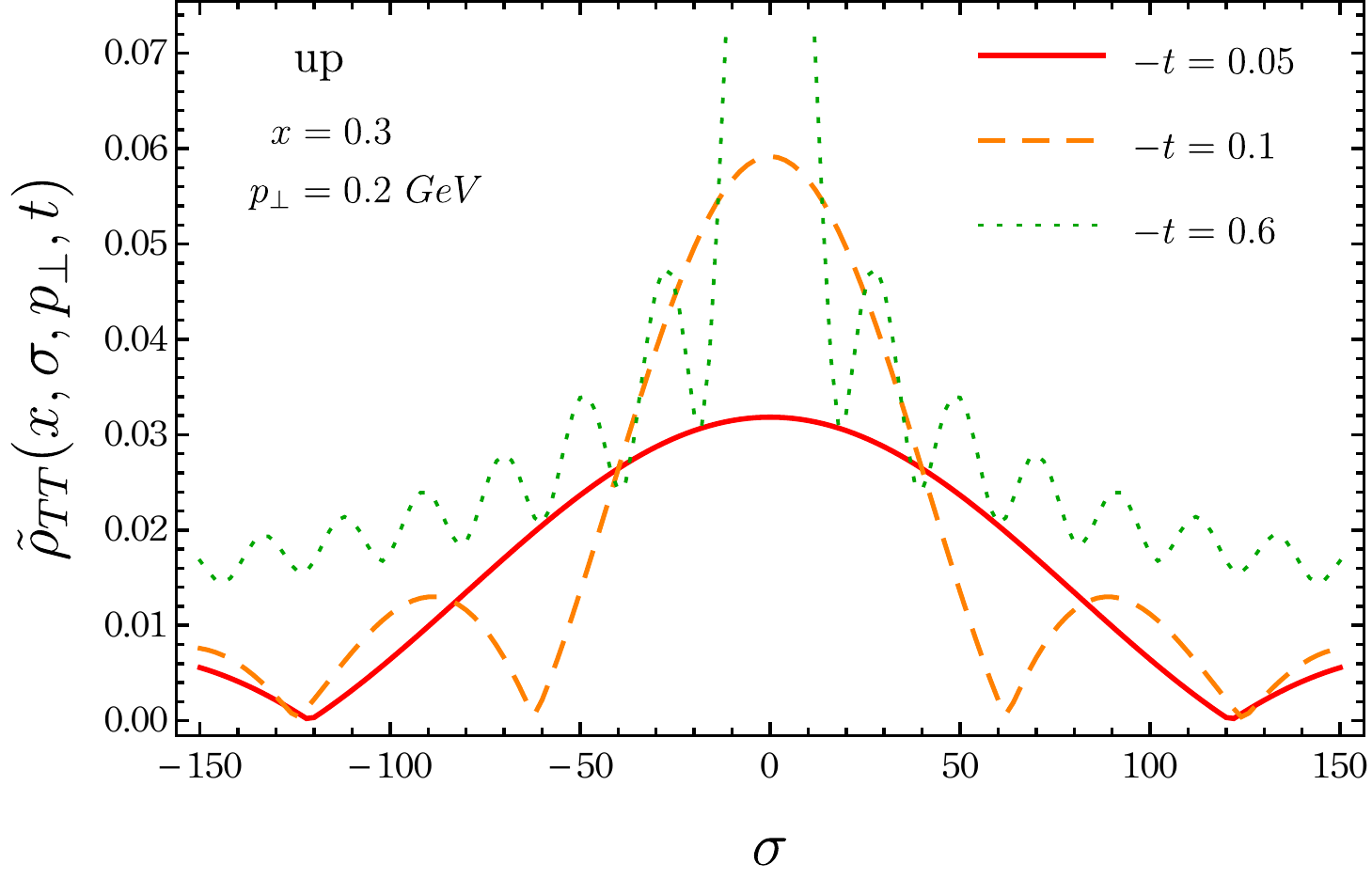}
\includegraphics[scale=.4]{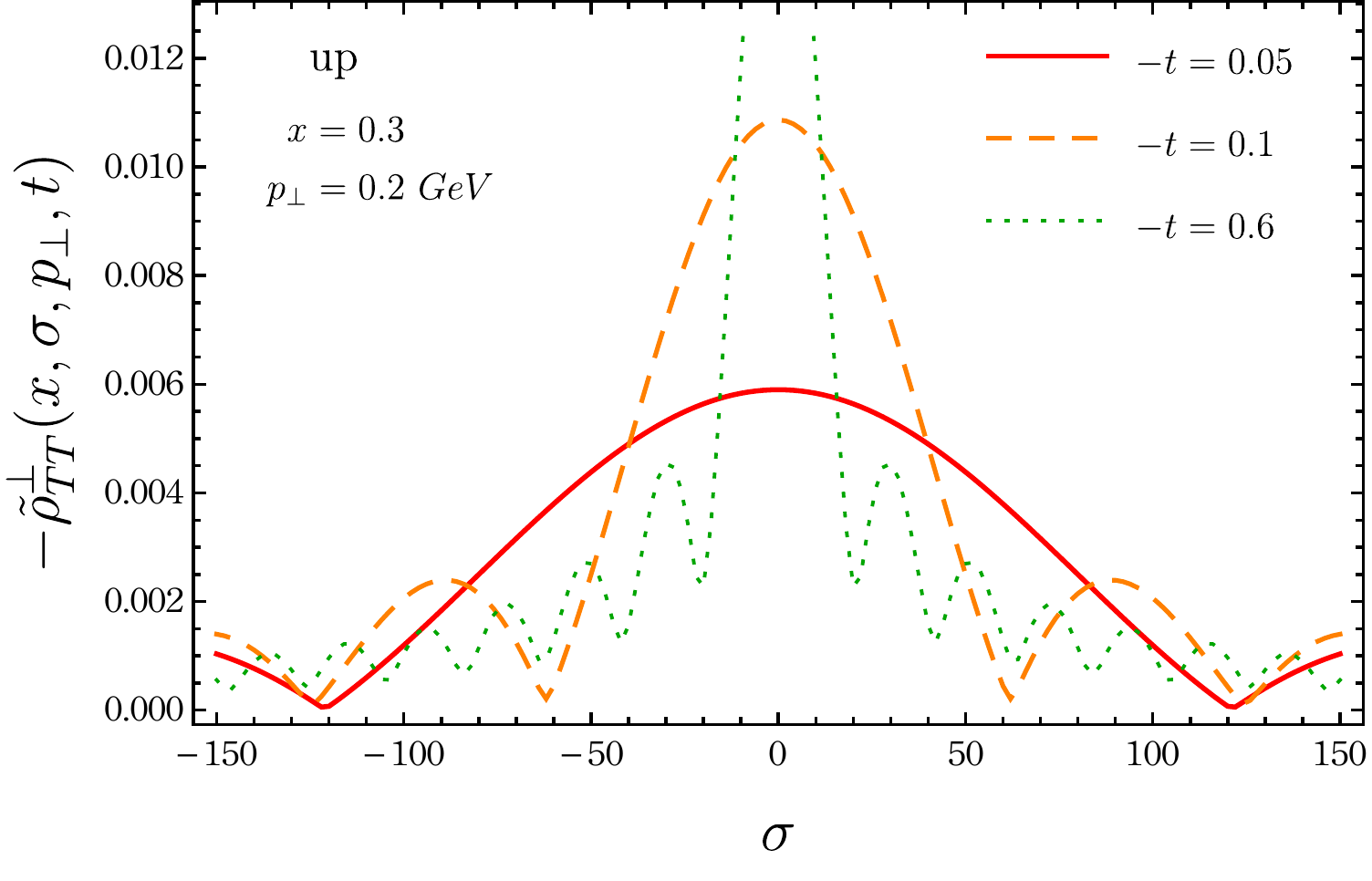} \\
\includegraphics[scale=.4]{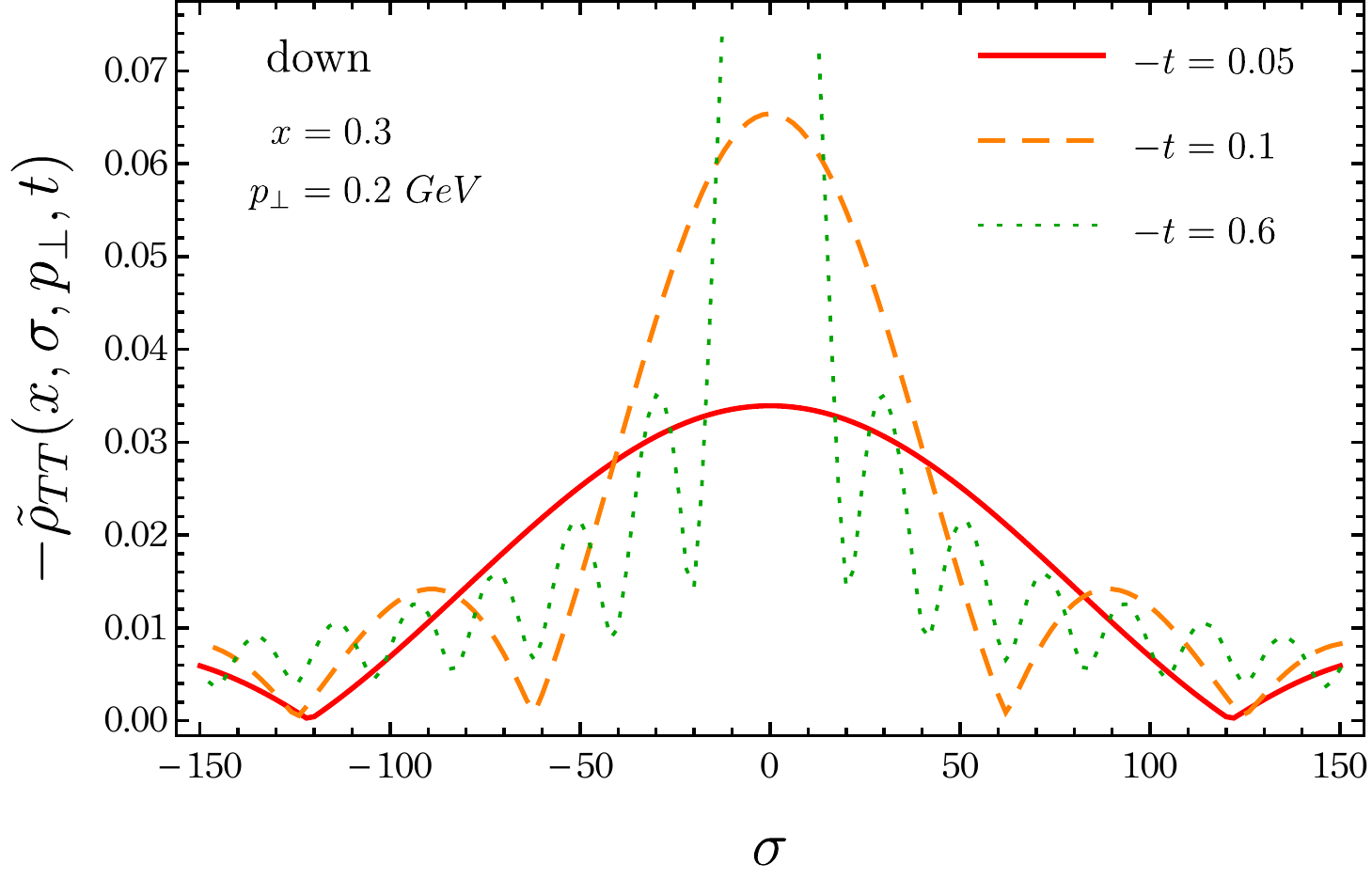} 
\includegraphics[scale=.4]{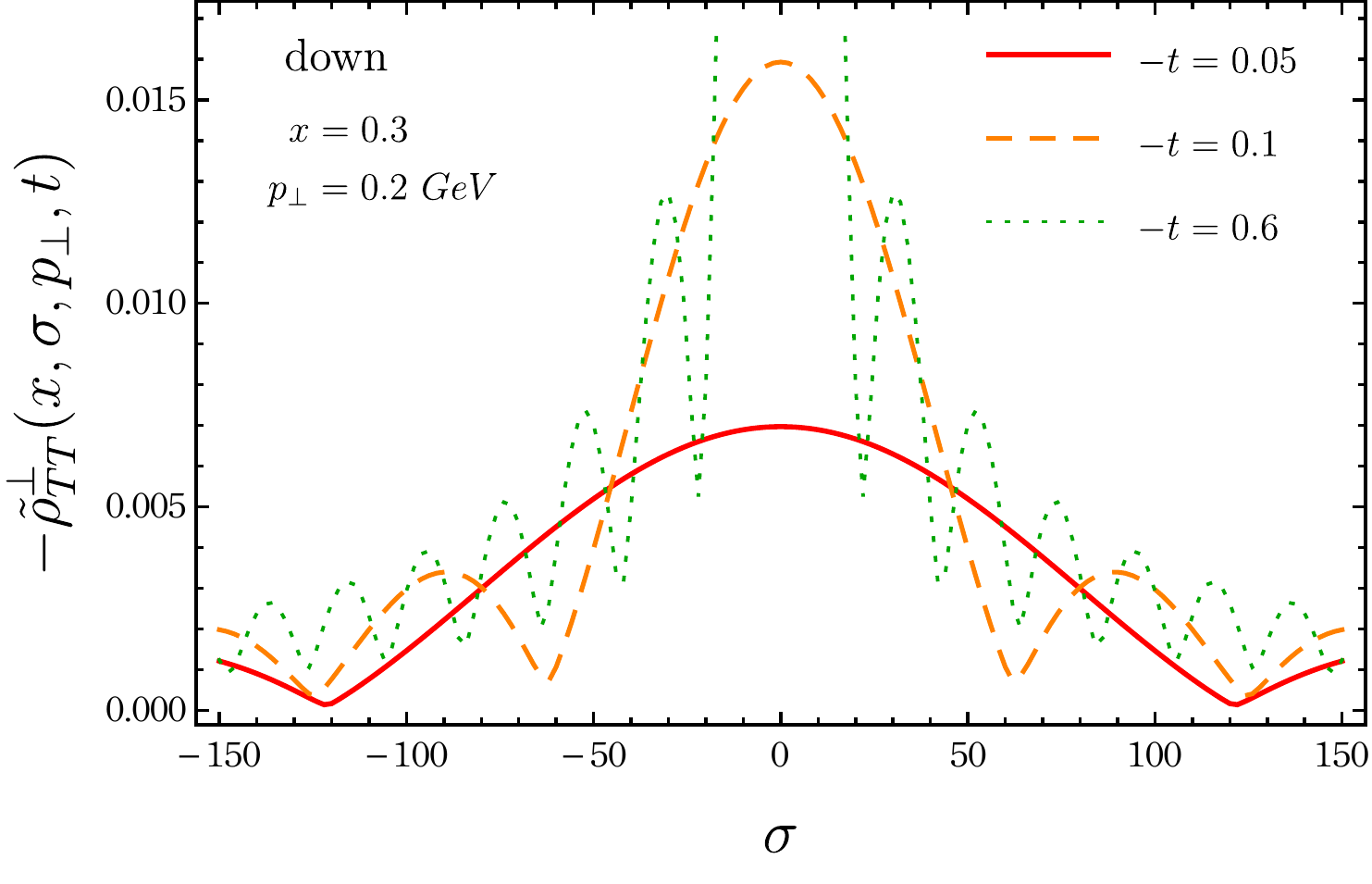}
\caption{\label{rhoTTp} The Wigner distribution  $ \tilde{\rho}^j_{TT}$ (left panel) and $  \tilde{\rho}^{\perp j}_{TT}$ (right panel)  in the boost invariant longitudinal position space at different values of $-t$ in GeV$^2$ for the $u$ (upper panel) and $d$ (lower panel) quarks.}
\end{figure}

The analytical results in Eqs.~(\ref{rhoUU_F})--(\ref{rhoTTperp_H}) are used for further numerical computation and few of the distributions in $\sigma$ space are shown in Fig.~\ref{rhoS3} and Fig.~\ref{rhoTTp}.  
All the distributions are functions of $\tilde{\rho}^{\nu}_{XY}(x,\sigma,\bfd,\bfp)$ and using the relation between $\bfd$ and total momentum transfer square $-t$,  as given in Eq.(\ref{mt_def}), the distributions are eventually expressed as  functions of $\tilde{\rho}^{\nu}_{XY}(x,\sigma, t,\bfp)$. In Fig.~\ref{rhoS3}, we illustrate the distributions $\tilde{\rho}^{\nu}_{UU}$ and $ \tilde{\rho}^{\nu}_{LL}$  when both the quark and proton are unpolarized and longitudinally polarized, respectively, as function of $\sigma$ at fixed $x=0.3$, $\bfp=0.2~ \hat{p}_y$ GeV, and $ -t=\{0.05,\,0.1,\,0.6\}$ GeV$^2$. 
The three different values of $-t$ correspond to the values of $\xi_{\rm max} \approx\{ 0.052,\, 0.101,\, 0.466\}$, respectively, which reflect the upper limit of the $\xi$ integration in Eq.~(\ref{zmax_def}), $\xi_s=0.052$ and $0.101$ for $-t=0.05$ and $0.1$, respectively, while for $-t=0.6$ the integration limit is $\xi_s=x=0.3$ since $\xi_{\rm max}>x$.  
Note that to get the non-vanishing contribution  of $\bfp.\bfb$, we prefer to choose $\bfd \parallel \bfp$. However, in some cases, for example, $\tilde{\rho}^{\nu}_{UL}$, $\tilde{\rho}^{\nu}_{LU}$, etc., presented in the Appendix~\ref{AppOtherWD}, which involve $\epsilon_\perp^{ij} p^i \Delta^j$, we consider $\bfd\perp\bfp$.  The $\rho_{UU}$ has the contribution from the GTMD $F_{1,1}$  and the $\rho_{LL}$ involves the GTMD $G_{1,4}$. For non-zero skewness, these two distributions, Eqs.~(\ref{F11}) and (\ref{G14}), have an additional contribution containing  $\bfp.\bfd$.  

Our results for the $\tilde{\rho}_{UU}$ and $\tilde{\rho}_{LL}$ in the longitudinal position space show an oscillatory behavior, which can be viewed as the diffraction pattern generated by the single slit experiment in optics. The size of the principle maxima in the diffraction pattern  is inversely proportional to the slit width. The
finite size of the $\xi$ in the Fourier transformation in  Eq.~(\ref{wig_rho_sig}) is responsible for producing the diffraction pattern, where  $\xi_s$ plays the role of the slit width of the single slit experiment. We should also mention here that the
Fourier transform with a finite range of $\xi$ of any arbitrary function does not provide the diffraction pattern~\cite{Manohar:2010zm}.  We observe that as $-t$ increases, $\xi_s$ also increases and the width of the principle maxima consequently decreases. In other words, the position of the first minima shifts towards the center with increasing $-t$. Note that a similar diffraction pattern in longitudinal position space has also been observed in DVCS aplitude~\cite{Brodsky:2006in,Brodsky:2006ku}, GPDs~\cite{Chakrabarti:2008mw,Manohar:2010zm,Kumar:2015fta,Mondal:2015uha,Chakrabarti:2015ama,Mondal:2017wbf,Kaur:2018ewq}, and the coordinate-space parton density~\cite{Miller:2019ysh}. Thus, this interesting feature of the Wigner distributions in $\sigma$ space is not very surprising.  For $\tilde{\rho}_{UU}$, the magnitude of the peak of the principal maxima increases gradually upto the limit $\xi_{\rm max}=x$ and beyond that region, e.g., $\xi_{\rm max}> x $, it decreases as shown in green dots for $-t=0.6$ GeV$^2$. Meanwhile, the magnitude of the maxima in $\rho_{LL}$ continuously increases as $-t$ increases. Expect in the magnitude, both the $u$ and $d$ quarks exhibit identical features for $\tilde{\rho}_{UU}$ and $\tilde{\rho}_{LL}$. 

The Wigner distributions $\tilde{\rho}_{TT}$ and $\tilde{\rho}_{TT}^\perp$ in longitudinal position space are illustrate in Fig.~\ref{rhoTTp}. Each of these distributions, Eqs.~(\ref{rhoTT_H}) and (\ref{rhoTTperp_H}), has contributions from several GTMDs with some prefactor of momentum structures e.g., $\epsilon^{ij}_\perp  p_\perp^i,\,\Delta_\perp^i$, $\epsilon^{ij}_\perp  p_\perp^i\Delta_\perp^j$,  $\epsilon^{ij}_\perp  p_\perp^ip_\perp^j$, $\epsilon^{ij}_\perp  \Delta_\perp^i\Delta_\perp^j$, etc. For the choice $\bfp$ and $\bfd$ both along the $y$-axis, the contributions from some of the GTMDs vanish irrespective of the choice of the quark polarization $j=1,\,2$. 
For example,  with $j=1$ in $\tilde{\rho}^j_{TT}$, the prefactors of $H_{1,4-8} $ become zero, while for  $j=2$, the prefactors of $H_{1,1}, H_{1,2} $ vanish.  
In case of $\tilde{\rho}^{\perp j}_{TT}$, the choice of $\bfp$ and $\bfd$ both along the same axis leads to $\tilde{\rho}_{TT}^\perp=0$. Even for the choice $\bfp \equiv (0,\, |\bfp|),  \bfd \equiv (|\bfd|,\,0) $,  with $j=1$, only the contributions from the $H_{1,1}, H_{1,5} $ survive and with  $j=2$, only the contributions from $H_{1,1}, H_{1,2} $ are nonzero. Therefore, to get the contributions from all the involved GTMDs, the preferable choice is $\bfp \equiv \left(|\bfp|/\sqrt{2},\, |\bfp|/\sqrt{2}\right)$ and $\bfd \equiv \left(|\bfd|/\sqrt{2},\, |\bfd|/\sqrt{2}\right)$  with $j=1$ or $2$. 

We observe that $\tilde{\rho}_{TT}$ and $\tilde{\rho}^{\perp}_{TT}$ show a similar diffraction as seen in $\tilde{\rho}_{UU}$ and $\tilde{\rho}^{\perp}_{LL}$. 
With increasing $-t$, the distributions shift along $y$-axis and trend toward overall single-peaked functions.
 For $-t=0.6$ GeV$^2$, the $\tilde{\rho}^u_{TT}$ exhibits distinctly different behavior, where the magnitudes of the secondary maxia and minima are comparatively higher than that in other distributions.  We also observe a sign flip from $u$ to $d$ quarks in $\tilde{\rho}_{TT}$, whereas $\tilde{\rho}^{\perp}_{TT}$ shows negative distributions for both the flavors. The numerical results of the other Wigner distributions in the boost invariant longitudinal space are presented in the Appendix~\ref{AppOtherWD}. For all the values of $-t$, the distributions do not show the prominent diffraction pattern. For example, $\tilde{\rho}_{TT}$ for the $u$ displays a central minima instead of a maxima for $-t=0.6$ GeV$^2$, it also does not show the prominent pattern for the $d$. These implies that the diffraction pattern is not solely due to the finite size of the $\xi$ integration, and the functional forms of the GTMDs are also important for this phenomenon. Notably, all the distributions in the boost invariant longitudinal space feature a long-distance tail as reported in Refs.~\cite{Miller:2019ysh,Weller:2021wog}. 



\section{Conclusions}\label{con}
We calculated all the leading twist quark GTMDs in the proton, when the momentum transfer is considered in both the transverse and the longitudinal directions. We presented the results in a light-front quark-diquark model motivated by soft-wall AdS/QCD considering the DGLAP region, i.e., for $x>\xi$. We then employed the skewness dependent GTMDs to investigate the quark Wigner distributions in the boost invariant longitudinal position space with all the possible polarization combinations of the quark and the proton. We observed that the Wigner distributions in the longitudinal position space for a fixed $x$ and $\bfp$ exhibit a diffraction pattern. The maxima of the distributions are sensitive to the amount of the square of momentum transfer, $-t$. The widths of the maxima become narrower and the positions of the minima move towards center with the increasing $-t$. In optics, the similar diffraction pattern is observed from a single slit experiment, where the size of the central maxima is inversely proportional to the  slit width. Our results are analogous to the diffractive scattering of a waves in optics and finiteness of $\xi$ integration (the upper limit, $\xi_s$) plays the role of the slit width. However, the diffraction pattern is not solely due to finiteness of $\xi$ integration and the functional behaviors of the GTMDs are crucial to have the phenomenon. A similar diffraction pattern has also been observed in several other observable such as DVCS aplitude, GPDs, and the parton density in longitudinal position space.

\begin{acknowledgments}
The work of T. M. and D. K. is supported by the National Key Research and Development Program of China under Contracts No. 2020YFA0406301, the National Natural Science Foundation of China (NSFC) through Grant Nos.  12150610461 and 11875112 and the China Postdoctoral Science Foundation through Grant No. KLH1512104. 
C. M. is supported by new faculty start up funding by the Institute of Modern Physics, Chinese Academy of Sciences, Grant No. E129952YR0. C. M. also thanks the Chinese Academy of Sciences Presidents International Fellowship Initiative for the support via Grants No. 2021PM0023. 


\end{acknowledgments}

\appendix

\section{}\label{App}

The bilinear decompositions of the quark-quark correlator of Eq.(\ref{Wdef}) relate to the leading twist GTMDs as~\cite{Meissner:2009ww}
\be 
W^{\nu [\gamma^+]}_{[\lambda^{\prime\prime}\lambda^{\prime}]}&=&\frac{1}{2M} \bar{u}(P^{\prime\prime},\lambda^{\prime\prime})\bigg[ F_{1,1} + \frac{i\sigma^{i+} \pp^i}{P^+}F_{1,2} +  \frac{i\sigma^{i+} \dpi^i}{P^+}F_{1,3}+ \frac{i\sigma^{ij} \pp^i \dpi^j}{M^2}F_{1,4}\bigg] u(P^\prime,\lambda^\prime)\label{WV_def}\,,\\
W^{\nu [\gamma^+\gamma^5]}_{[\lambda^{\prime\prime}\lambda^{\prime}]}&=&\frac{1}{2M} \bar{u}(P^{\prime\prime},\lambda^{\prime\prime})\bigg[-\frac{i\epsilon^{ij}_\perp \pp^i \dpi^j}{M^2} G_{1,1} + \frac{i\sigma^{i+} \gamma^5 \pp^i}{P^+}G_{1,2} \nonumber \\
&& \hspace{4cm} +  \frac{i\sigma^{i+} \gamma^5 \dpi^i}{P^+}G_{1,3}+ i\sigma^{+-} \gamma^5 G_{1,4}\bigg] u(P^\prime,\lambda^\prime)\,, \label{WA_def}\\
W^{\nu [i \sigma^{j+}\gamma^5]}_{[\lambda^{\prime\prime}\lambda^{\prime}]}&=&\frac{1}{2M} \bar{u}(P^{\prime\prime},\lambda^{\prime\prime})\bigg[-\frac{i\epsilon^{ij}_\perp \pp^i }{M} H_{1,1} - \frac{i\epsilon^{ij} \dpi^i}{M}H_{1,2} + \frac{M i\sigma^{j+} \gamma^5}{P^+} H_{1,3} +  \frac{\pp^j i\sigma^{k+} \gamma^5 \bfp^k}{M P^+} H_{1,4} \nonumber \\
&& + \frac{\dpi^j i\sigma^{k+} \gamma^5 \bfp^k}{M P^+} H_{1,5}+ \frac{\dpi^j i\sigma^{k+} \gamma^5 \dpi^k}{M P^+} H_{1,6} +  \frac{\pp^j i\sigma^{+-} \gamma^5 }{M}H_{1,7}+  \frac{\dpi^j i \sigma^{+-} \gamma^5 }{M}H_{1,8}\bigg] u(P^\prime,\lambda^\prime)\,,\nonumber\\ \label{WT_def}
\ee
where the spinors $u(k,\lambda) $ with the momentum $k$ and the helicity $\lambda\,(=\pm )$ are given by
\begin{center}
$u(k, +)=\frac{1}{\sqrt{2 k^+}}\left( \begin{matrix} 
  k^+ + m_F\\
  k^1 + i k^2\\
  k^+ - m_F\\
  k^1 + i k^2\\
\end{matrix} \right) $, \hspace{1cm}
$u(k, -)=\frac{1}{\sqrt{2 k^+}}\left( \begin{matrix} 
    - k^1 + i k^2\\
    k^+ + m_F\\
  k^1- i k^2\\
  - k^+ + m_F\\
\end{matrix} \right) $
\end{center}
with $m_F$ being the mass of the fermion.
Using the kinematics given in  Eqs.~(\ref{Pp}), (\ref{Ppp}), one can find out the spinors $u(P^\prime,\lambda^\prime) $  and $u(P^{\prime \prime},\lambda^{\prime \prime}) $ and compute the matrix elements of $\bar{u}(k,\lambda)\Gamma u(k,\lambda)$, where $\Gamma$ represents the Dirac matrix structure.

The unpolarized ($H$ and $E$) and the helicity dependent ($\tilde{H}$ and $\tilde{E}$) quark GPDs are connected to the unpolarized and the longitudinally polarized quark GTMDs via
\be
H(x,\xi,t) &=& \int d^2\bfp \bigg[F_{1,1} + 2 \xi^2 \bigg(\frac{\bfp.\bfd}{\bfd^2}F_{1,2}+F_{1,3}\bigg)\bigg]\,, \label{H}\\
E(x,\xi,t)&=& \int d^2\bfp \bigg[-F_{1,1} + 2 (1-\xi^2) \bigg(\frac{\bfp.\bfd}{\bfd^2}F_{1,2}+F_{1,3}\bigg)\bigg]\,, \label{E}\\
\tilde{H}(x,\xi,t) &=& \int d^2\bfp \bigg[ 2 \xi \bigg(\frac{\bfp.\bfd}{\bfd^2}G_{1,2}+G_{1,3}\bigg)+G_{1,4}\bigg]\label{Ht}\,,\\
\tilde{E}(x,\xi,t) &=& \int d^2\bfp \bigg[ \frac{2(1-\xi^2)}{\xi}\bigg(\frac{\bfp.\bfd}{\bfd^2} G_{1,2} + G_{1,3}\bigg)-G_{1,4}\bigg]\,. \label{Et}
\ee

\begin{figure}[t]
\includegraphics[scale=.32]{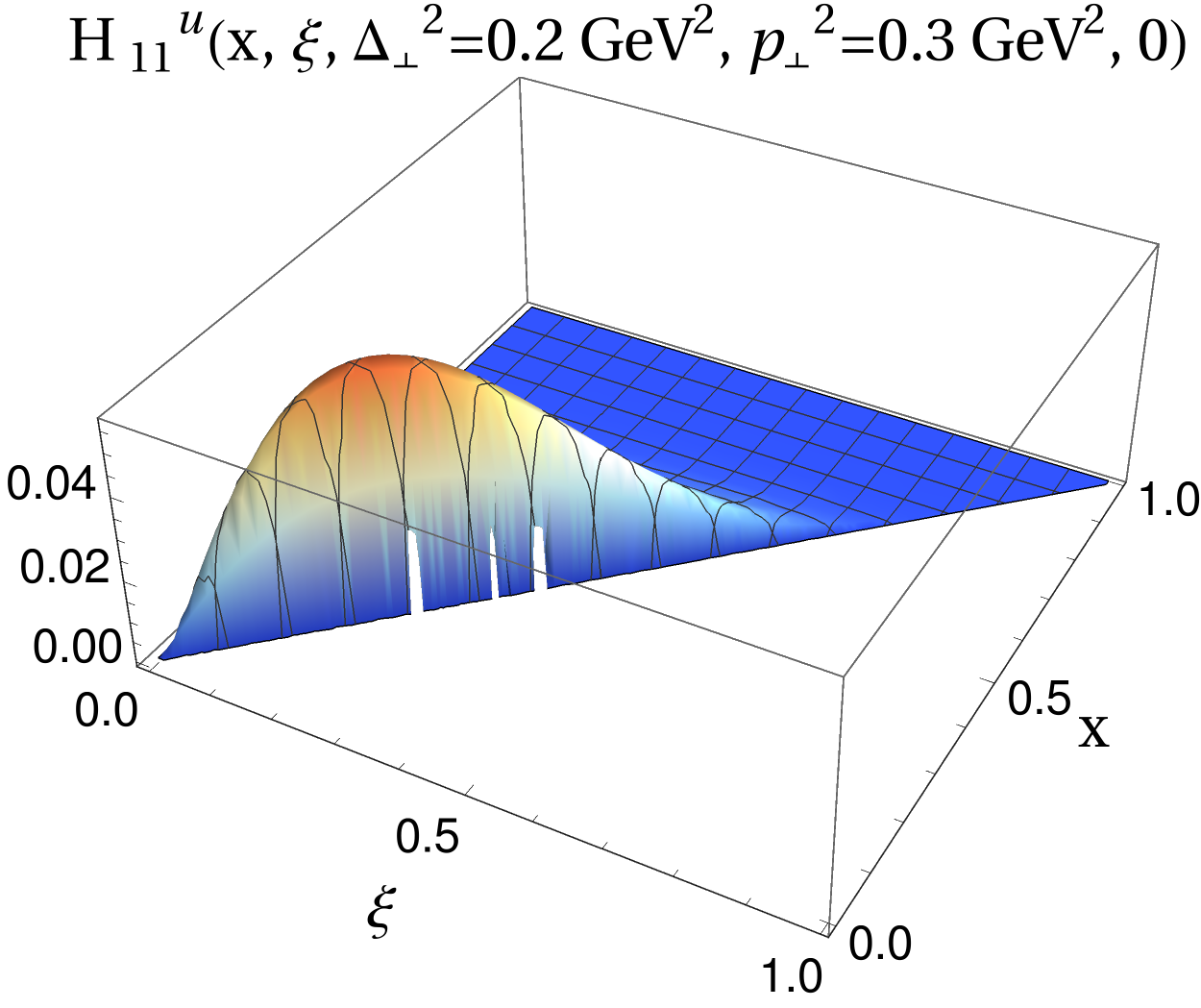}
\includegraphics[scale=.32]{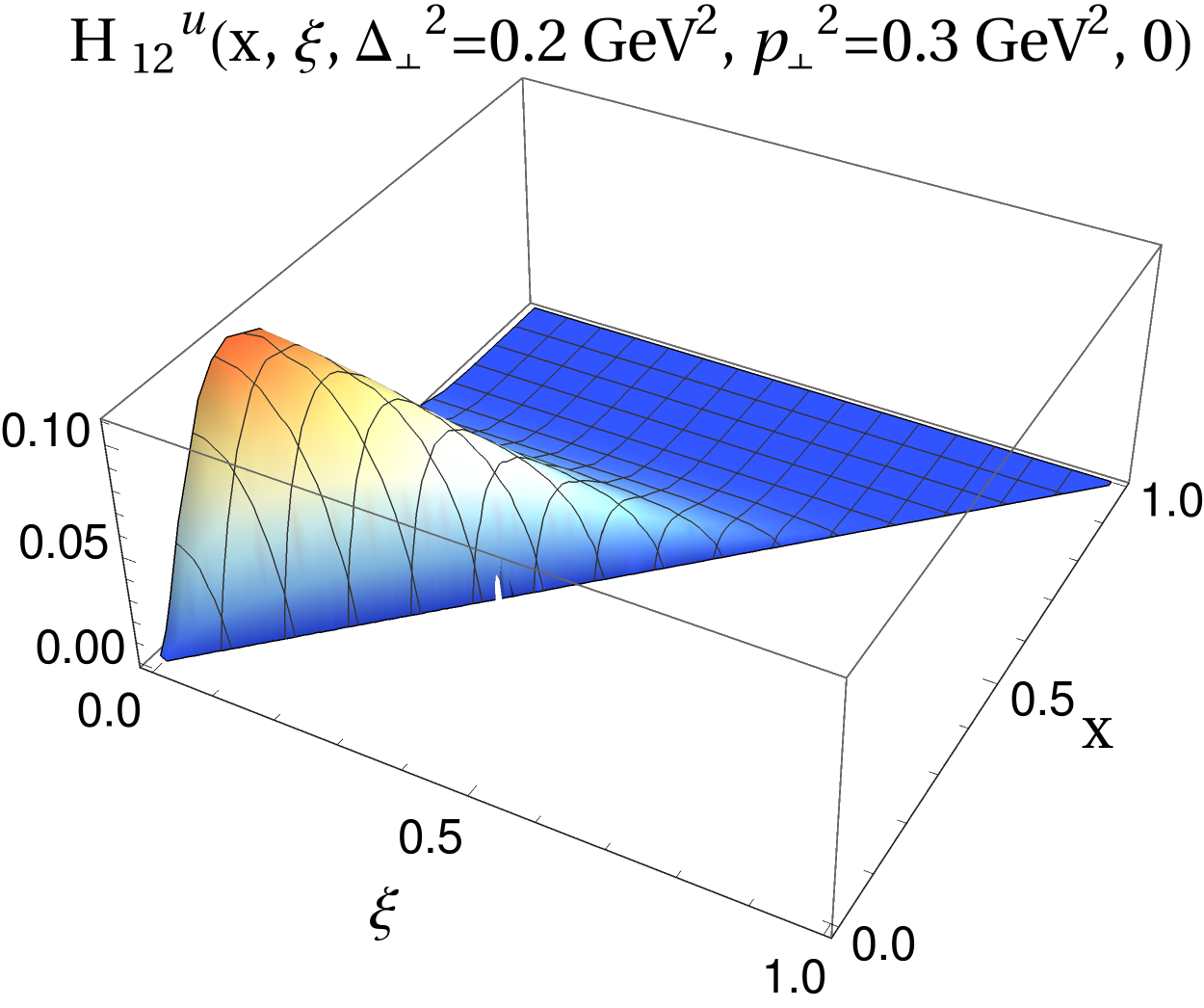} 
\includegraphics[scale=.32]{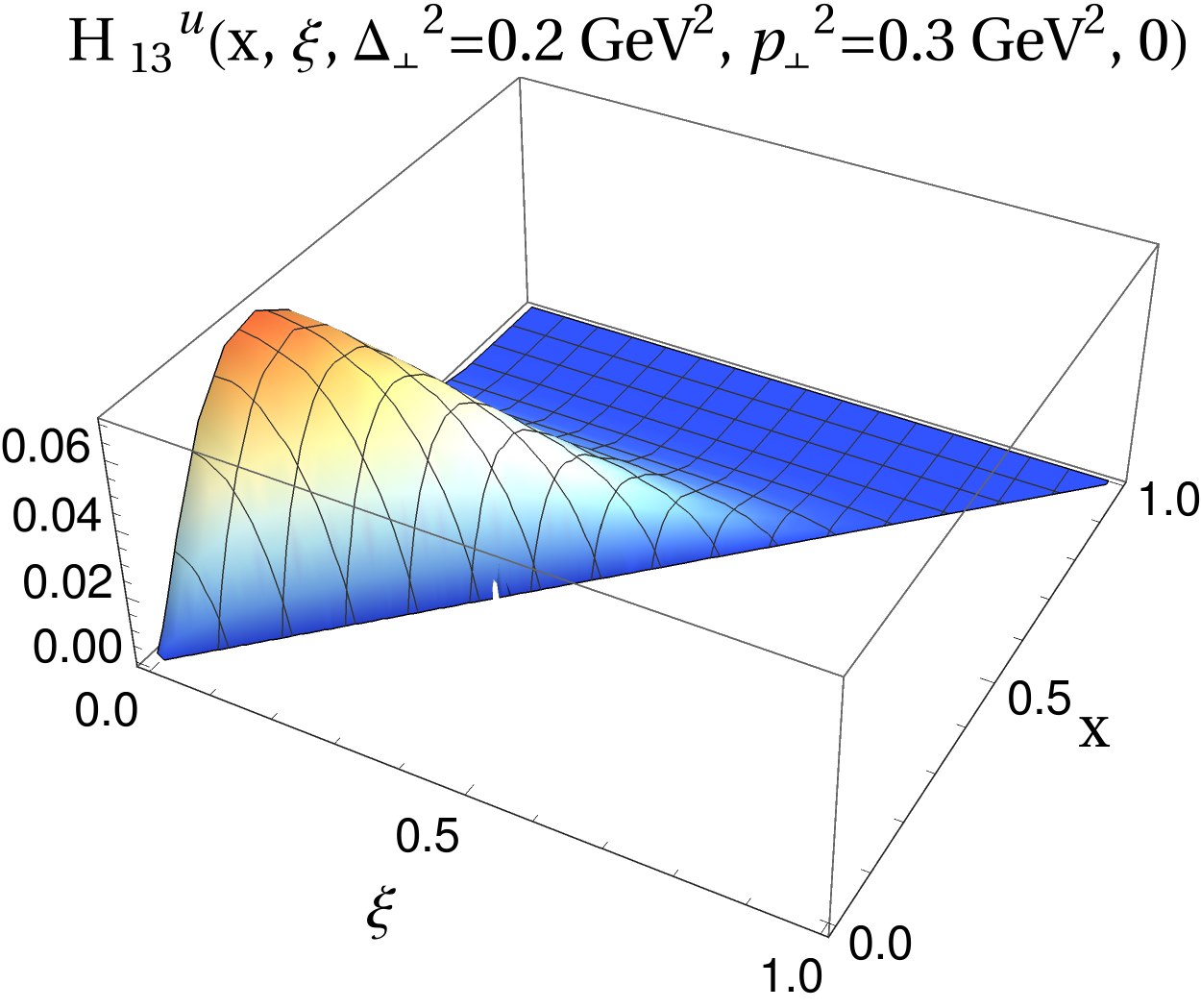} 
\includegraphics[scale=.32]{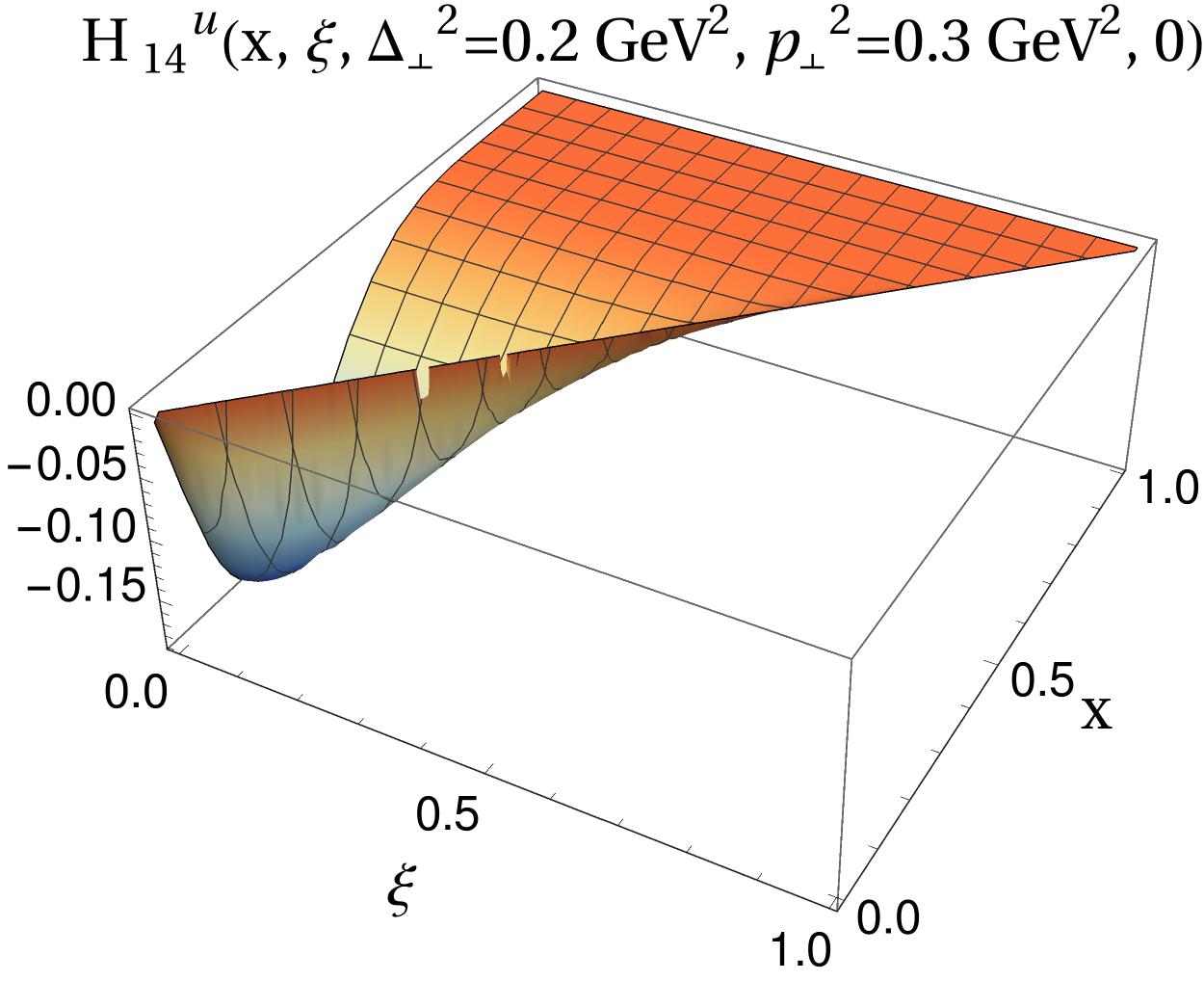} \\
\includegraphics[scale=.32]{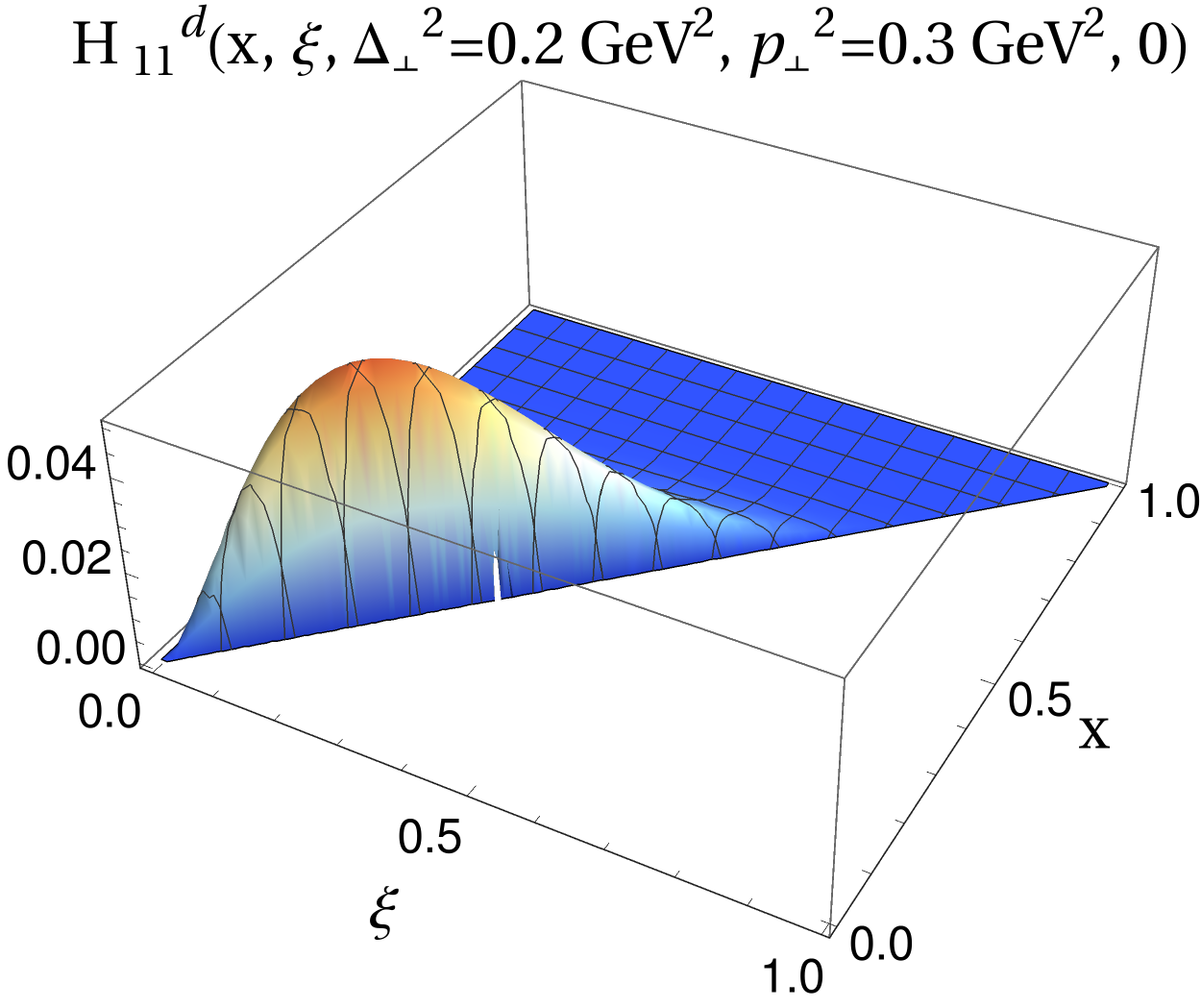} 
\includegraphics[scale=.32]{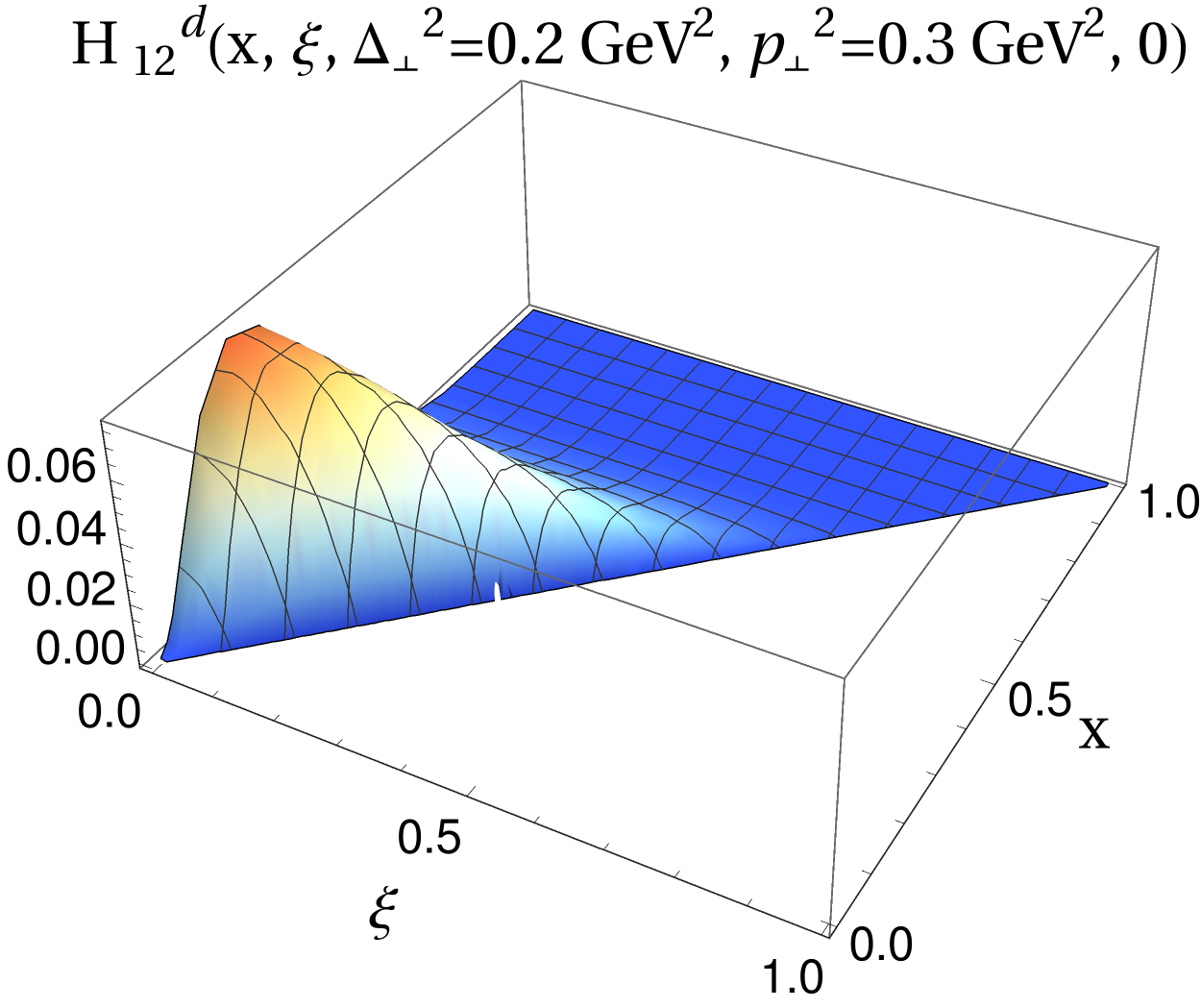} 
\includegraphics[scale=.32]{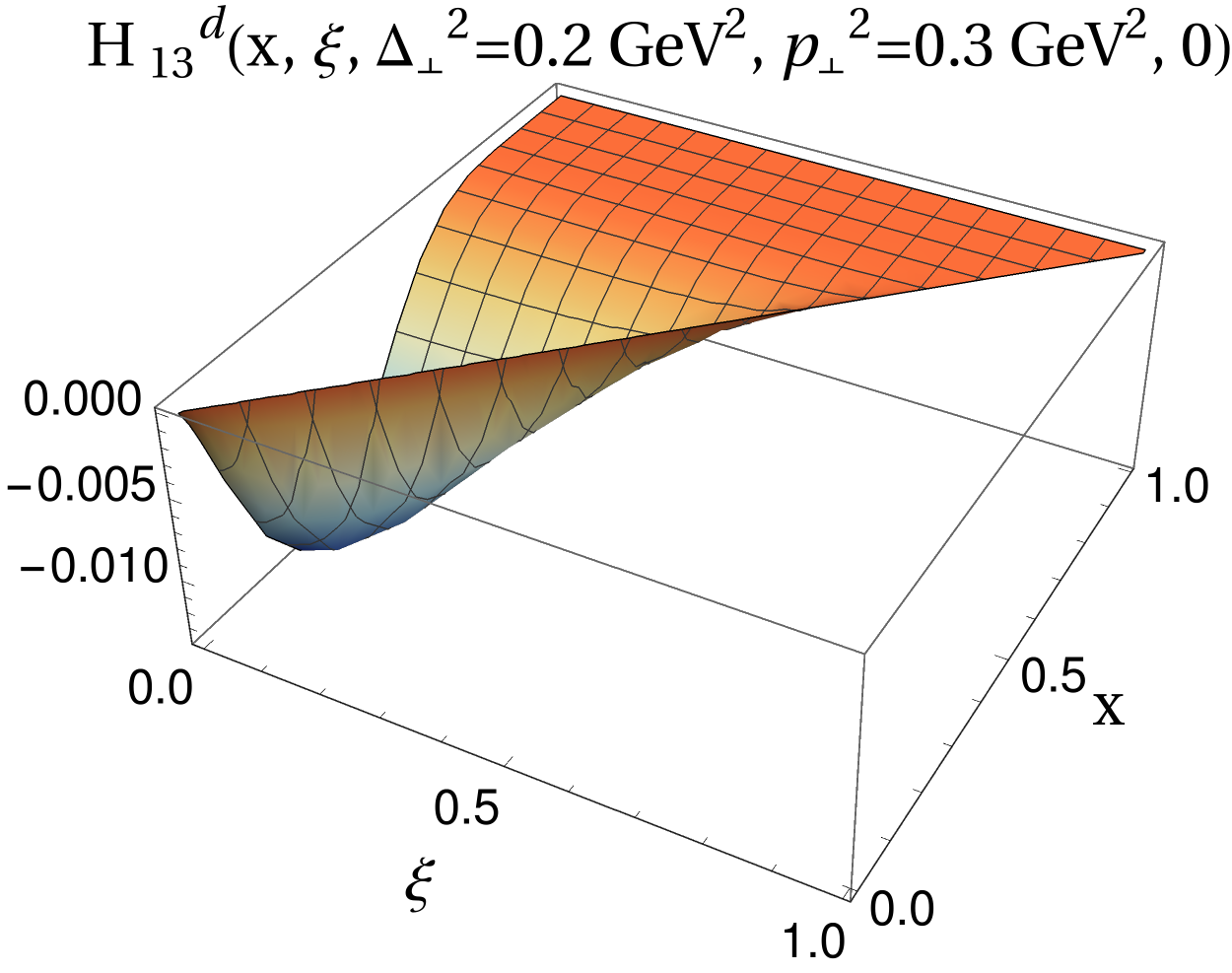} 
\includegraphics[scale=.32]{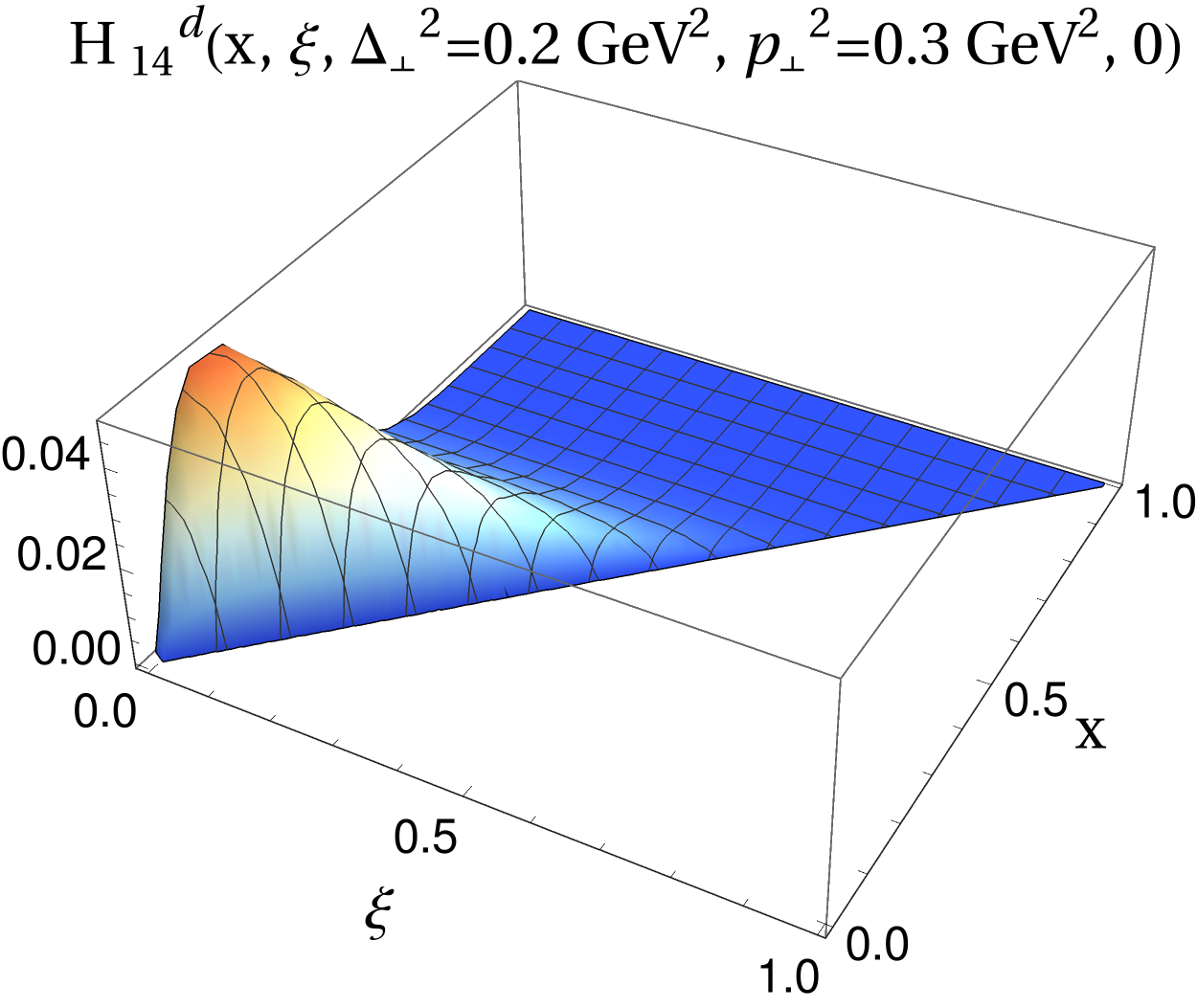} \\
\includegraphics[scale=.32]{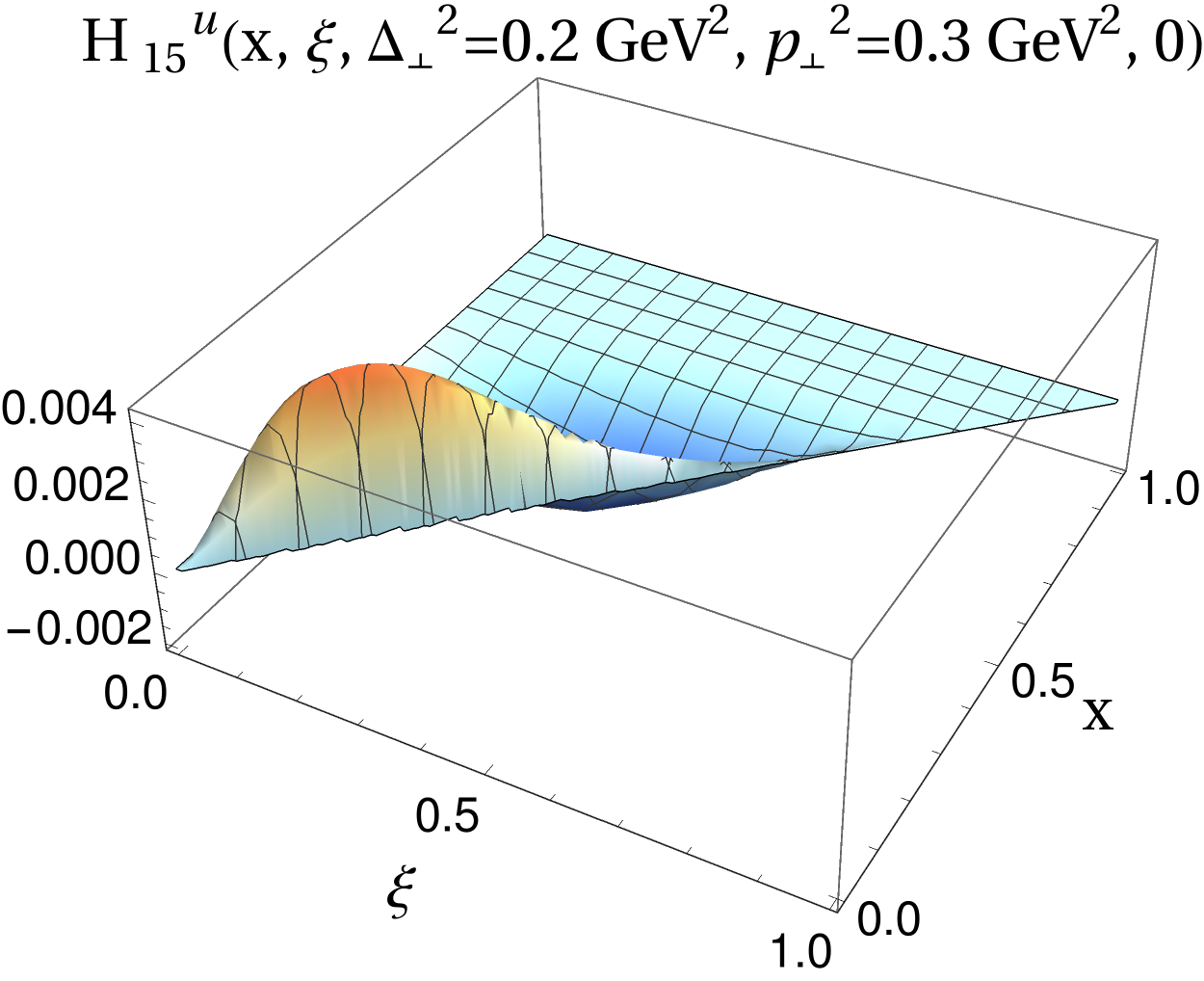}
\includegraphics[scale=.32]{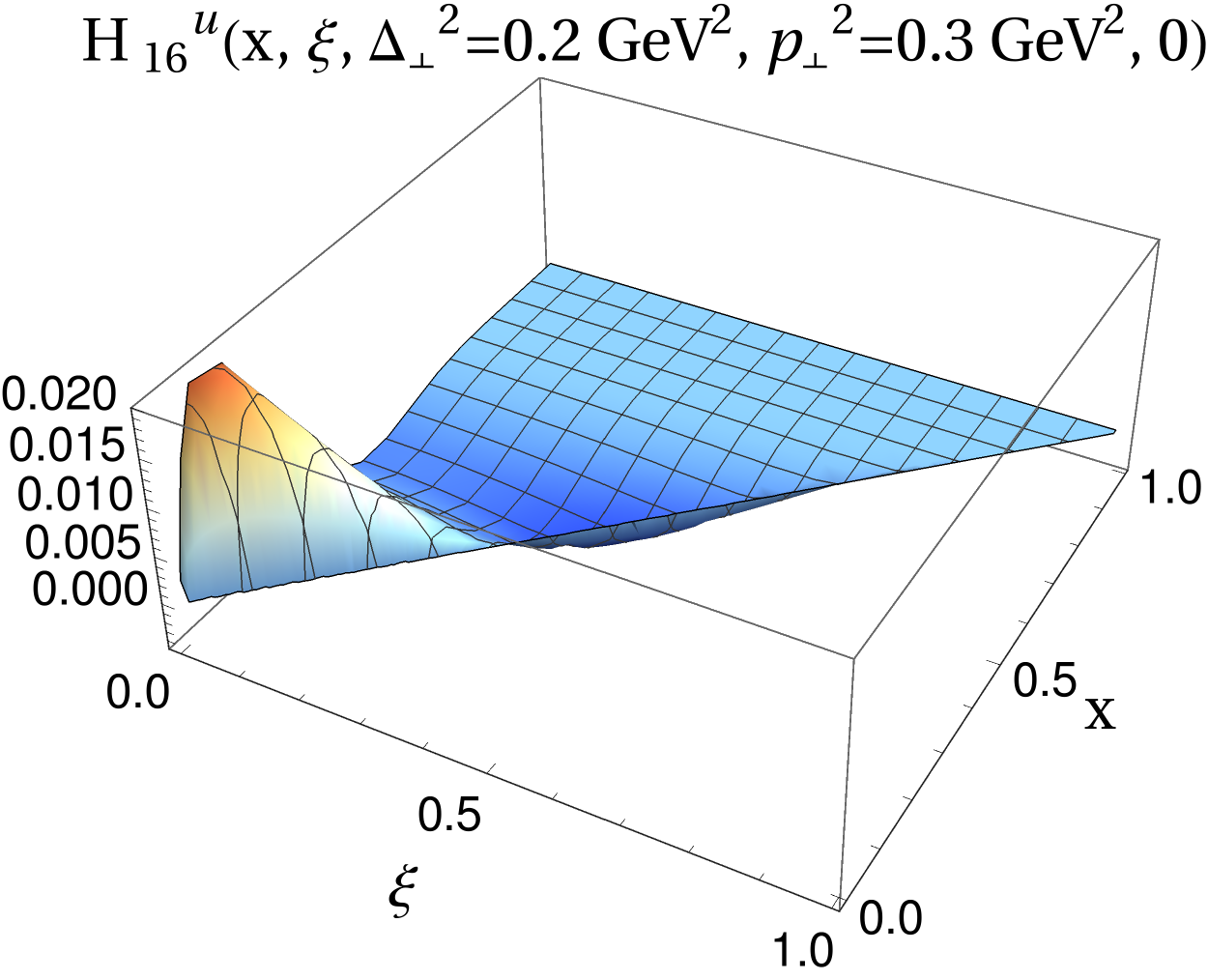} 
\includegraphics[scale=.32]{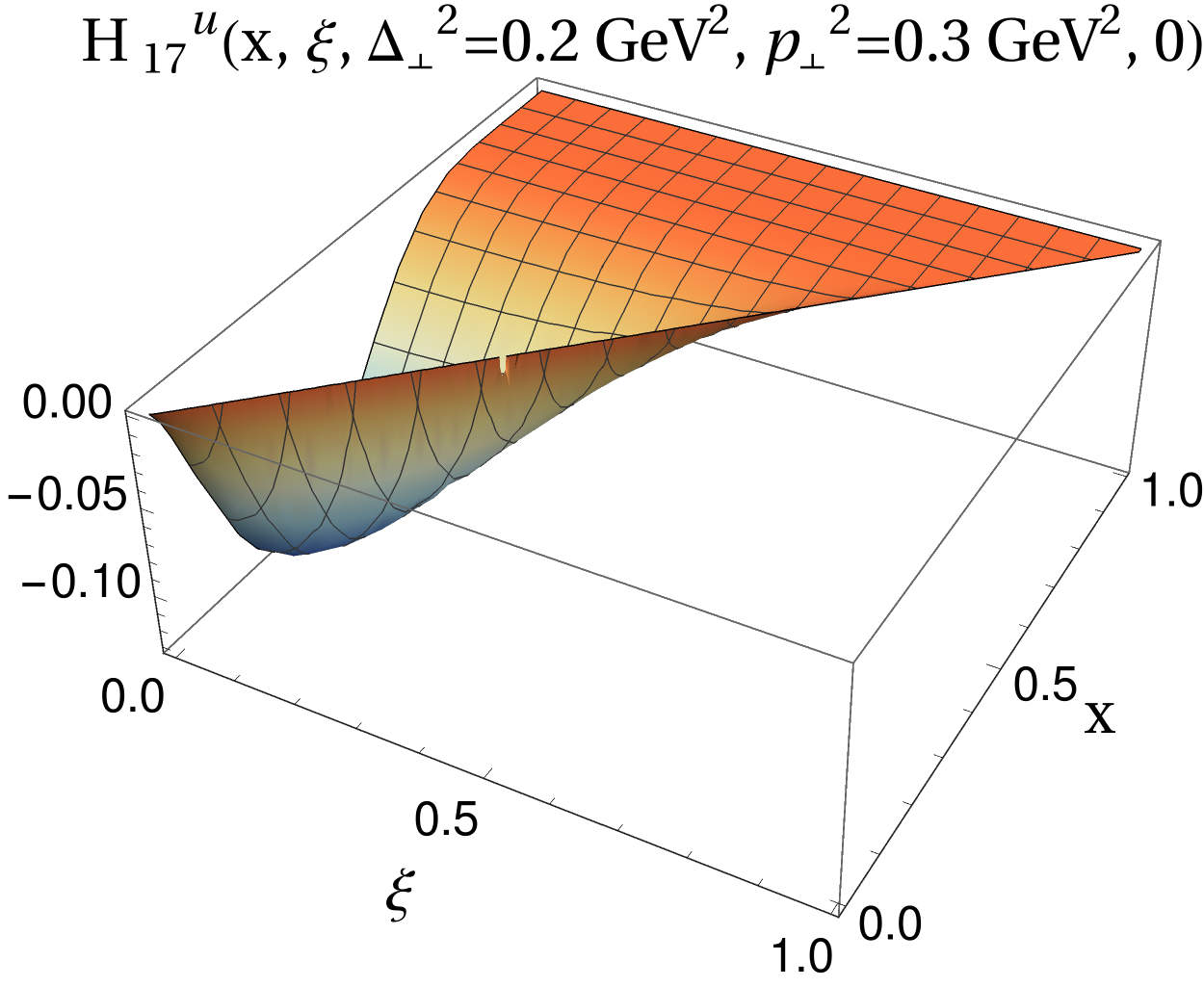} 
\includegraphics[scale=.32]{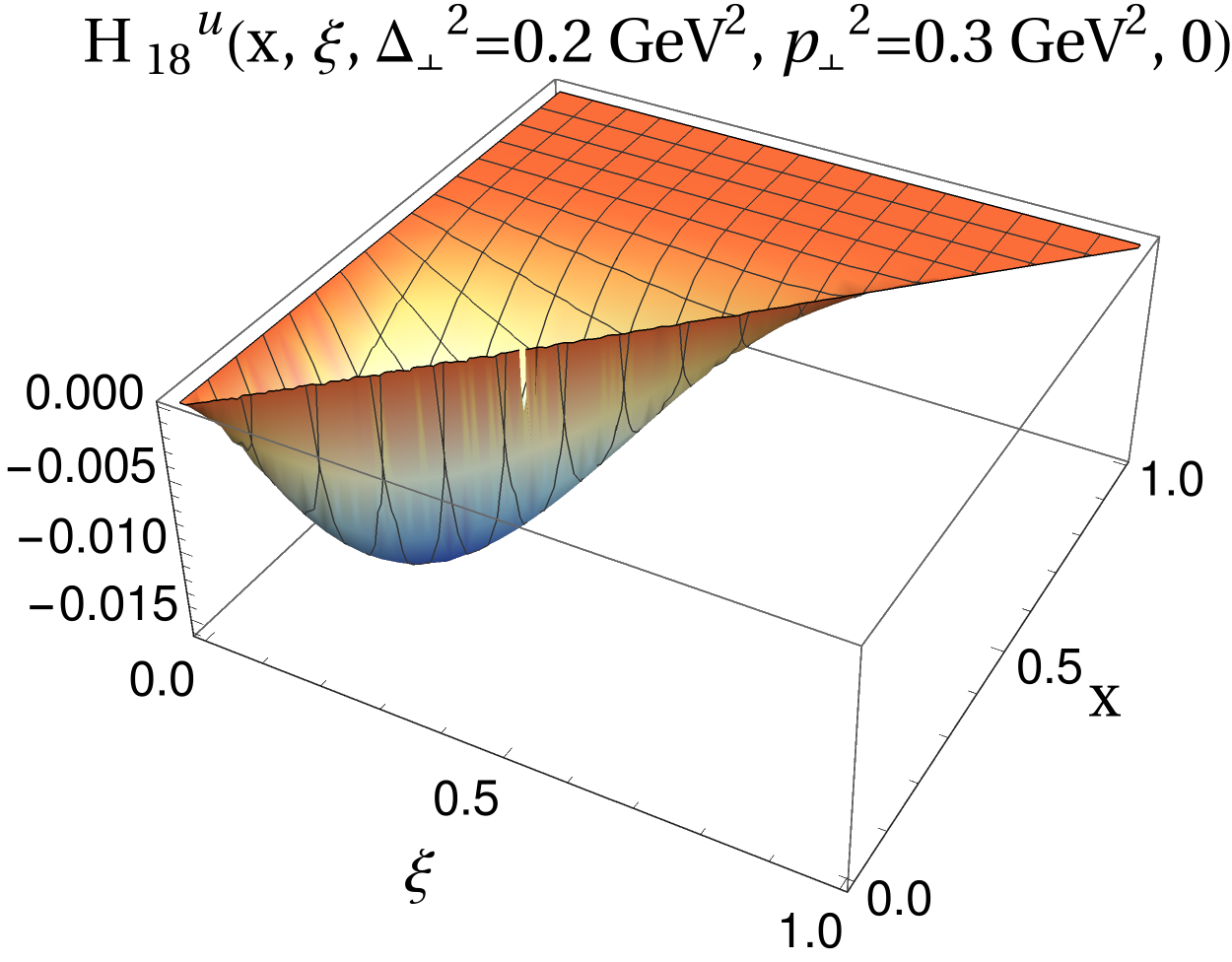} \\
\includegraphics[scale=.32]{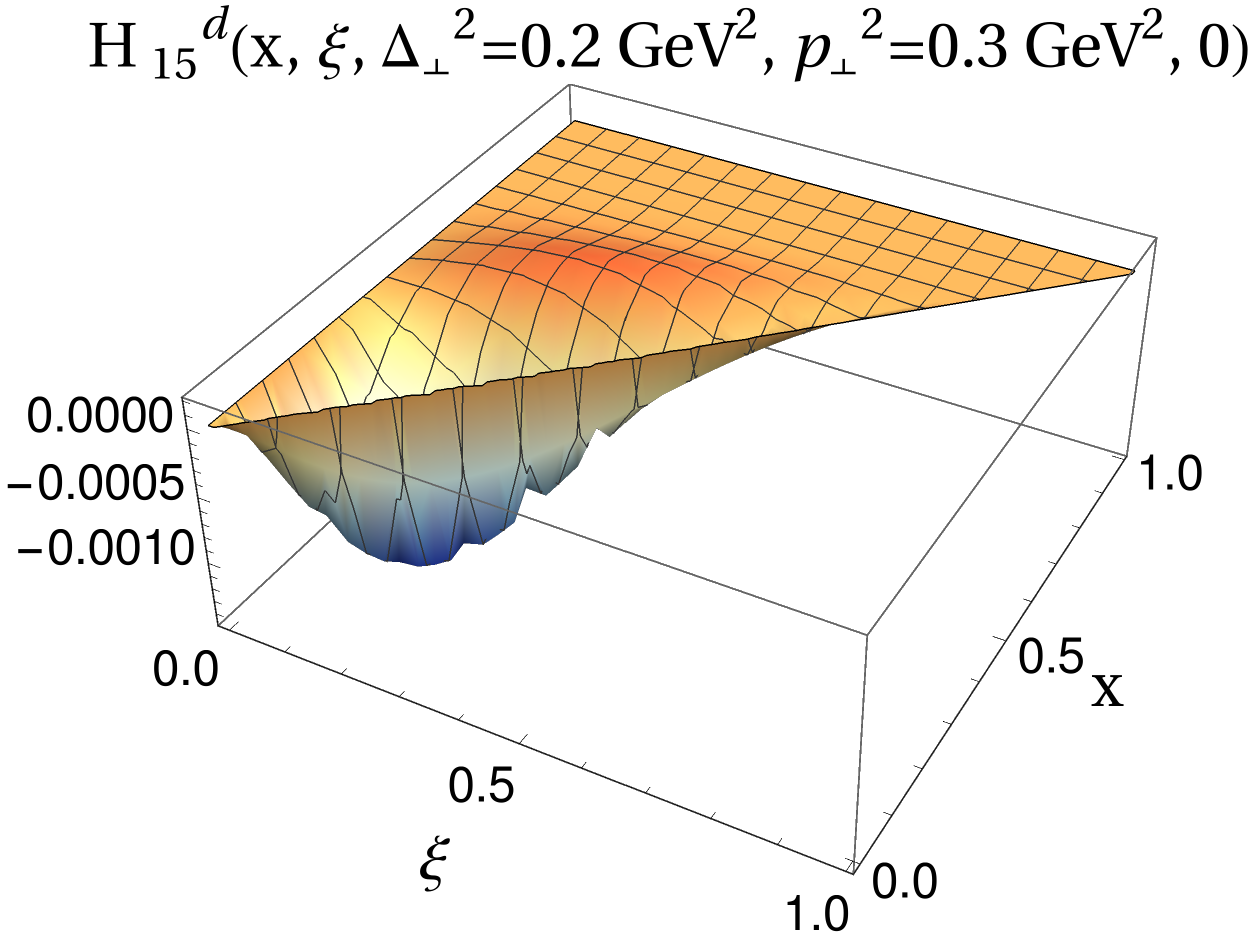} 
\includegraphics[scale=.32]{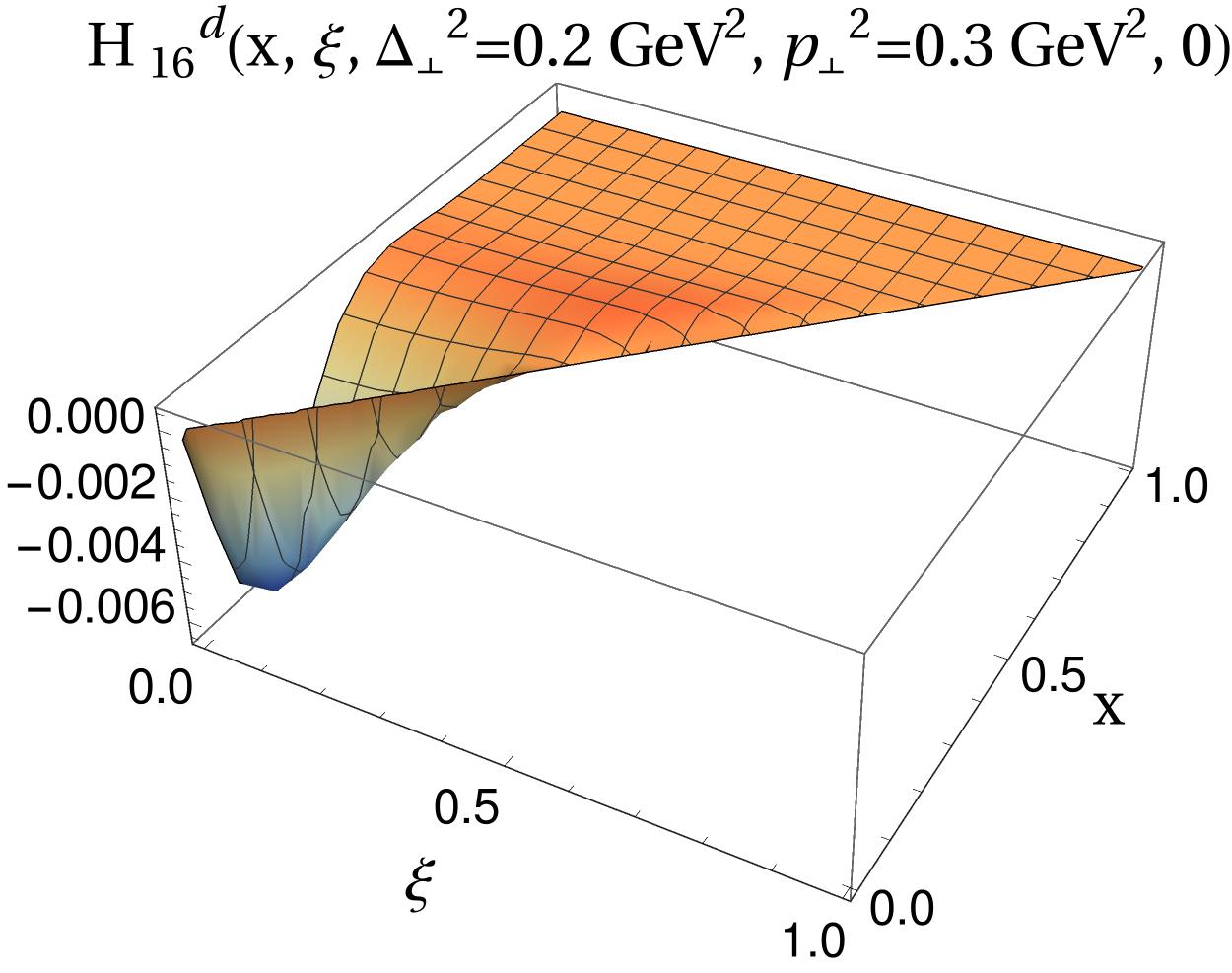} 
\includegraphics[scale=.32]{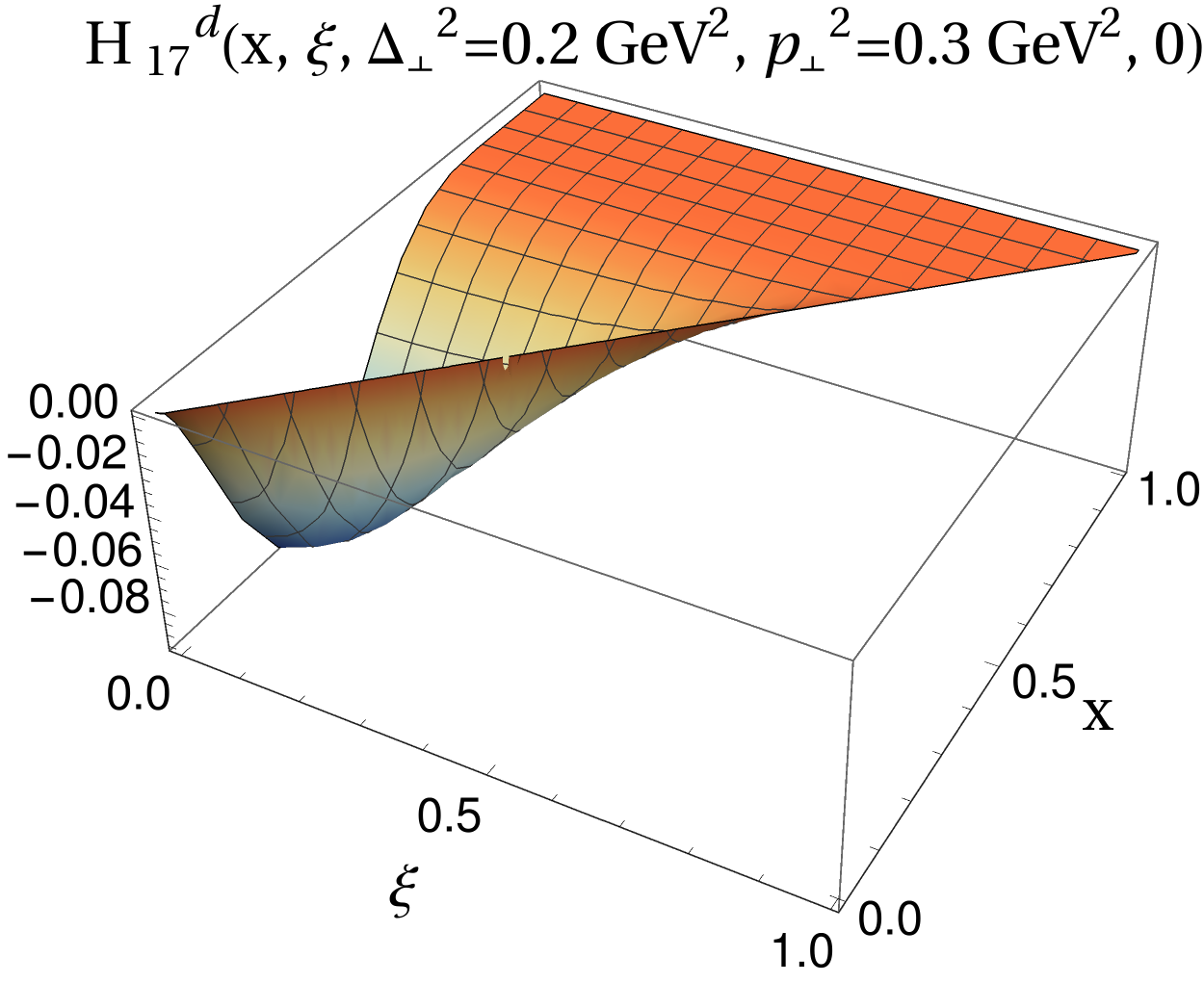} 
\includegraphics[scale=.32]{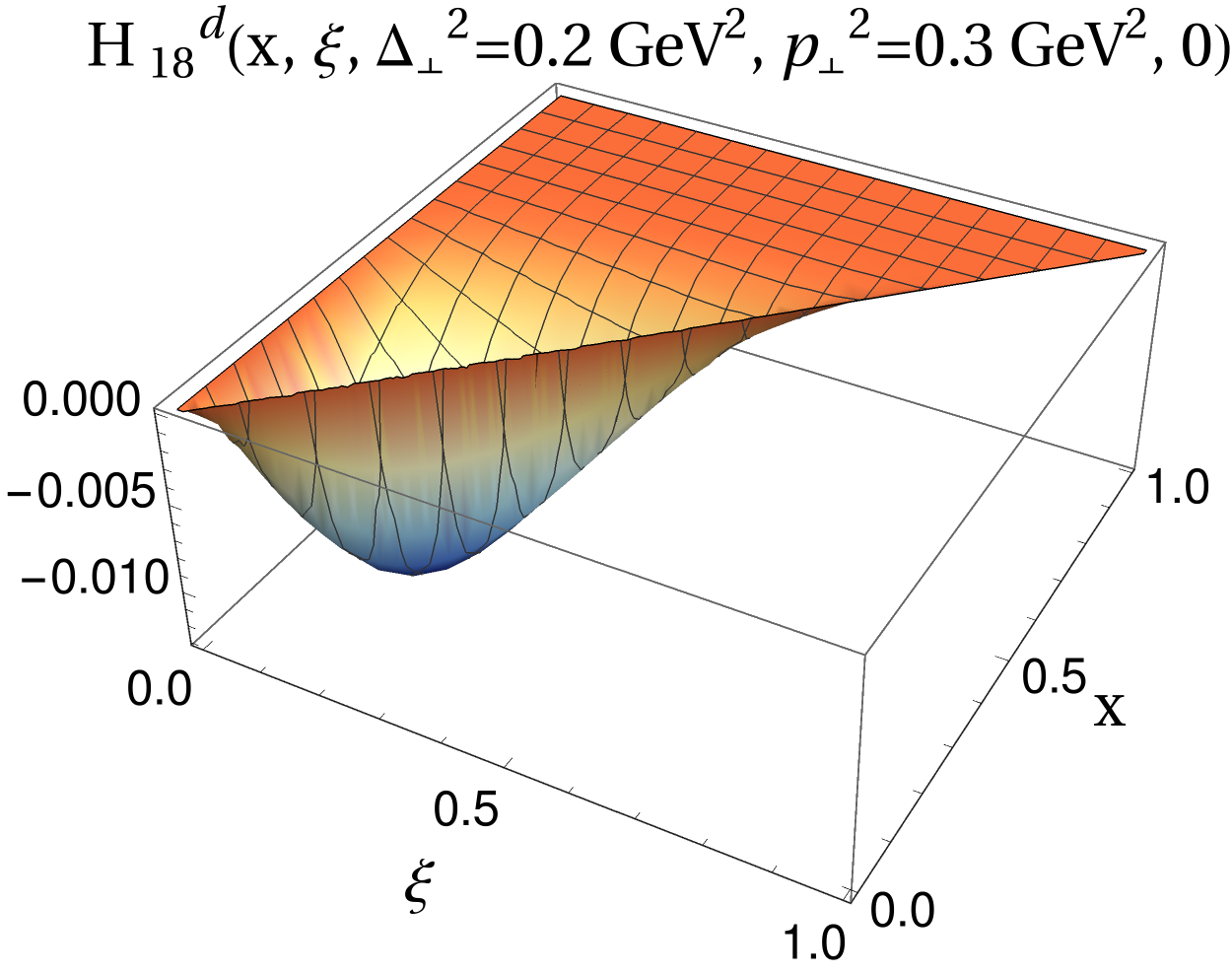} 
\caption{\label{tranxz} The leading twist GTMDs as functions of $x$ and $\xi$ when the quark is transversely polarized. }
\end{figure}
\begin{figure}[h]
\includegraphics[scale=.32]{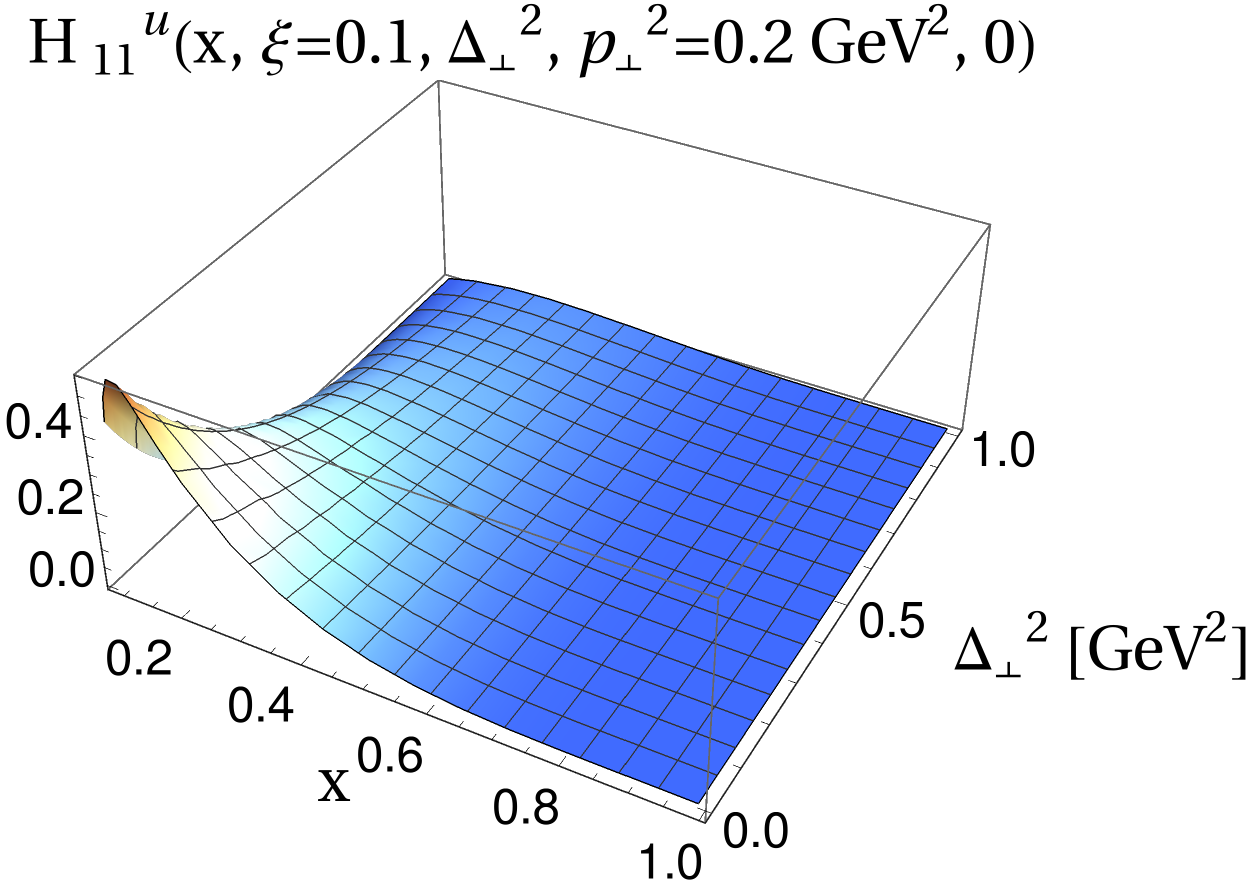}
\includegraphics[scale=.32]{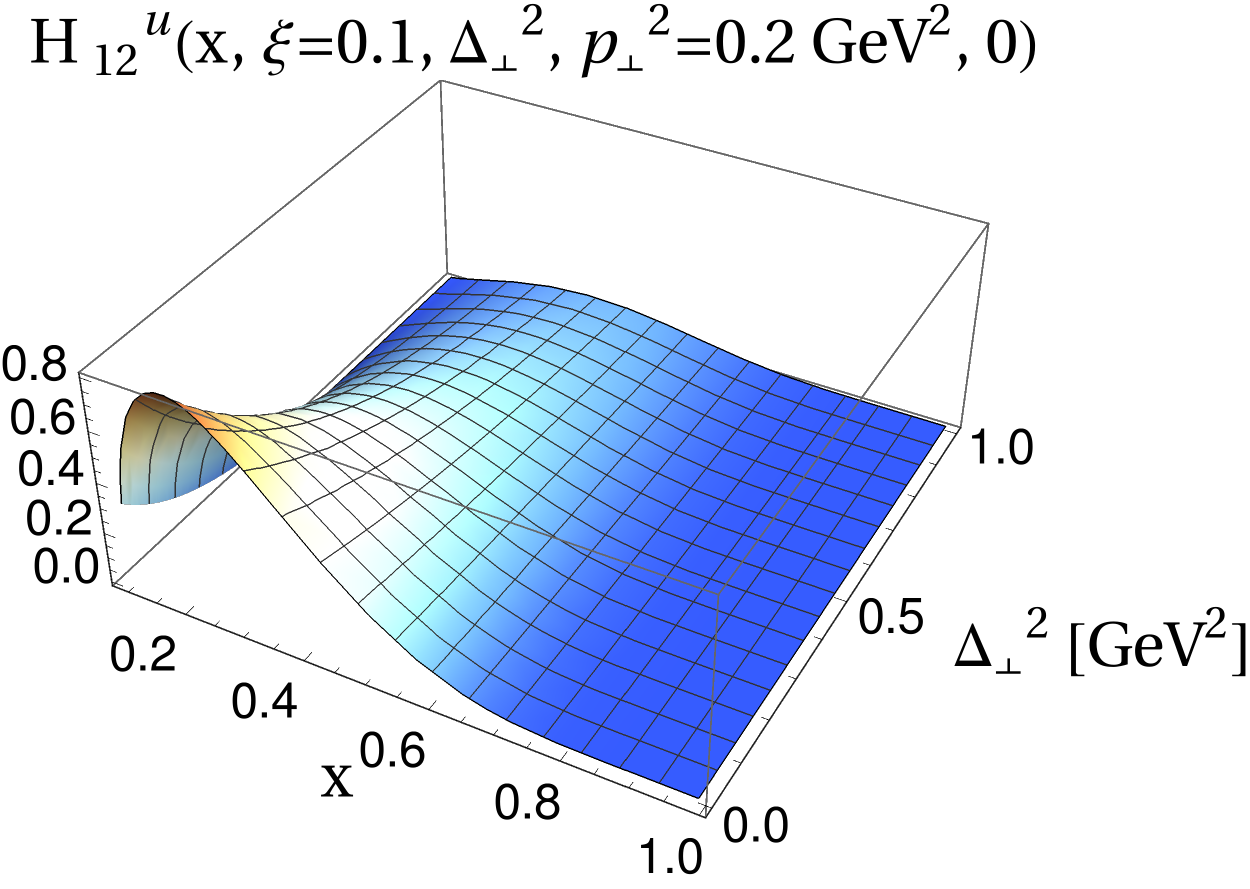} 
\includegraphics[scale=.32]{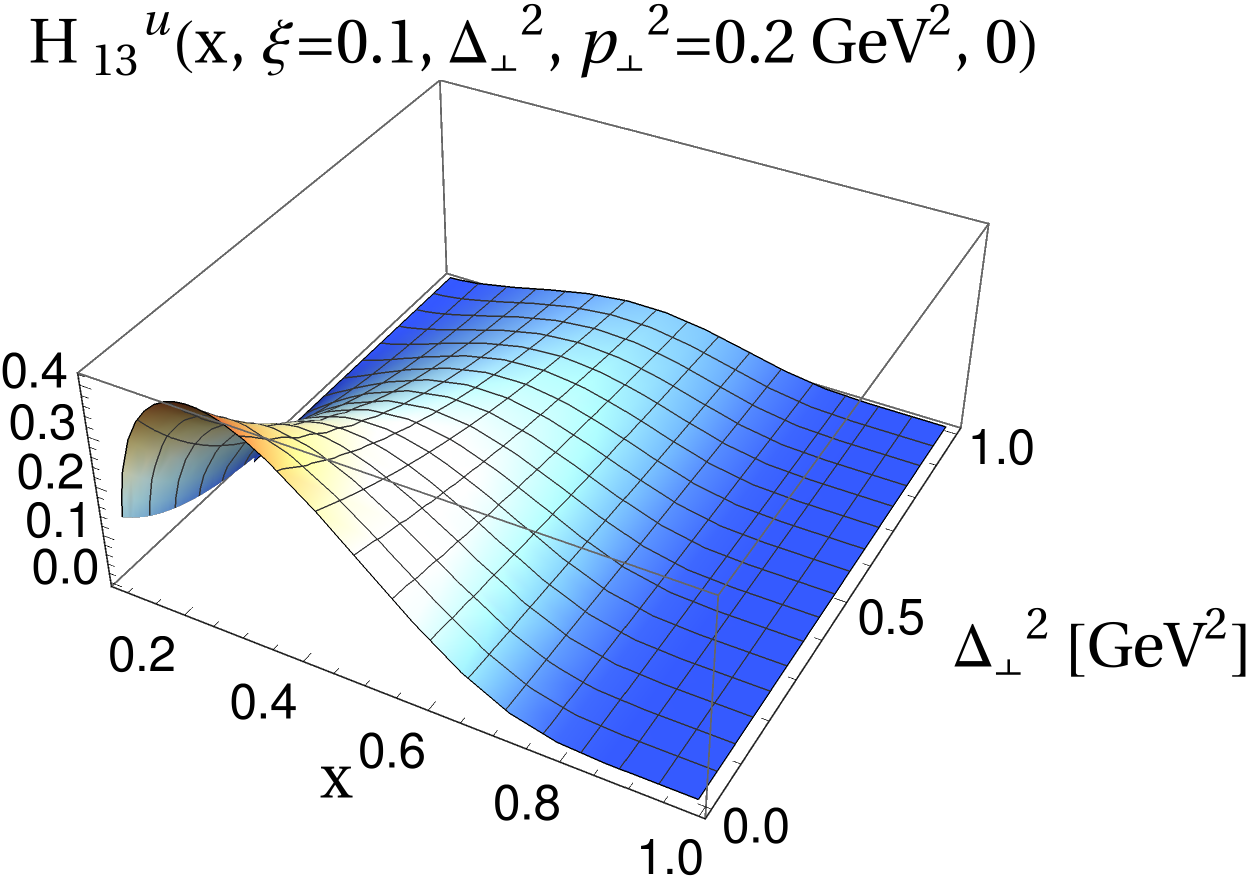} 
\includegraphics[scale=.32]{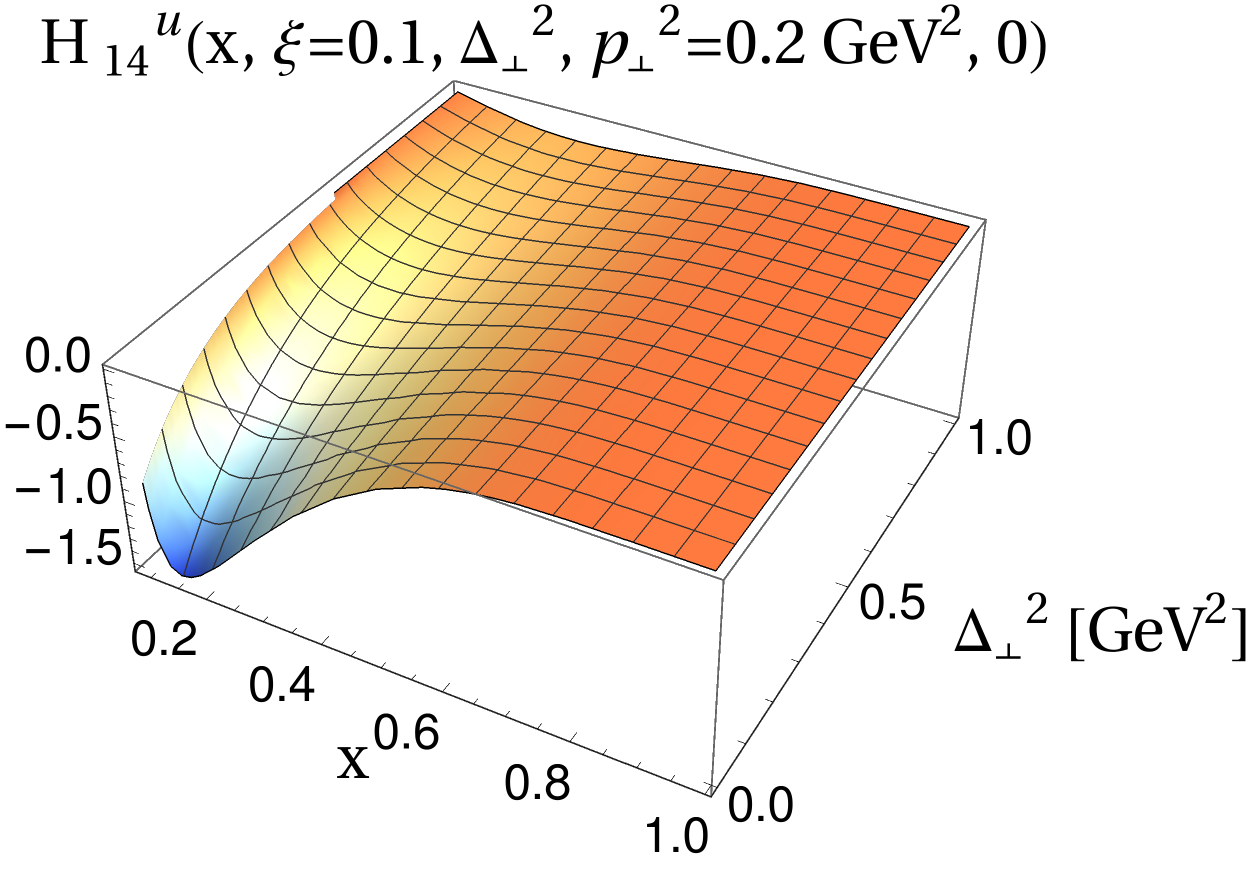} \\
\includegraphics[scale=.32]{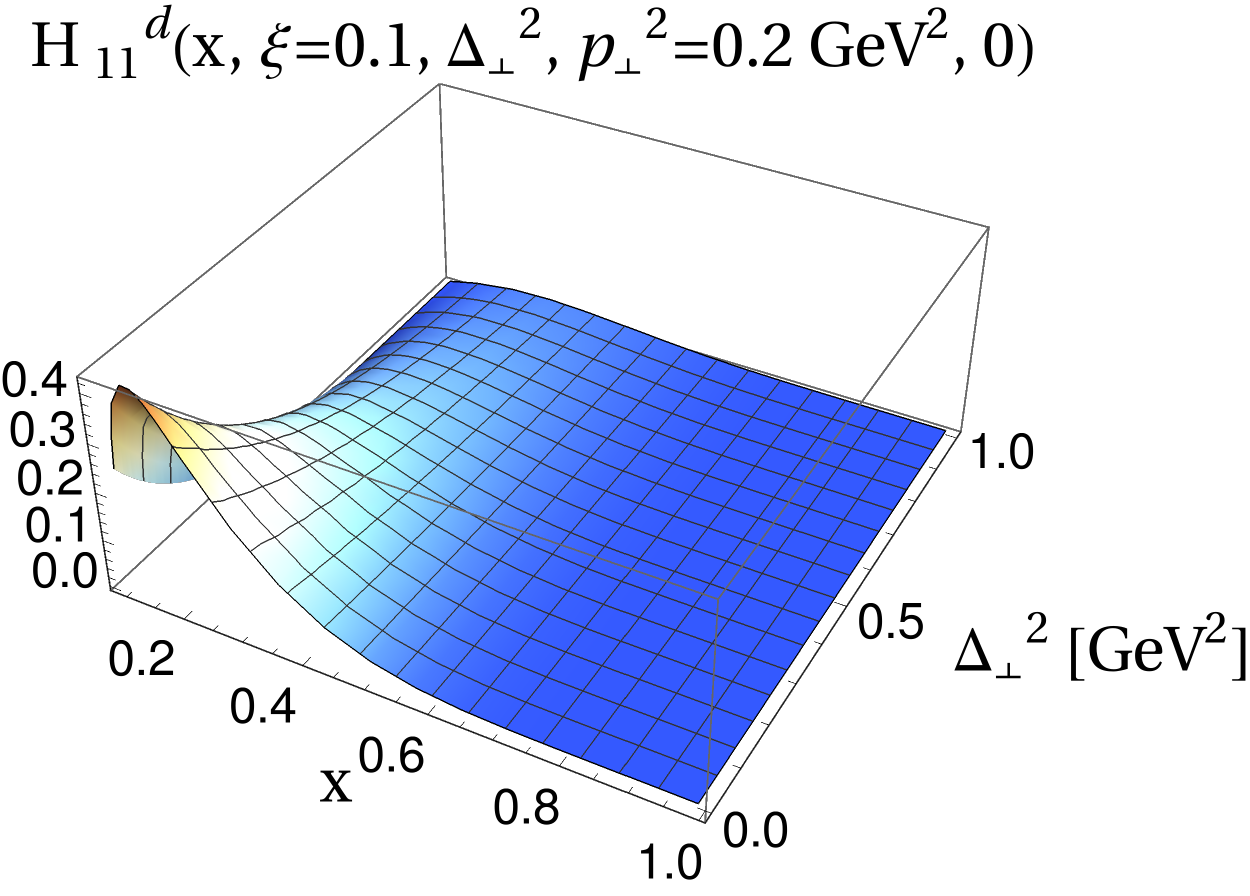} 
\includegraphics[scale=.32]{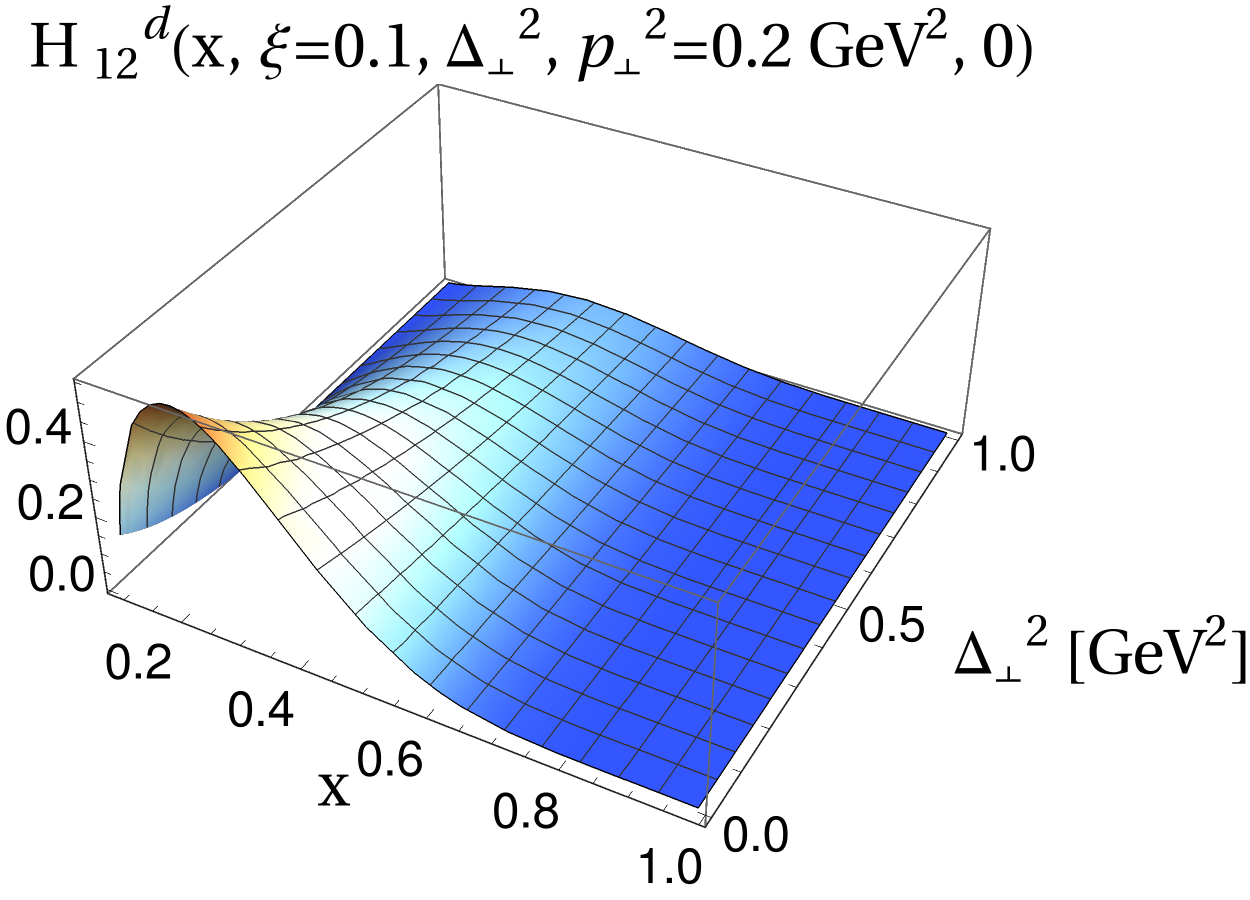} 
\includegraphics[scale=.32]{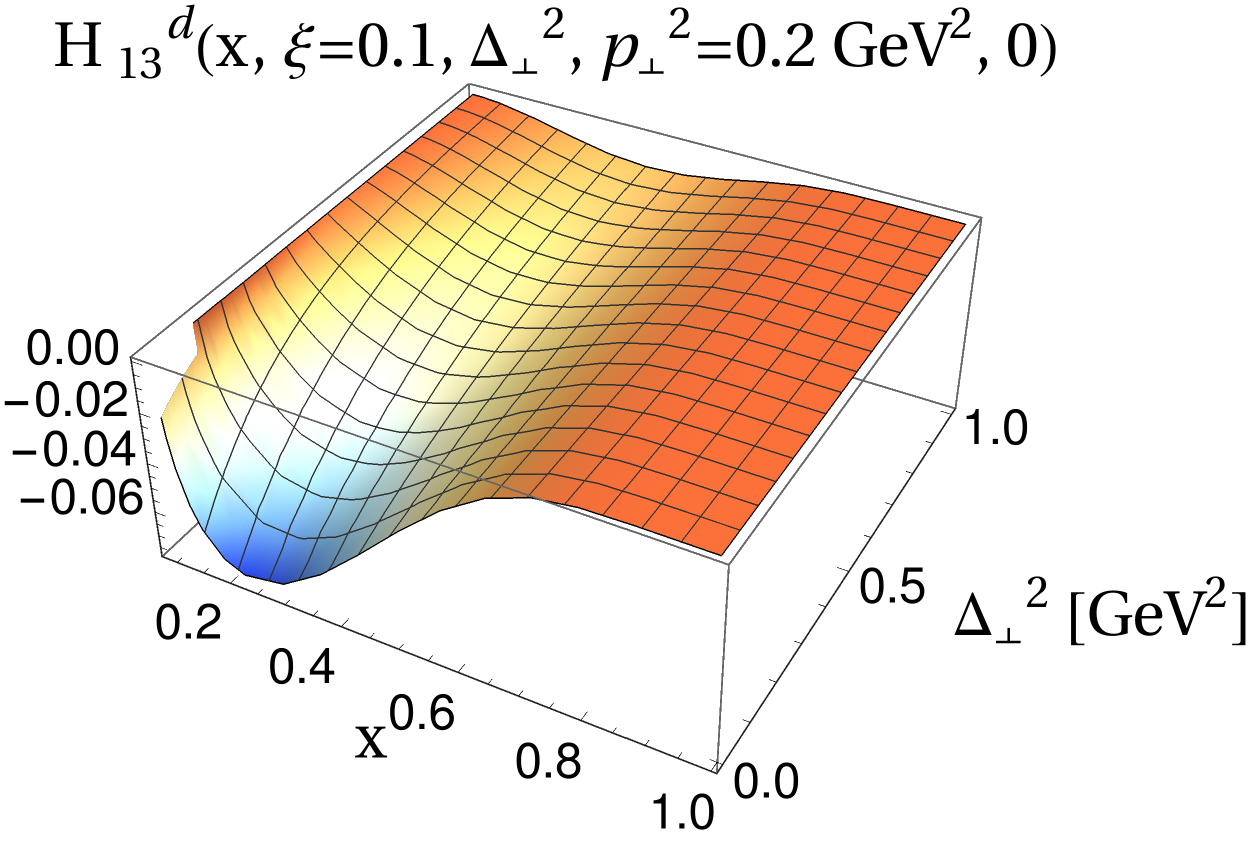} 
\includegraphics[scale=.32]{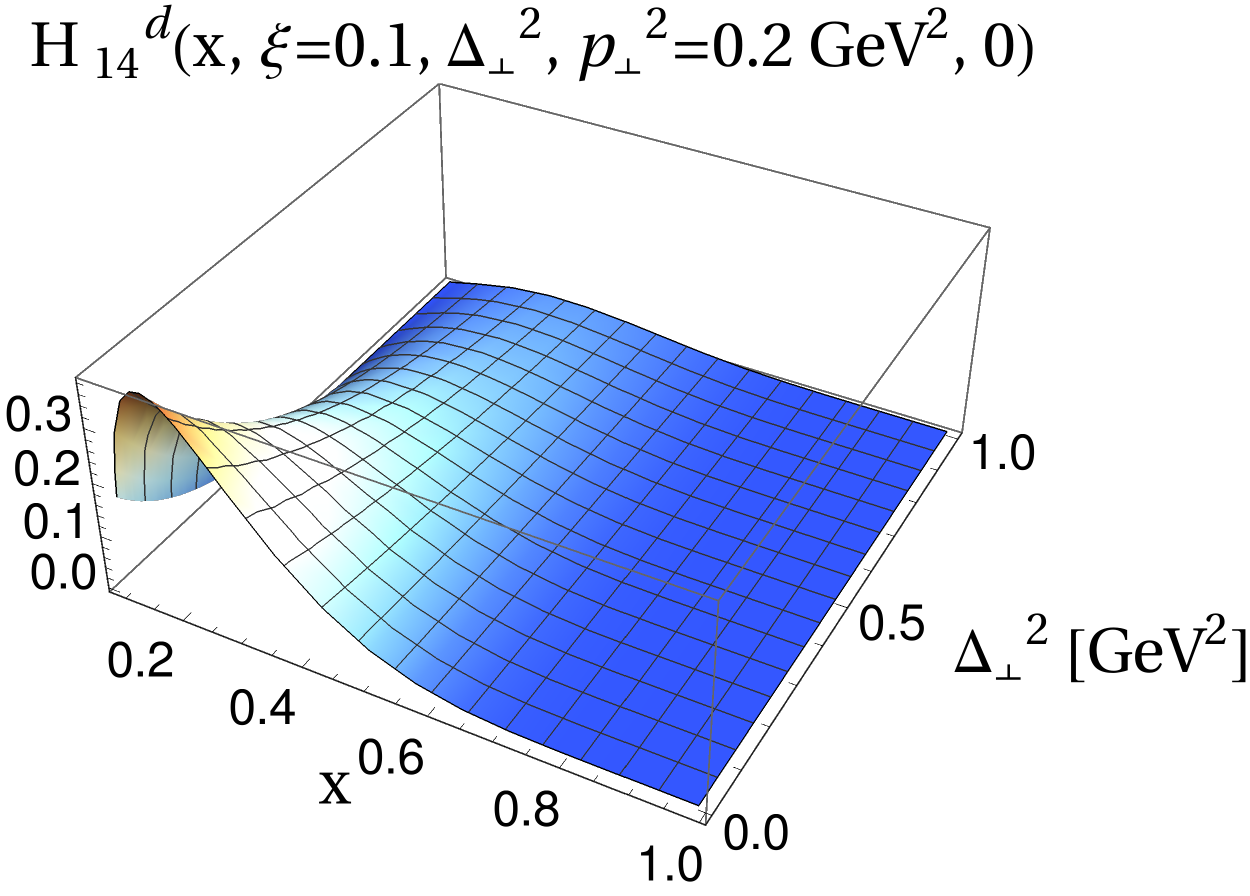} \\
\includegraphics[scale=.32]{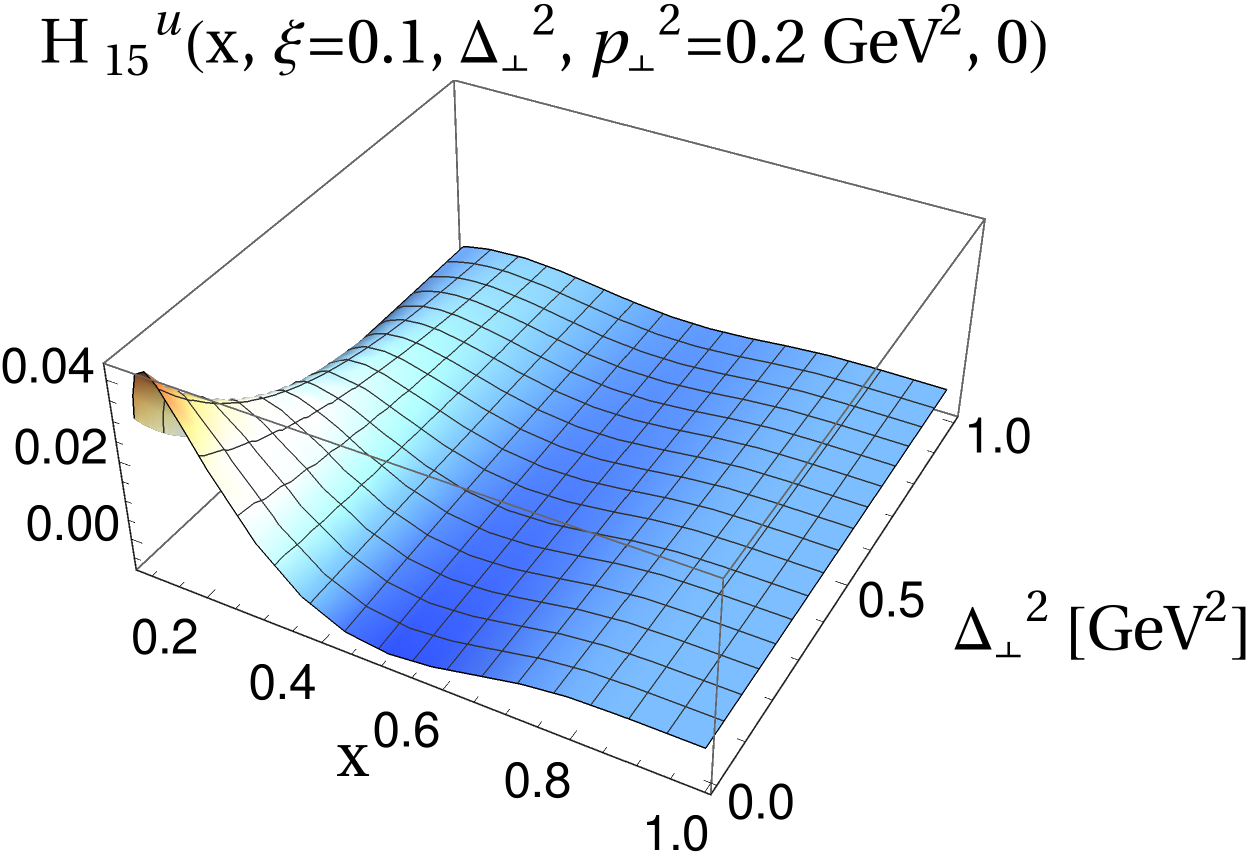}
\includegraphics[scale=.32]{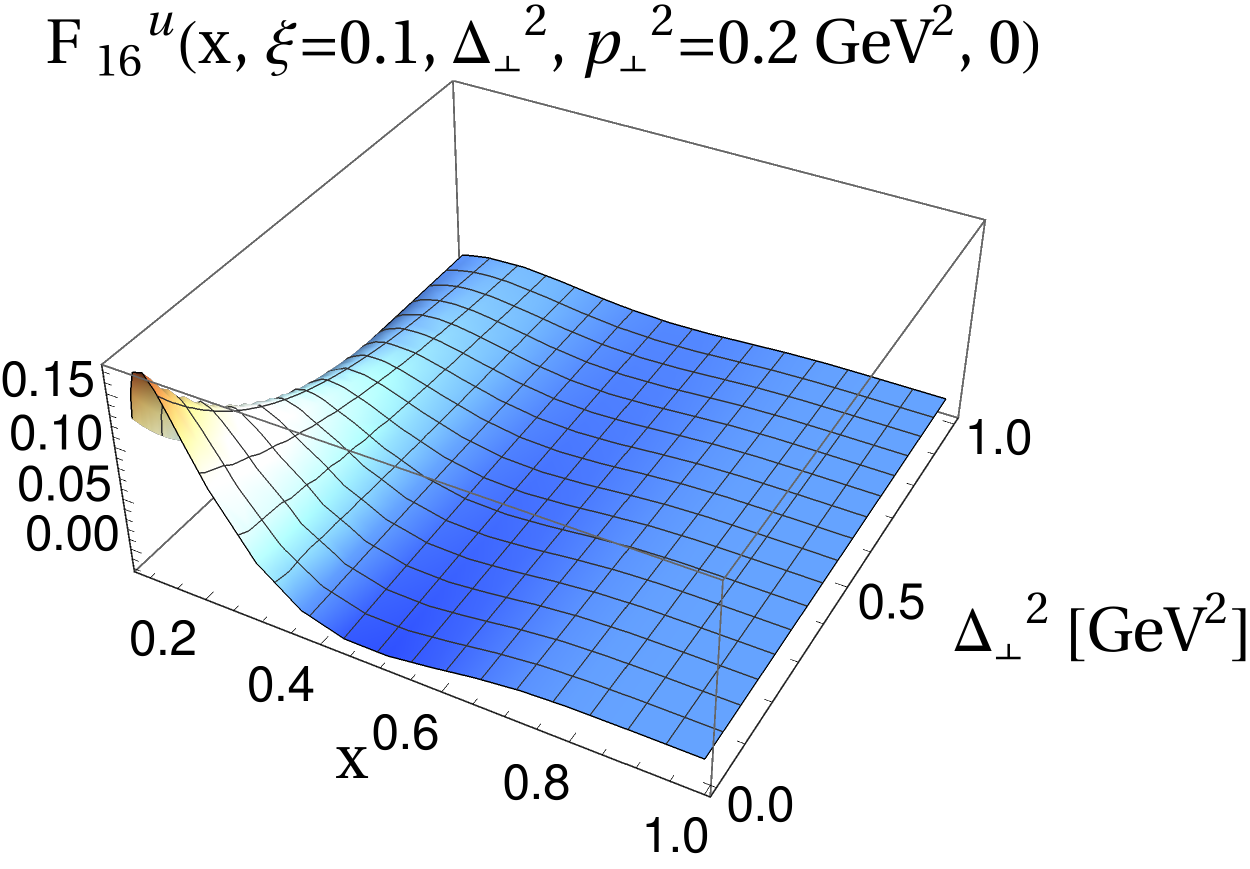} 
\includegraphics[scale=.32]{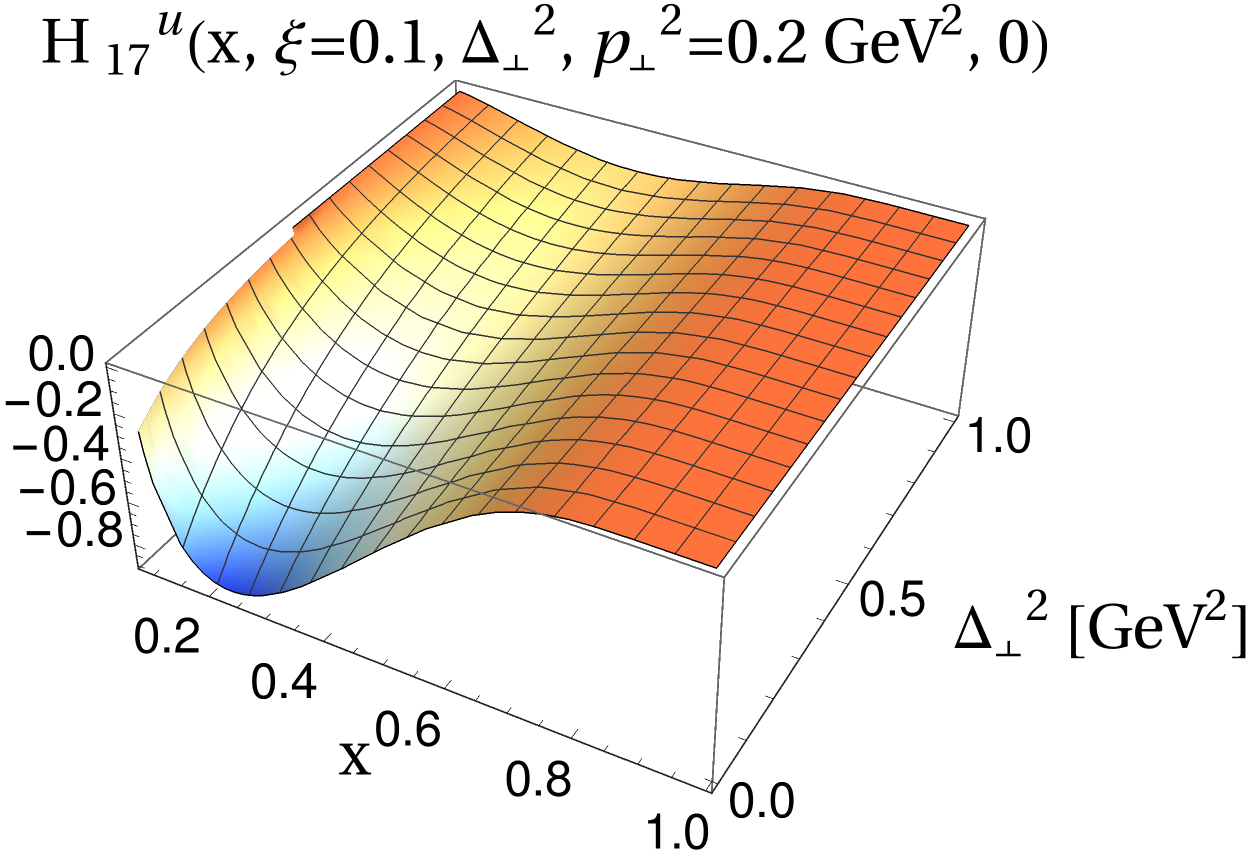} 
\includegraphics[scale=.32]{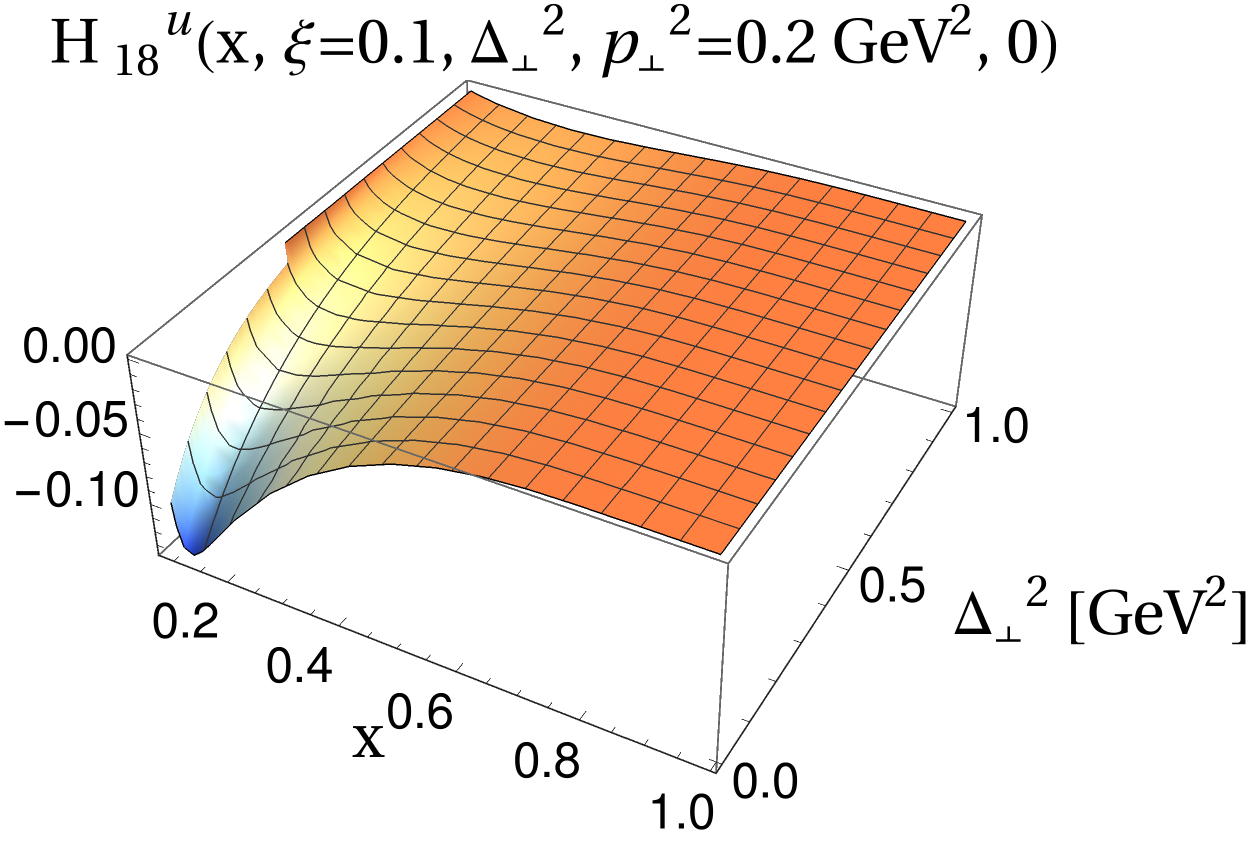} \\
\includegraphics[scale=.32]{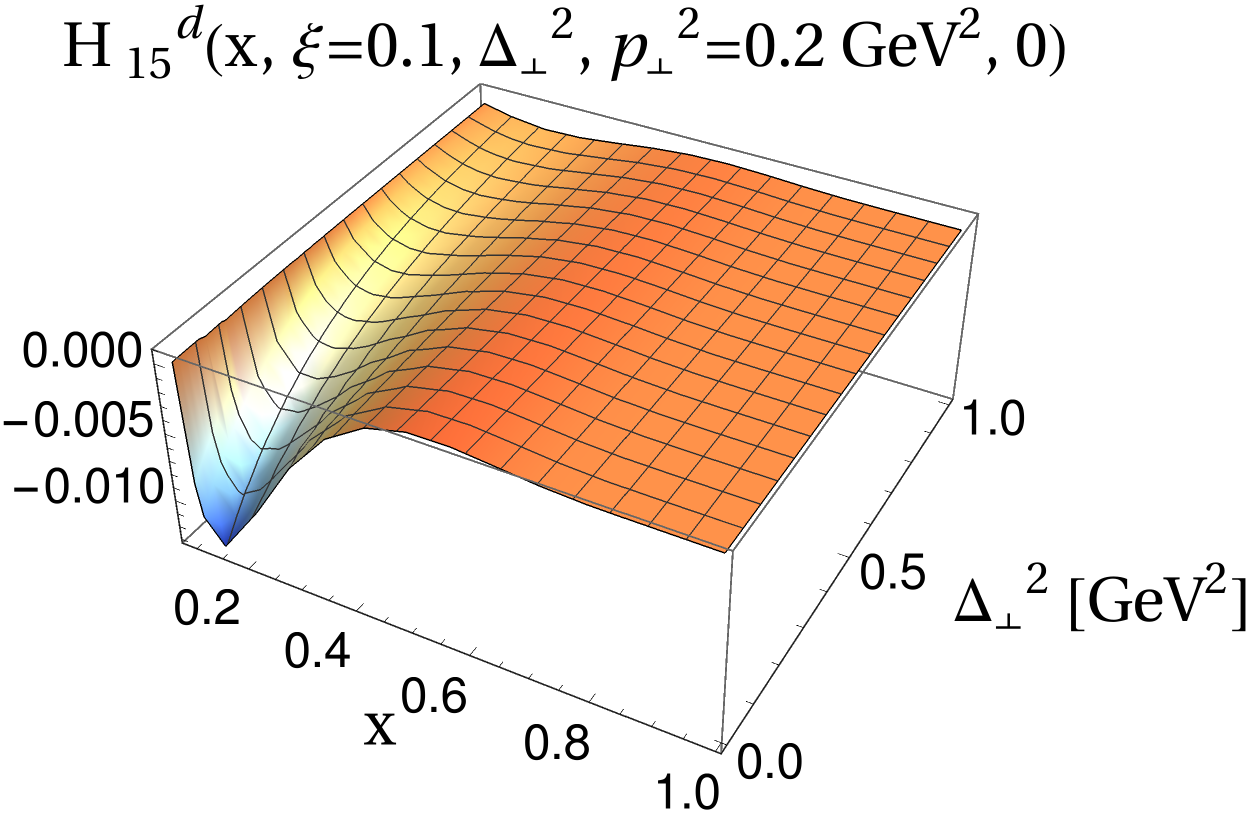} 
\includegraphics[scale=.32]{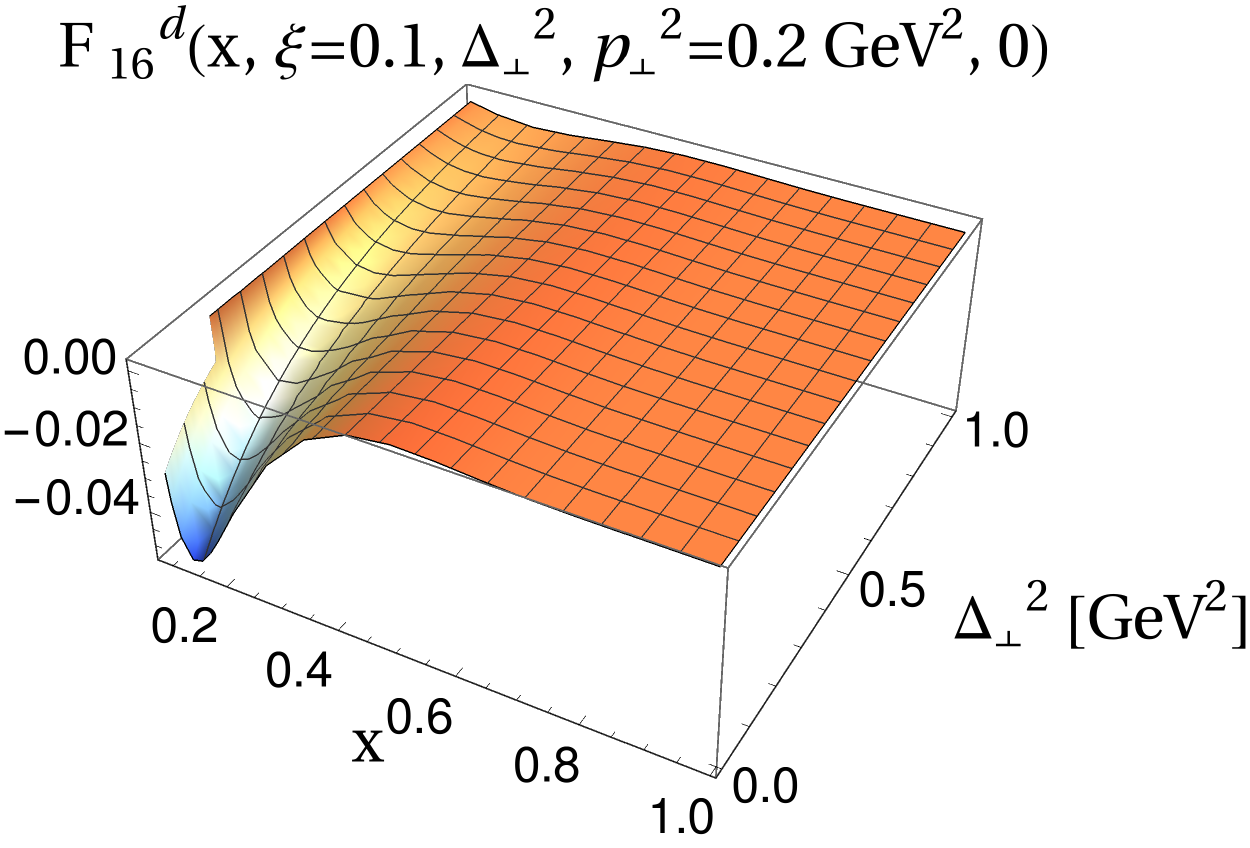} 
\includegraphics[scale=.32]{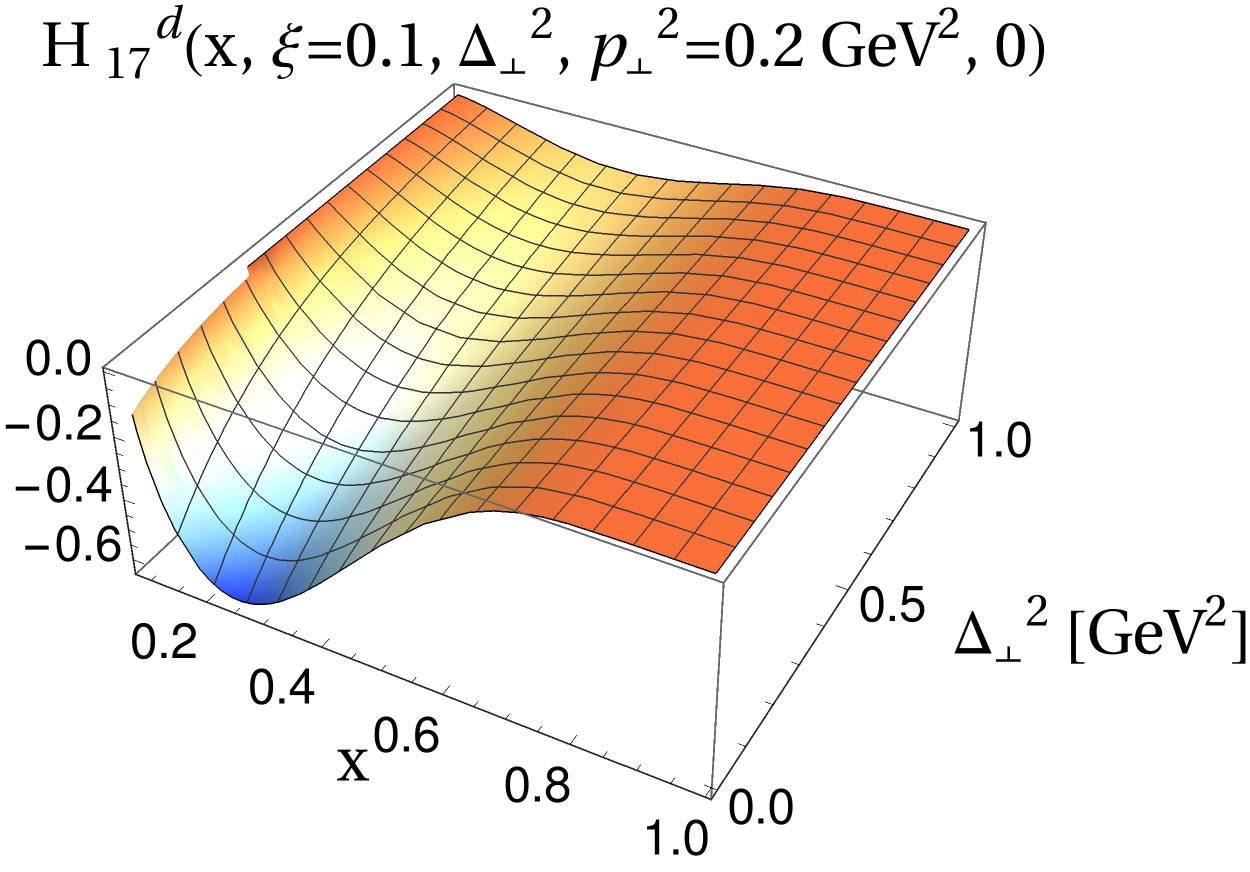} 
\includegraphics[scale=.32]{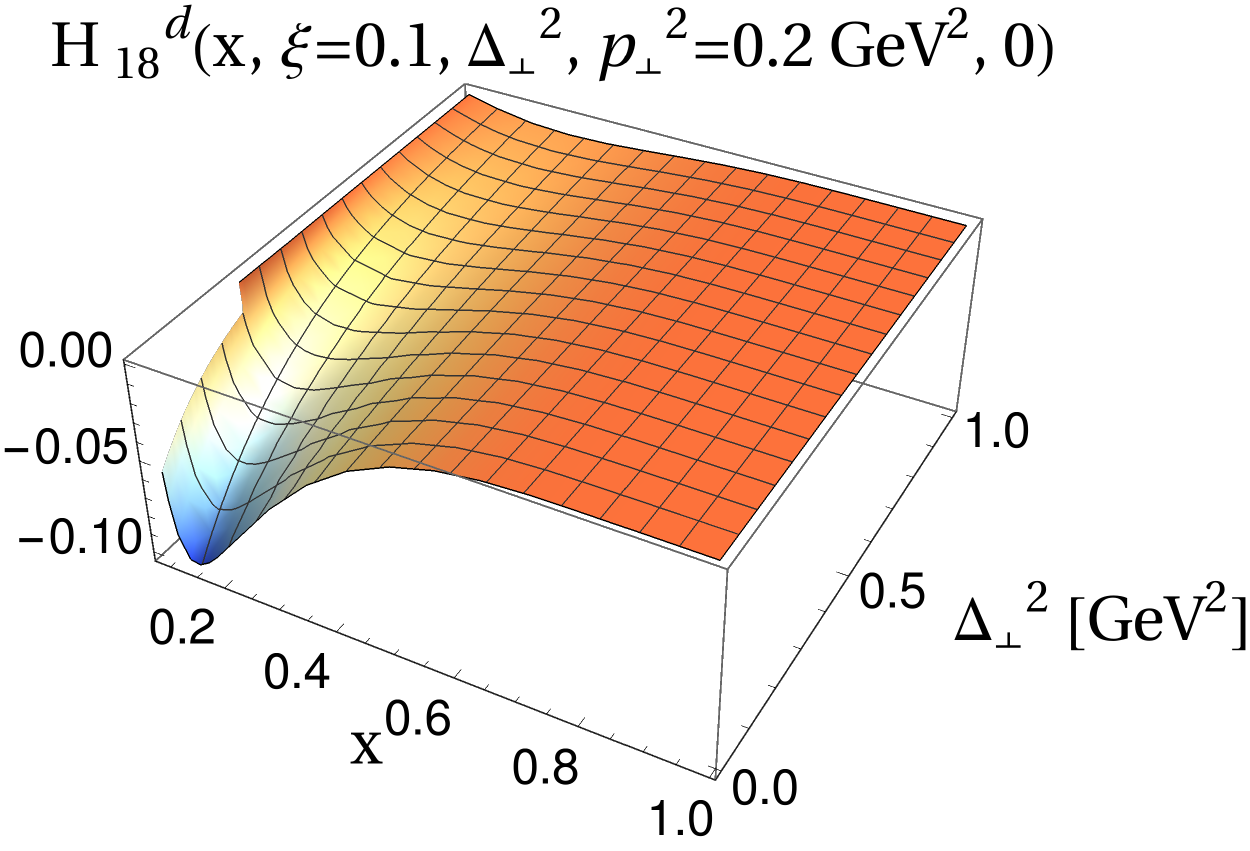} 
\caption{\label{tranxd} The leading twist GTMDs as functions of $x$ and $\bfd$ when the quark is transversely polarized. }
\end{figure}

\section{GTMDs for transversely polarized quark} \label{AppB}
The analytical expressions for the GTMDs with transversely polarized quark are given in Eqs.~(\ref{H11})--(\ref{H18}). We list the numerical results of those GTMDs as functions of $x$ and $\xi$ in Fig.~\ref{tranxz} and  as functions of $x$ and $\bfd^2$ in Fig.~\ref{tranxd}. We observe that $H_{1,1}, H_{1,2}$ show positive distributions and $H_{1,7}, H_{1,8}$ show negative distributions for both the flavors. The distributions $H_{1,3},\, H_{1,5}$, and $H_{1,6}$ are positive for  the $u$ quark and negative for the $d$ quark. Meanwhile, $H_{1,4}$ exhibits negative distribution for the $u$ quark and positive for the $d$ quark. The polarities of these GTMDs play important roll in their different combinations that contribute to the Wigner distributions $\rho_{TY}$ with $Y=U,L,T$ as discussed in Sec.~\ref{sec_WD_sigma}. The  tensor charge $g_T$ can be expressed in terms of $H_{1,3}$ and $H_{1,4}$ at $\bfd=0$ and zero skewness as $g^\nu_T = \int {\rm d}x\,{\rm d}^2\bfp\,  \left[ H^{\nu}_{1,3}(x,0,\bfp^2,0,0) + \frac{\bfp^2}{2 M^2}\,H_{1,4}^\nu(x,0,\bfp^2,0,0) \right]$. In this model the polarity flip in the distribution $H_{1,3}$ and $H_{1,4}$ against the flavors give rise to the positive tensor charge for $u$ and negative for $d$ quarks.

\begin{figure}[t]
\includegraphics[scale=.28]{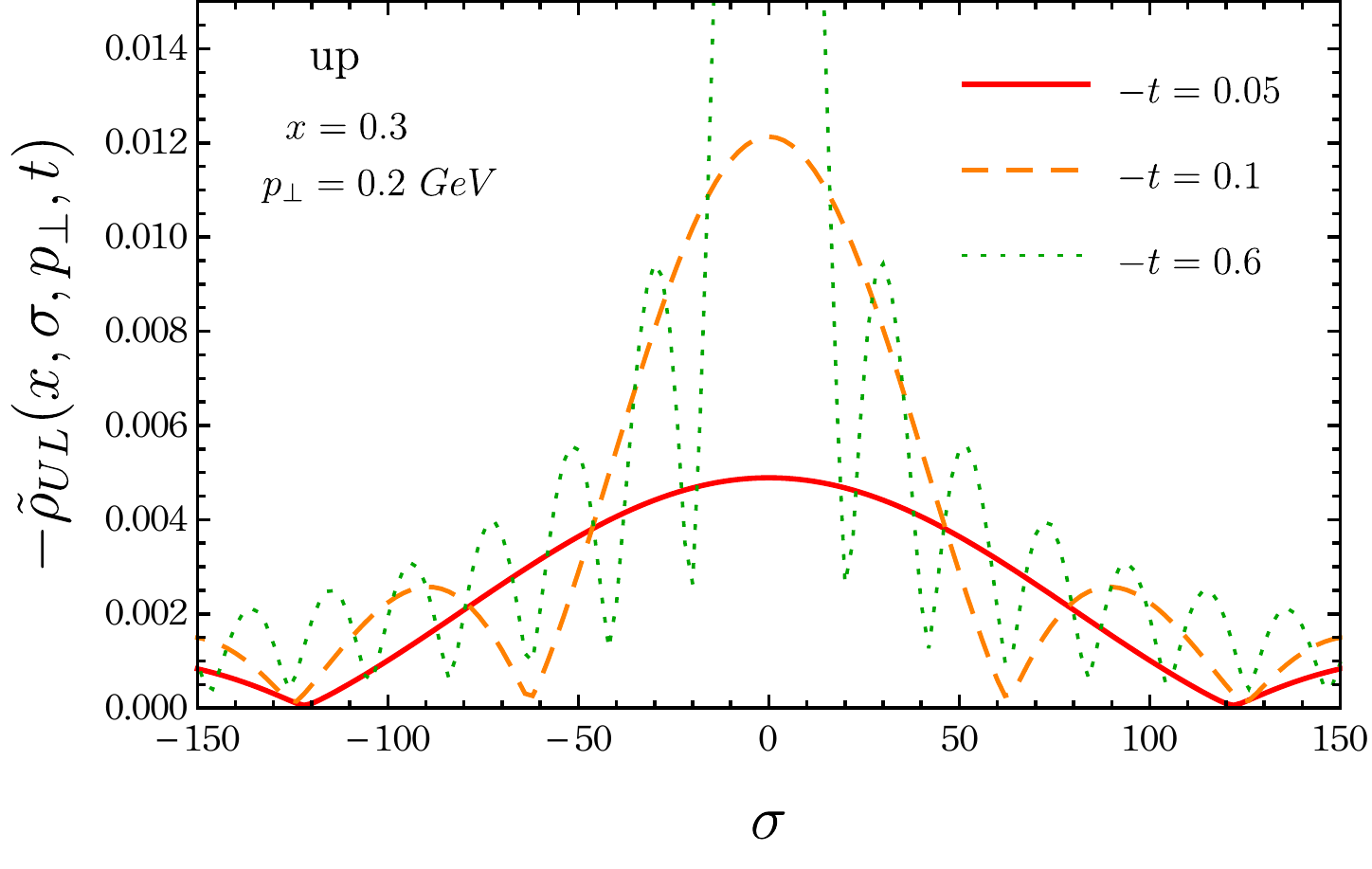} 
\includegraphics[scale=.32]{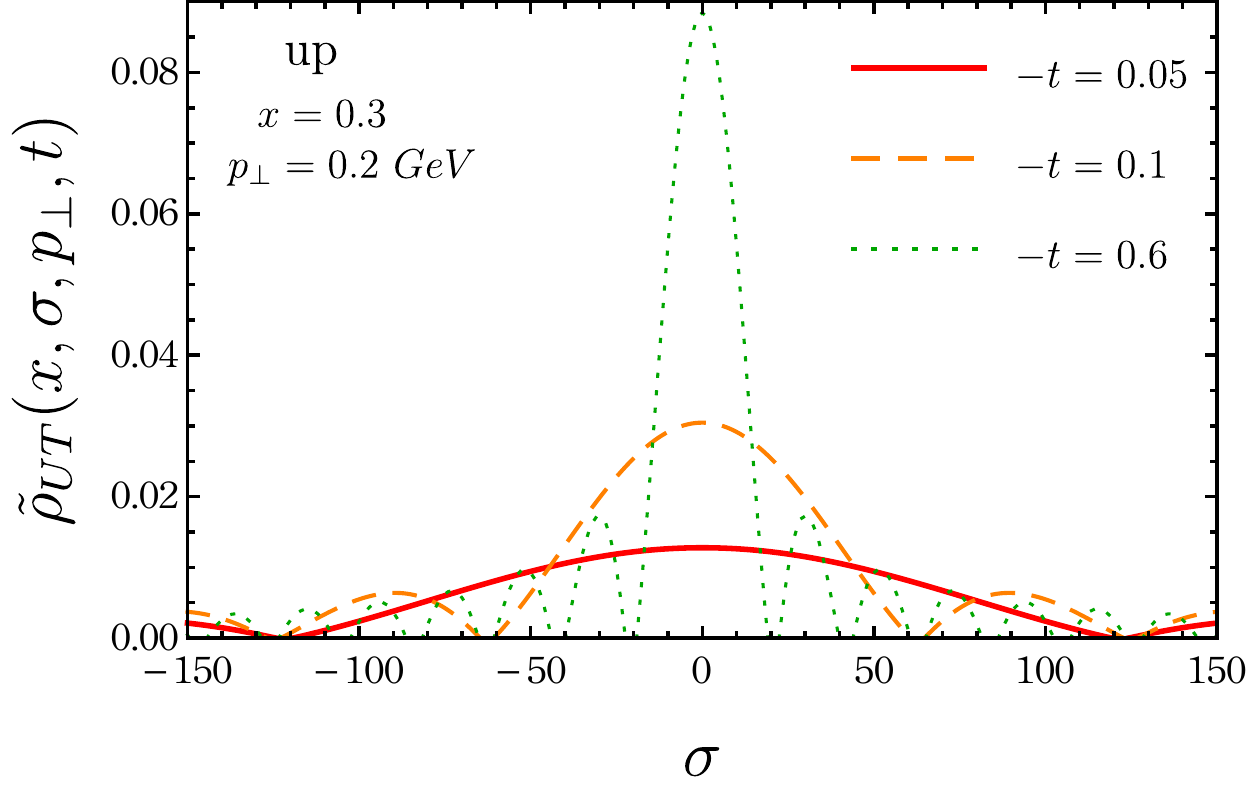} 
\includegraphics[scale=.32]{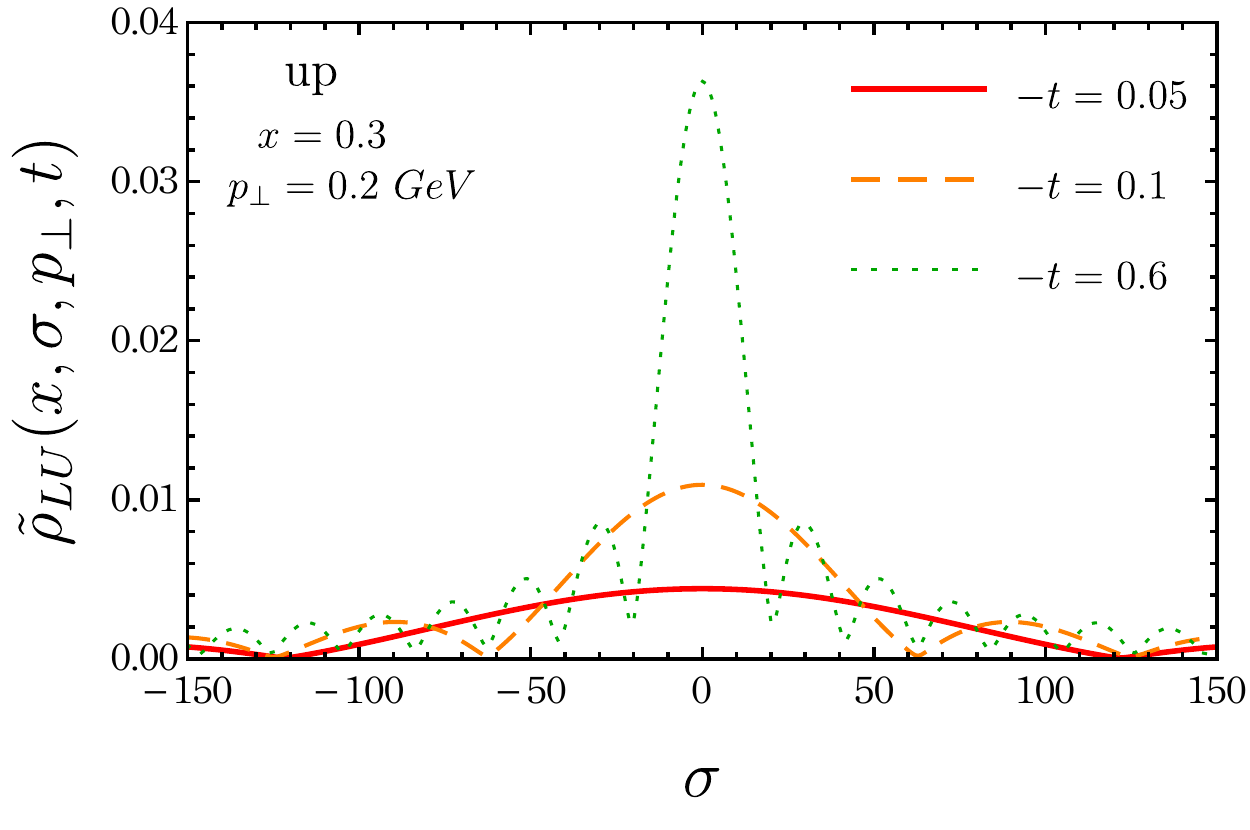} 
\includegraphics[scale=.28]{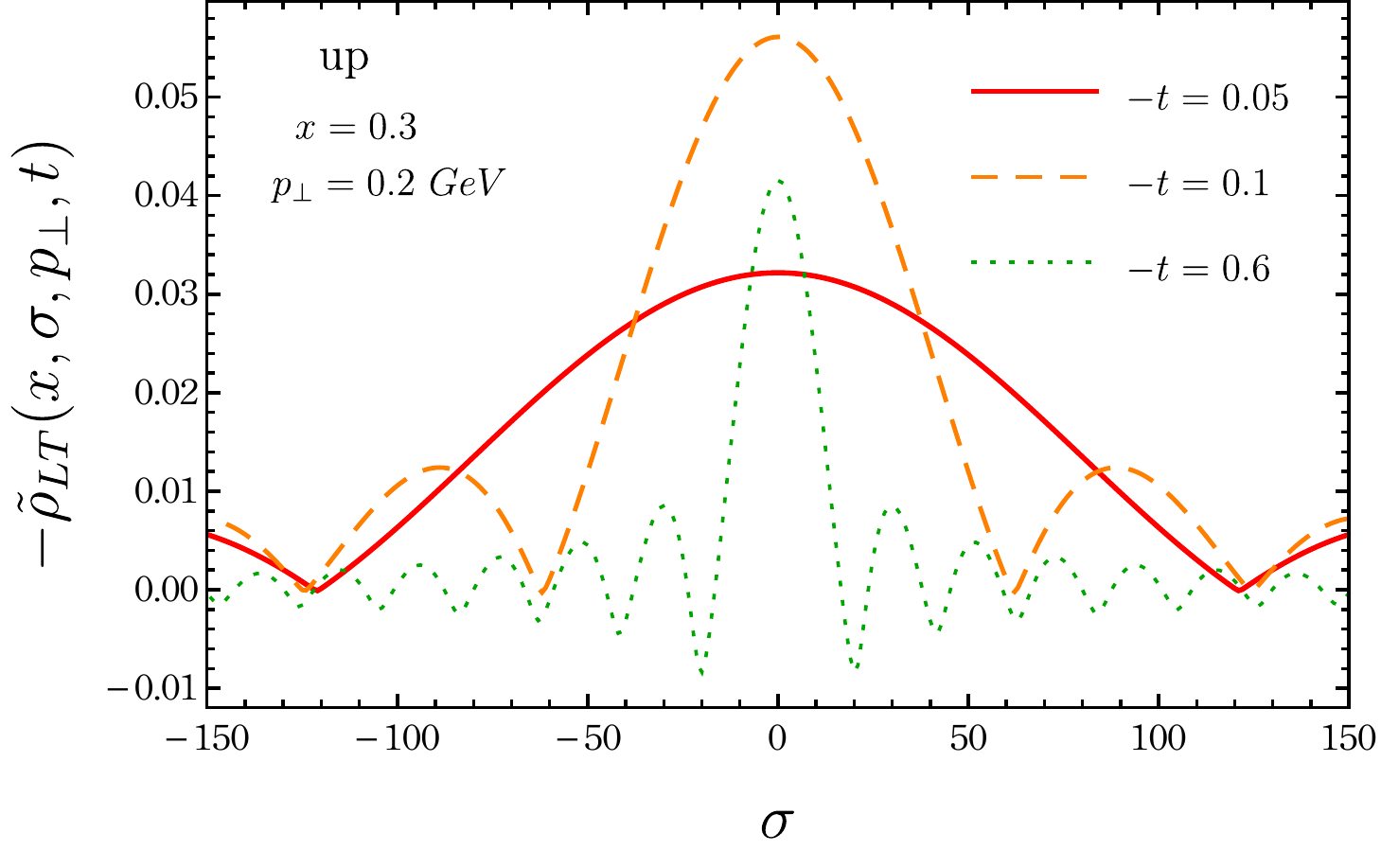} \\
\includegraphics[scale=.28]{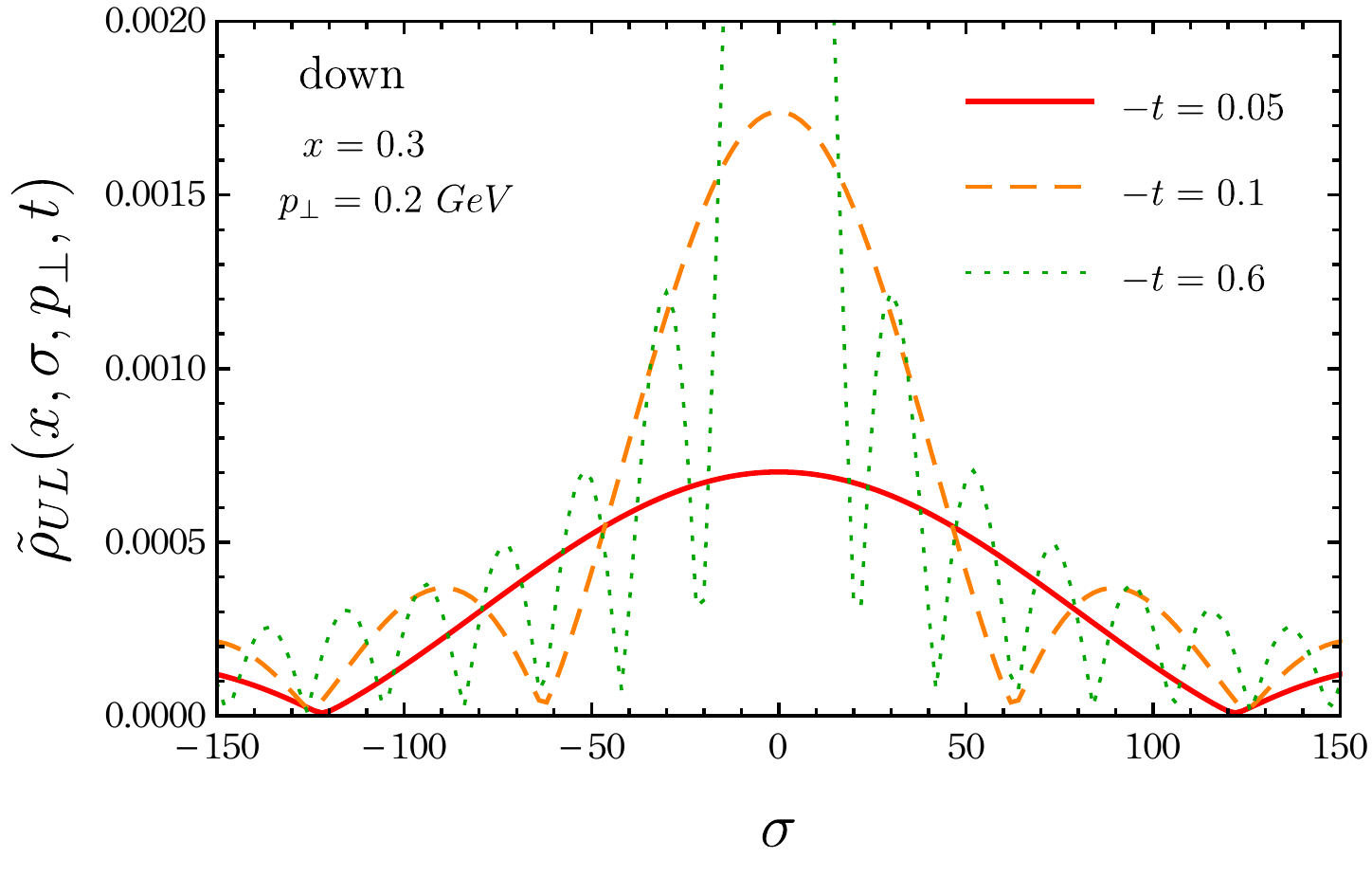} 
\includegraphics[scale=.32]{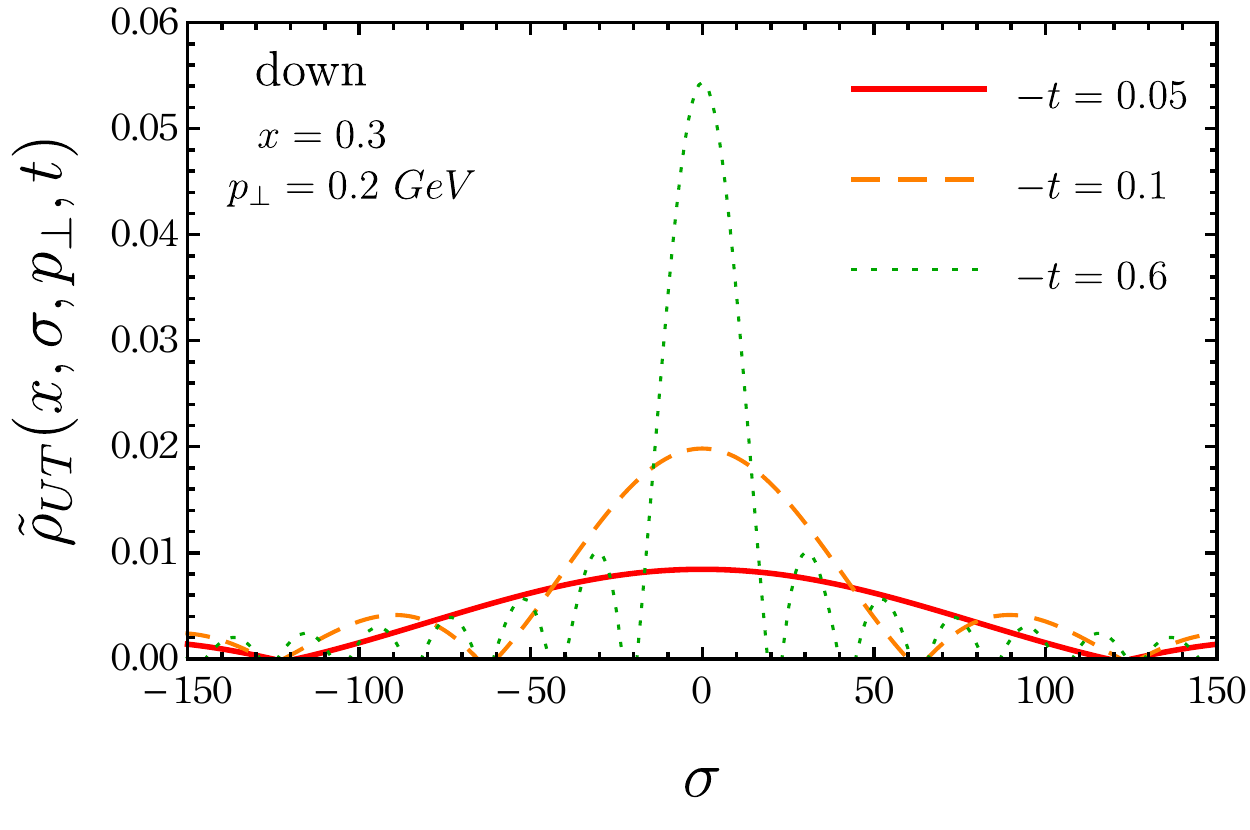} 
\includegraphics[scale=.32]{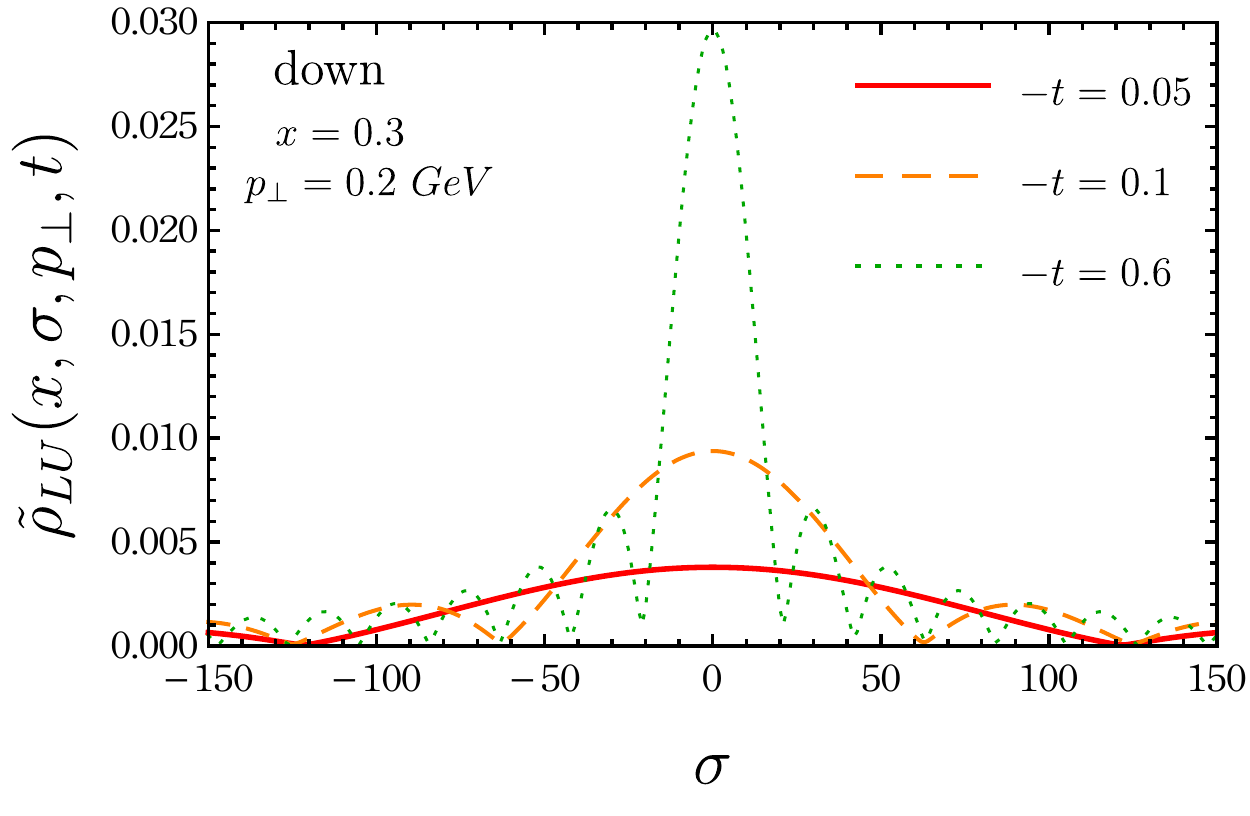} 
\includegraphics[scale=.28]{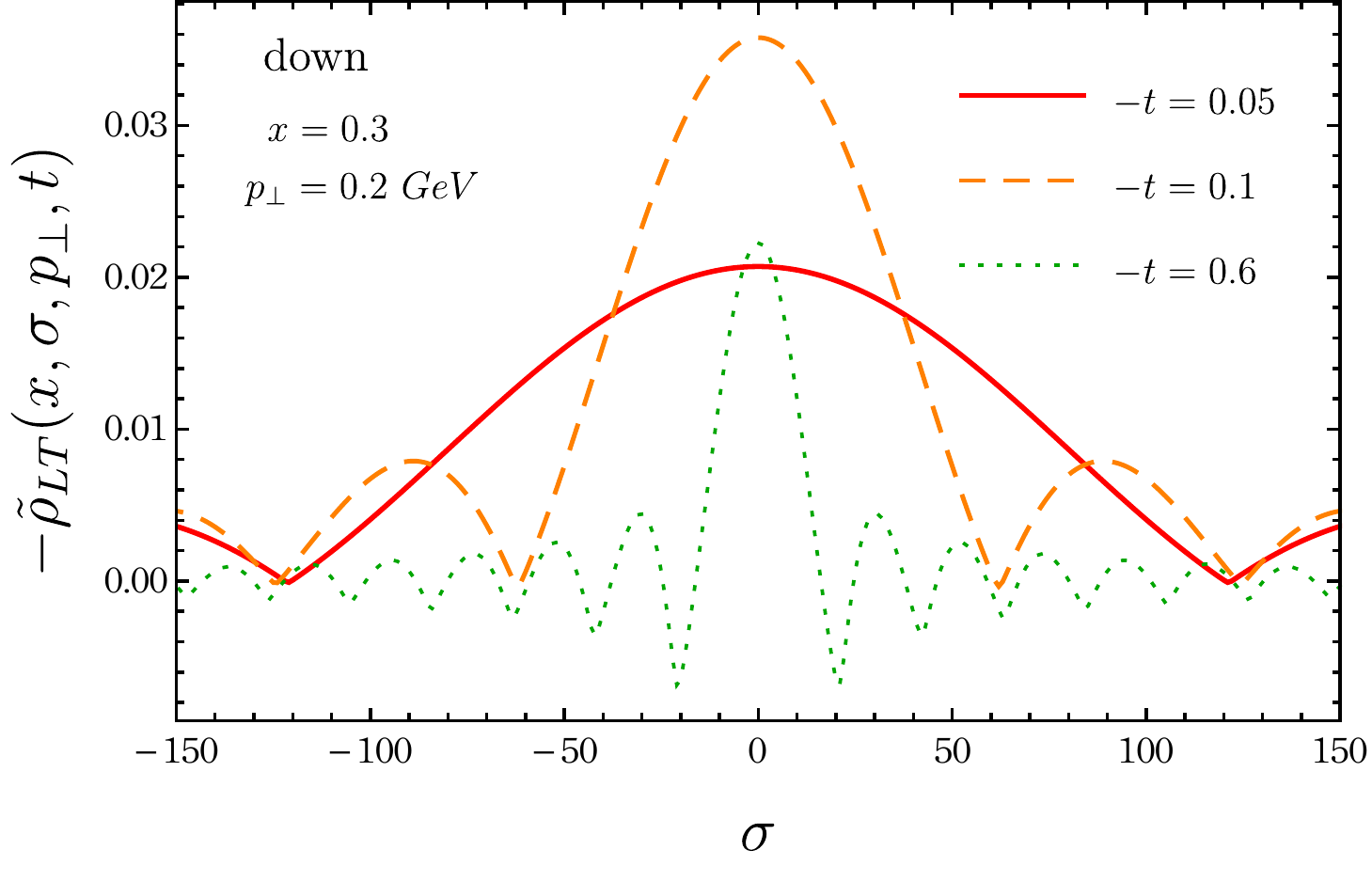} 
\caption{\label{WD_rest} The Wigner distributions $\tilde{\rho}_{UL}\,, \tilde{\rho}_{UT}\,, \tilde{\rho}_{LU}$, and  $\tilde{\rho}_{LT}$ in the boost invariant longitudinal position space. The upper panel is for the $u$ quark and the lower panel is for the $d$ quark.}
\end{figure}
\begin{figure}[t]
\includegraphics[scale=.35]{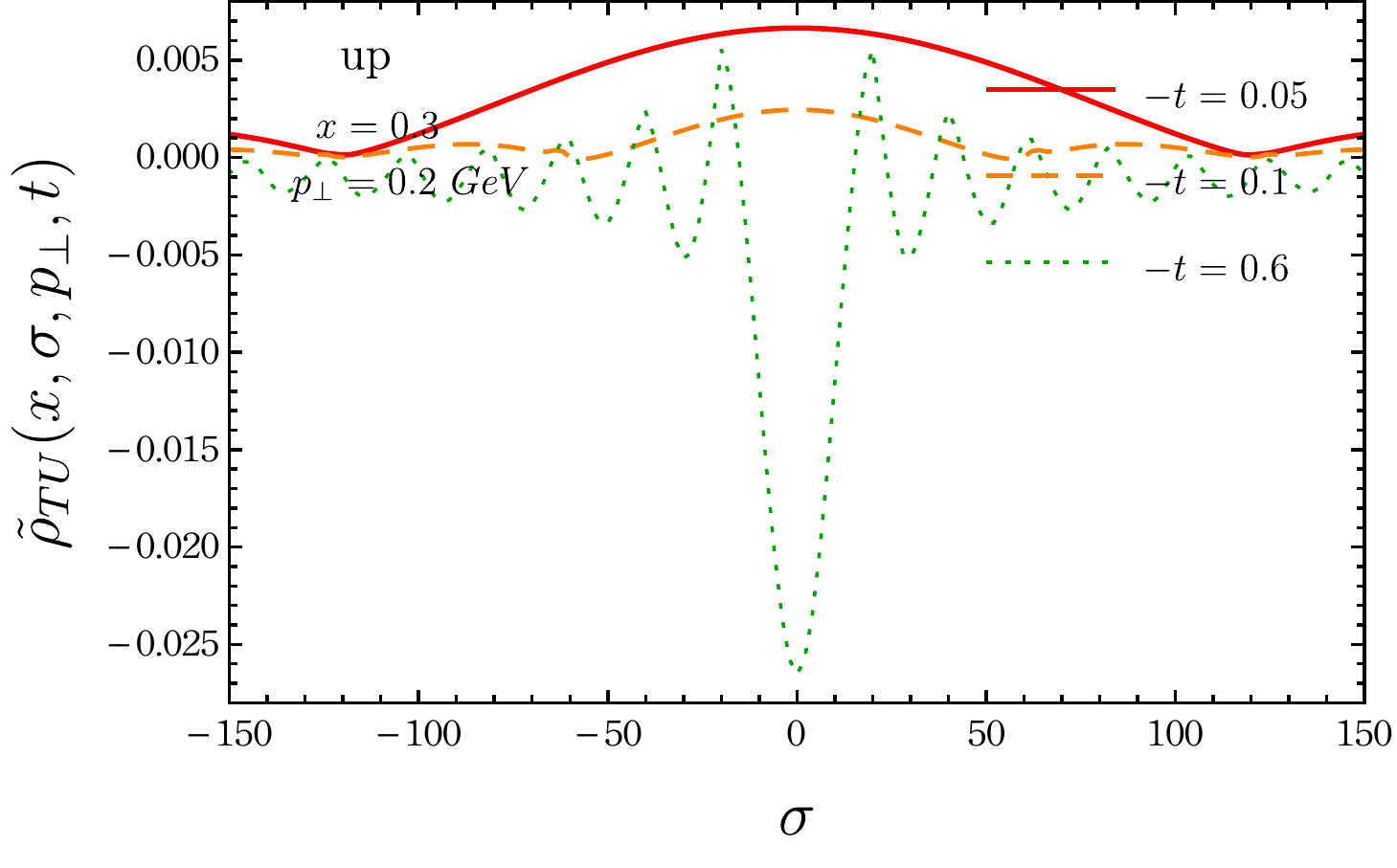} 
\includegraphics[scale=.35]{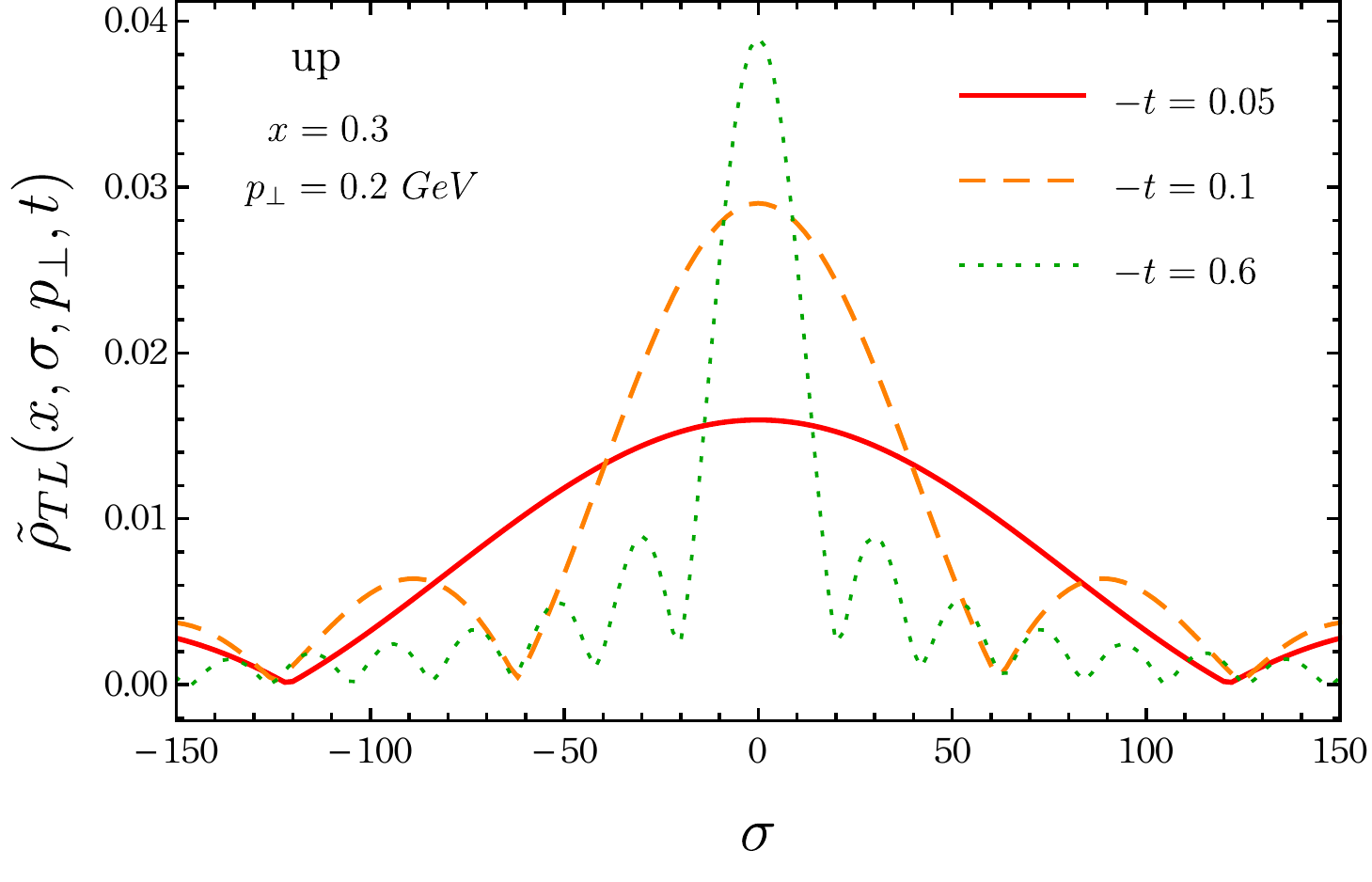} \\
\includegraphics[scale=.35]{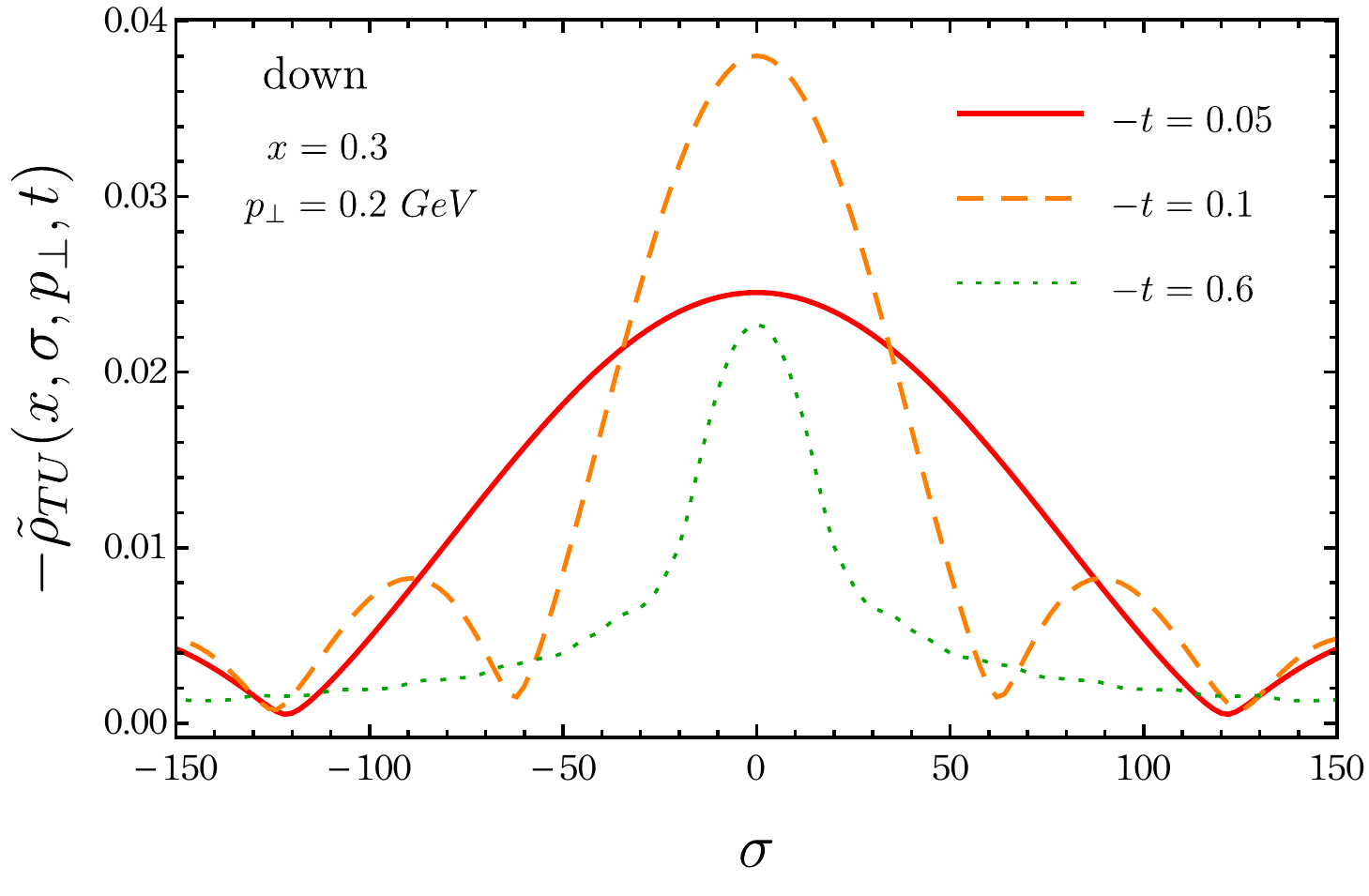} 
\includegraphics[scale=.35]{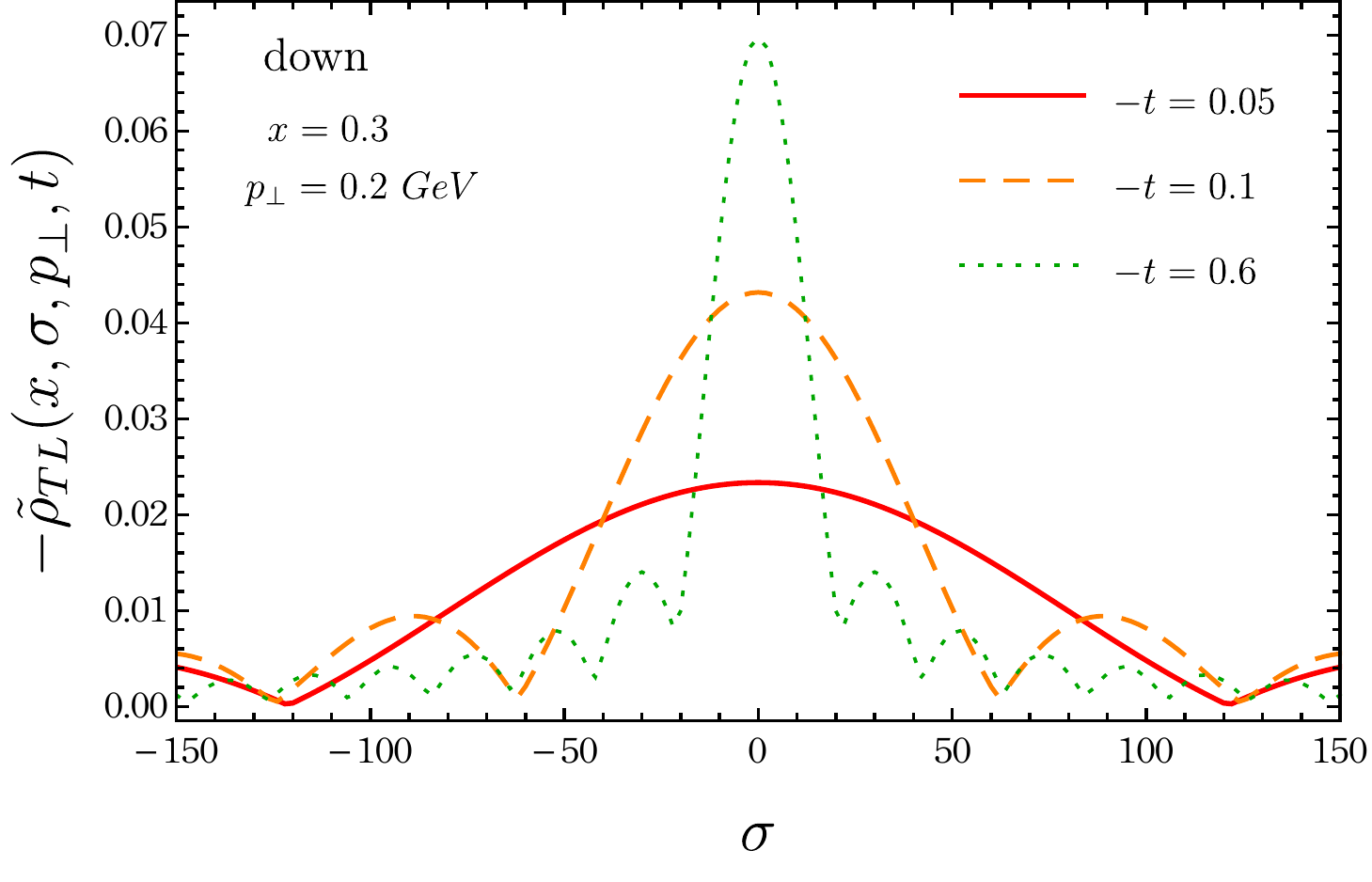} 
\caption{\label{WD_restT} The Wigner distributions $\tilde{\rho}_{TU}$ and  $\tilde{\rho}_{TL}$ in the boost invariant longitudinal position space. The upper panel is for the $u$ quark and the lower panel is for the $d$ quark.}
\end{figure}
\section{Other WDs in the $\sigma-$space} \label{AppOtherWD}
For completeness, here we present the numerical results for the Wigner distributions having the cross-polarization combinations in the longitudinal impact parameter space. The Wigner distributions $\tilde{\rho}_{UL}\,, \tilde{\rho}_{UT}\,, \tilde{\rho}_{LU}$, and  $\tilde{\rho}_{LT}$ are shown in Fig.~\ref{WD_rest}. The upper panel is for the $u$ quark, while the lower panel is for the $d$ quark. For all the distributions, we take $x=0.3$ and $|\bfp|=0.2$ GeV. In Eqs.~(\ref{rhoUL_G}) and (\ref{rhoLU_F}), the  momentum structure $\epsilon^{ij}_\perp p_\perp^i \Delta^j_\perp$ restricts us to  choose $\Delta_\perp$ along the $x$-axis, which provides the non-vanishing distributions. The transverse polarization distributions $\tilde{\rho}_{TU}$, and $\tilde{\rho}_{TL}$ are presented separately in Fig.~\ref{WD_restT}. The momentum structure of the prefactors in Eqs.~(\ref{rhoTU_F}) and  (\ref{rhoTL_G}) indicate that all the involved GTMDs survive only for the choice $\bfp \equiv \left(|\bfp|/\sqrt{2}, |\bfp|/\sqrt{2}\right)$, $\bfd \equiv \left(|\bfd|/\sqrt{2}, |\bfd|/\sqrt{2}\right)$,  and $i=1$. Except the distribution $\tilde{\rho}_{TU}$, the qualitative behavior of all other distributions is more or less very similar. For all values of $-t$, $\tilde{\rho}_{TU}$ does not show the prominent diffraction pattern. For $-t=0.6$ GeV$^2$, it shows a central minima instead of maxima for the $u$ quark, while the pattern is not eminent for the $d$ quark. 
\bibliography{WD_Zeta_v3}
\end{document}